\documentclass[apj]{emulateapj}
\slugcomment{{\sc Accepted to ApJ:} September 3, 2015} 

\pdfoutput=1 

\usepackage{graphicx}
\usepackage{epstopdf}
\usepackage{morefloats}
\usepackage{multirow}
\usepackage{lscape}
\usepackage{booktabs}
\usepackage{xfrac}

\shorttitle{VLA observations of Double-Peaked AGNs}
\shortauthors{M\"uller S\'anchez et al.}

\begin{document}


\title{The Origin of Double-Peaked Narrow Lines in Active Galactic Nuclei I: Very Large Array Detections of Dual AGNs and AGN Outflows\footnotemark[1]
}


\author{F. M\"uller-S\'anchez$^{1}$, J. Comerford$^{1}$, R. Nevin$^{1}$, S. Barrows$^{1}$, M. Cooper$^{2}$, J. Greene$^{3}$}

\affil{$^1$ Department of Astrophysical and Planetary Sciences, University of Colorado, Boulder, CO 80309, USA}


\affil{$^2$ Center for Cosmology, Department of Physics and Astronomy, University of California, Irvine, CA 92697, USA} 

\affil{$^3$ Department of Astrophysical Sciences, Princeton University, Princeton, NJ 08544, USA}



\footnotetext[1]{Based on observations at the NRAO Karl G. Jansky VLA (program 12A-103).}






\begin{abstract} 

We have examined a subset of 18 active galactic nuclei (AGNs) drawn from a sample of 81 galaxies that possess
double-peaked narrow optical emission line spectra in the Sloan Digital Sky Survey, have two optical AGN emission components separated by $>0.2\arcsec$, and are detected in the Faint Images
of the Radio Sky at Twenty-centimeters survey. Without follow-up observations, the sources of the double-peaked narrow emission lines are uncertain, and may be produced by 
kpc-scale separation dual active supermassive black holes, AGN outflows, or disk rotation. 
In this work, we propose a new methodology to characterize double-peaked narrow emission-line galaxies based on optical long-slit spectroscopy and high resolution multi-band Very Large Array observations. The nature of the radio emission in the sample galaxies is varied. Of the 18 galaxies, we detect two compact flat-spectrum radio cores with projected spatial separations on
the sky between $0.6-1.6$ kpc in three galaxies: J$1023+3243$, J$1158+3231$, and J$1623+0808$. 
The two radio sources are spatially coincident with the two optical components of ionized gas with AGN-like line ratios, 
which confirms the presence of dual AGNs in these three galaxies. 
Dual AGNs account for only $\sim15\%$ ($\sfrac{3}{18}$) of the double-peaked AGNs in our sample. Gas kinematics produce $\sim75\%$ ($\sfrac{13}{18}$) of the double-peaked narrow emission lines, distributed in the following way: $7$ AGN wind-driven outflows, $5$ radio-jet driven outflows, and one rotating narrow-line region. The remaining $10\%$ ($\sfrac{2}{18}$) are ambiguous cases. 
Our method demonstrates the power of spatially resolved spectroscopy and high resolution radio observations for the identification of AGN outflows and AGN pairs with angular separations as small as $0.18\arcsec$. 

 
\end{abstract}



\section{Introduction}\label{introduction}


Active galactic nuclei (AGNs) have been recognized as key drivers of galaxy evolution, as is illustrated in the two main processes of their life cycles: feeding and feedback. 

The stochastic accretion of gas and galaxy merger-driven gas inflows are both known triggers of supermassive black hole (SMBH) growth and nuclear activity, but the relative contributions of each is still unclear. Simulations of galaxy mergers show that they drive gas to the centers of merger-remnant galaxies (e.g., Springel et al. 2005, Hopkins \& Hernquist 2009), predicting that merger-driven SMBH mass growth occurs when the black hole nears the center of the merger-remnant. Observations have shown that the AGN fraction does increase with decreasing distance between two merging galaxies (Ellison et al. 2011, Koss et al. 2012, Ellison et al. 2013), 
but this has not been well tested at the very centers of merger-remnant galaxies because of the observational difficulty of detecting and resolving two AGNs with separations $<10$ kpc. This is known as the ``dual AGN'' phase\footnote{The separation scale expected for dual AGNs is between 0.1 to 10 kpc. The SMBHs in a merger stay at these separations for a few hundred Myr before evolving into a gravitationally bound, parsec-scale separation binary AGN system (Begelman et al. 1980).}. Hundreds of AGN pairs with $>10$ kpc separations have been discovered (Myers et al. 2008, Hennawi et al. 2010, Liu et al. 2011). However, there are only a few confirmed kpc-scale dual AGNs (Junkkarinen et al. 2001, Komossa et al. 2003, Rodriguez et al. 2006, Hudson et al. 2006, Bianchi et al. 2008, Koss et al. 2011, Fu et al. 2011a, Koss et al. 2012, Mazzarella et al. 2012, Liu et al. 2013, Comerford et al. 2015). 
Dual AGNs are an intermediate evolutionary stage between 
first encounter and final coalescence of two merging gas-rich galaxies (e.g., Comerford et al. 2009, Liu et al. 2012), in which strong tidal interactions are more likely to influence the nuclear accretion and star formation in both galaxies (Barnes \& Hernquist 1996). 
Indeed, galaxy merger simulations and observations clearly show that the dual AGN phase is the critical stage when SMBH growth and star formation activity are the most vigorous (e.g., Van Wassenhove et al. 2012, Koss et al. 2012, Blecha et al. 2013). 


In the various proposed feedback models (e.g., di Matteo et al. 2005), once the AGN is triggered, it injects energy into the surrounding interstellar medium (ISM) via jets, winds and radiation, inhibiting the build-up of massive galaxies and suppressing star formation.
AGN feedback can potentially starve the black hole, giving rise to the strong observed correlation between SMBH mass and bulge stellar velocity dispersion: the $M_{\mathrm{BH}}-\sigma^*$ relation (e.g., Ferrarese \& Merrit 2000, Greene \& Ho 2006, McConnell \& Ma 2013). On the other hand, some models also require `positive' AGN feedback, to reproduce some of the observed black hole-galaxy relations (e.g., King 2005, Ishibashi \& Fabian 2012, Silk 2013) and some cases of this type of feedback have been found (Croft et al. 2006, Evans et al. 2008). However, the real impact of AGN feedback (positive or negative) in the evolution of galaxies is still one of the main debated topics in extragalactic astronomy. 

Both processes in the lifecycle of AGNs (feeding and feedback) can be studied using AGNs that exhibit disturbed gas kinematics in the narrow-line region (NLR, which extends from a few hundreds of parsecs to $\sim30$ kpc; Schmitt et al. 2003, Hainline et al. 2013). These disturbed gas kinematics are traced by multiple velocity components in the spectral profiles of emission lines from the NLR, such as [O III] $\lambda 5007$ (hereafter [O III]). In recent years, galaxies with double-peaked narrow AGN emission lines have gained popularity  because they indicate complex dynamical processes related to feeding and feedback, and also because they can be used as targets in dual AGN searches. 

Chronologically, double-peaked [O III] line profiles were first seen in early NLR studies (e.g., Sargent 1972, Heckman et al. 1981, Veilleux 1991), the most famous cases being Mrk 78 and NGC 1068, in which the NLR has a bipolar outflowing structure, and is apparently aligned with the radio jet (Crenshaw \& Kraemer 2000, Whittle et al. 2005, Fischer et al. 2011). Spatially resolved studies with the Hubble Space Telescope ($HST$) and Adaptive Optics-assisted integral-field spectrographs have revealed more such cases, where the NLR is dominated by outflows, and double-peaked line profiles arise in spatially-integrated spectra (e.g., Crenshaw et al. 2010, M\"uller-S\'anchez et al. 2011). In addition, some NLRs exhibit a rotational component (gas ionized by the AGN rotating in the galaxy disk; see e.g., Greene \& Ho 2005, M\"uller-S\'anchez et al. 2011), and this type of motion can also produce double-peaked narrow emission lines\footnote{Several physical processes can produce double peaks in spatially-integrated spectra of rotating NLRs. 
These include partial obscuration of a simple rotating disk, anisotropic illumination of the disk, and dynamical processes such as disk instabilities (e.g., bars and spiral patterns) or merger-induced gas inflows. Since rotation is the dominant kinematic component in all these cases, we use the terms rotating disk, NLR rotation or disk rotation interchangeably throughout the paper. 
We do not consider [O III] peaks caused by star-forming rings/nuclear star clusters since the optical line ratios of the galaxies in our sample are consistent with AGN-dominated emission rather than star formation (Wang et al. 2009, Liu et al. 2010a).}
 (Shen et al. 2011, Smith et al. 2012).    

On the other hand, systematic searches for dual AGNs have focused on AGNs with two peaks of [O III] emission in spatially integrated spectra (Comerford et al. 2009, Wang et al. 2009, Liu et al. 2010a, Smith et al. 2010, Ge et al. 2012, Comerford et al. 2013, Shi et al. 2014). The working hypothesis is similar to the case of spectroscopic binary stars. The two velocity components of the [O III] line originate from the orbital motion of two AGNs along with their own NLR gas on kpc-scales. 
These studies have yielded hundreds of candidates. Yet, very few of these candidates have been confirmed to host two actively accreting SMBHs, 
because with the optical data alone, it is difficult to rule out outflows or rotating disks. In fact, subsequent high-resolution imaging and spatially resolved spectroscopy have found that most of these double-peaked AGNs are produced by NLR kinematics (Shen et al. 2011, Fu et al. 2011b).

The origin of the double-peaked emission lines remains unclear, with the following three main possibilities: bipolar outflows, rotating disks, and dual AGNs. 
Robust observational evidence is needed to distinguish between these three scenarios.  
High-resolution radio or X-ray observations can provide unambiguous evidence for actively accreting SMBHs. In fact, all of the confirmations of kpc-scale dual AGNs mentioned above have been produced either through radio or X-ray detections, calling for systematic studies at these wavelengths of AGNs with double-peaked optical narrow emission-line spectra. 

Very few studies to date have undertaken spatially resolved radio observations of double-peaked AGNs. 
Three notable examples are the studies of SDSS J$151709+335324$ by Rosario et al. (2010), SDSS J$150243+111557$ by Fu et al. (2011a), and 3C 316 by An et al. (2013). While SDSS J$151709+ 335324$ and 3C 316 have a radio jet that suggests that a radio jet-driven outflow produces the double-peaked [O III] lines (Rosario et al. 2010, An et al. 2013), Very Large Array (VLA, Perley et al. 2011) observations of SDSS J$150243+111557$ revealed two radio cores that confirmed it as a dual AGN system (Fu et al. 2011a). Interestingly, follow-up Very Long Baseline Interferometry (VLBI) observations of this galaxy revealed two marginally resolved flat-spectrum radio sources within the spatially unresolved VLA component J1502S. These were interpreted as binary SMBHs, suggesting that SDSS J$150243+111557$ is in fact a triple SMBH system (Deane et al. 2014). However, recent Very Long Baseline Array (VLBA) observations with three times better spatial resolution challenge this interpretation and suggest the presence of an obscured AGN with two symmetric hotspots and an unobscured AGN (Wrobel et al. 2014b). However, the flat radio spectra of the sources are rather unusual for symmetric hotspots or jet knots. 
Further, VLBA observations of 11 Type 2 Seyferts that have double-peaked [O III] lines as well as Faint Images of the Radio Sky at Twenty-cm (FIRST, White et al. 1997) detections reveal compact radio emission in only two galaxies (one of them is  SDSS J151709 + 335324; Tingay \& Wayth 2011). Both objects show only single components of compact radio emission at 1.4 GHz and average rms image noise of 0.5 mJy beam$^{-1}$, giving a $3\sigma$ detection threshold of 1.5 mJy beam$^{-1}$. However, the sensitivity of these observations is not as good as what is possible now with the VLA ($\sim10$ $\mu$Jy, see below).  
Also, they note the need for sensitive spatially resolved imaging at radio wavelengths for a larger sample of double-peaked AGNs (see also  Frey et al. 2012, Wrobel et al. 2014a and Gabanyi et al. 2014). 
In this work we use, for the first time together, optical long-slit spectroscopy and high resolution ($\sim0.2\arcsec$) VLA multi-band observations in the range $8-12$ GHz to further clarify the ambiguities concerning the nature of double-peaked narrow emission-line galaxies\footnote{Recently, Fu et al. (2015) 
combined radio images at 1.4 GHz with SDSS spectroscopy (and in some cases with long-slit spectroscopy) to determine the nature of the radio sources in the VLA Stripe 82 Survey. 
This study, however, is not focused on the nature of double-peaked AGNs. Furthermore, the VLA survey cannot identify double nuclei with angular separations $<1.8\arcsec$ and cannot confirm dual AGNs through radio spectral analysis since the survey provides data at only one frequency.}. 
The newly-enhanced sensitivity of the VLA allowed us to obtain average rms image noises of $0.01$ mJy beam$^{-1}$ in 18 double-peaked narrow emission-line galaxies from the Sloan Digital Sky Survey (SDSS).  
An unambiguous confirmation of dual AGNs separated by a few kpc is dual compact radio cores as revealed by high-resolution radio imaging. Compact cores with flat spectra would not only directly indicate the existence of SMBHs, but also easily allow the measurement of their projected physical separation. On the other hand, a single radio core with extended emission that is aligned with the emission-line gas visible in the long-slit spectra would provide confirmation of a radio jet-driven AGN outflow. Finally, if we detect a single flat-spectrum radio core and no steep-spectrum jets (or the jets are not aligned with the ionized gas emission), then the double-peaked AGN signature is produced either by an AGN outflow that is not radio jet-driven or by a rotating NLR (or by a radio structure that is smaller than our resolution or fainter than the detection limit in our images, 
see Section~\ref{step1}). In this case the long-slit spectra can be used to distinguish between these scenarios, since ionized gas emission from a disk would be located in the plane of the galaxy and trace the galactic rotation curve. 
This paper is structured as follows. In Section 2, we describe the selection criteria and the resulting sample of radio targets, the VLA observations that were undertaken, and the data processing. Here we also include a reanalysis of published optical long-slit spectroscopy for the estimation of the size of the NLR. 
In Section 3 we combine the new radio observations with our previous long-slit optical spectroscopy, and discuss the origin of the double-peaked narrow AGN emission lines in each galaxy (dual AGNs, AGN outflows, or disk rotation). 
We have also investigated the size of the [O III] emitting region for each of the double-peaked AGNs in our sample and how the size of the NLR scales with the luminosity of the central source. The implications for each of the scenarios that give rise to the double-peaked narrow lines are discussed in Section~\ref{size}. In addition to the classifications presented in Section 3, some galaxies show interesting radio morphologies that could be indicative of small separation binary AGNs. These cases are discussed in Section~\ref{cso}. The properties of the dual AGNs identified in systematic searches of double-peaked AGNs are discussed in Section~\ref{mergers}. 
Finally, we discuss the attributes and limitations of the radio approach compared to optical and X-ray searches for the confirmation of dual AGNs in Section~\ref{limitations}, and summarize our conclusions in Section~\ref{conclusions}. 

Throughout this paper, we assume a concordance cosmology with $\Omega_m=0.3$, $\Omega_\Lambda=0.7$, and $H_0= 70$ km s$^{-1}$ Mpc$^{-1}$.



\begin{table*}
\caption[List of Double-Peaked AGNs Observed with the VLA]{List of Double-Peaked AGNs Observed with the VLA}
\begin{center}
{\small
\begin{tabular}{c c c c c c c}
\hline
\hline \noalign{\smallskip}
 Source & R.A. (J2000) & Dec. (J2000) & $z$ & kpc / $\arcsec$ &  Flux cal. & Phase cal. \\
 \hline
SDSS J$000249+004504$ & $00\,02\,49.07$ & $+00\,45\,04.8$ & $0.0868$ & 1.62 & 3C48 & J$0016-0015$ \\
SDSS J$000911-003654$ & $00\,09\,11.58$ & $-00\,36\,54.7$ & $0.0733$ & 1.39 & 3C48 & J$0016-0015$ \\
SDSS J$073117+452803$ & $07\,31\,17.55$ & $+45\,28\,03.2$ & $0.0835$ & 1.57 & 3C138 & J$0735+4750$ \\
SDSS J$073656+475946$ & $07\,36\,56.47$ & $+47\,59\,46.8$ & $0.0962$ & 1.78 & 3C138 & J$0735+4750$ \\
SDSS J$080218+304622$ & $08\,02\,18.65$ & $+30\,46\,22.7$ & $0.0766$ & 1.45 & 3C138 & J$0751+3313$ \\ 
SDSS J$084624+425842$ & $08\,46\,24.51$ & $+42\,58\,42.8$ & $0.2192$ & 3.54 & 3C138 & J$0836+4125$ \\
SDSS J$085841+104122$ & $08\,58\,41.76$ & $+10\,41\,22.1$ & $0.1480$ & 2.58 & 3C138 & J$0839+0319$ \\
SDSS J$091646+283526$ & $09\,16\,46.03$ & $+28\,35\,26.7$ & $0.1423$ & 2.50 & 3C138 & J$0915+2933$ \\ 
SDSS J$093024+343057$ & $09\,30\,24.84$ & $+34\,30\,57.3$ & $0.0610$ & 1.18 & 3C286 & J$0927+3902$ \\
SDSS J$102325+324348$ & $10\,23\,25.57$ & $+32\,43\,48.4$ & $0.1271$ & 2.27 & 3C286 & J$1022+3041$ \\ 
SDSS J$102727+305902$ & $10\,27\,27.90$ & $+30\,59\,02.4$ & $0.1245$ & 2.23 & 3C286 & J$1022+3041$ \\ 
SDSS J$111201+275053$ & $11\,12\,01.78$ & $+27\,50\,53.8$ & $0.0472$ & 0.93 & 3C286 & J$1150+2417$ \\
SDSS J$115249+190300$ & $11\,52\,49.33$ & $+19\,03\,00.3$ & $0.0967$ & 1.79 & 3C286 & J$1150+2417$ \\ 
SDSS J$115822+323102$ & $11\,58\,22.58$ & $+32\,31\,02.2$ & $0.1658$ & 2.84 & 3C286 & J$1150+2417$ \\ 
SDSS J$155619+094855$ & $15\,56\,19.30$ & $+09\,48\,55.6$ & $0.0678$ & 1.30 & 3C286 & J$1608+1029$ \\ 
SDSS J$162345+080851$ & $16\,23\,45.20$ & $+08\,08\,51.1$ & $0.1992$ & 3.30 & 3C286 & J$1608+1029$ \\ 
SDSS J$171544+600835$ & $17\,15\,44.05$ & $+60\,08\,35.7$ & $0.1569$ & 2.71 & 3C286 & J$1657+5705$ \\ 
SDSS J$225420-005134$ & $22\,54\,20.99$ & $-00\,51\,34.1$ & $0.0795$ & 1.50 & 3C48 & J$2257+0243$ \\
\hline
\hline
\end{tabular}
}
\end{center}
\label{table1}
\end{table*}

\begin{figure*}
\epsscale{.99}
\plotone{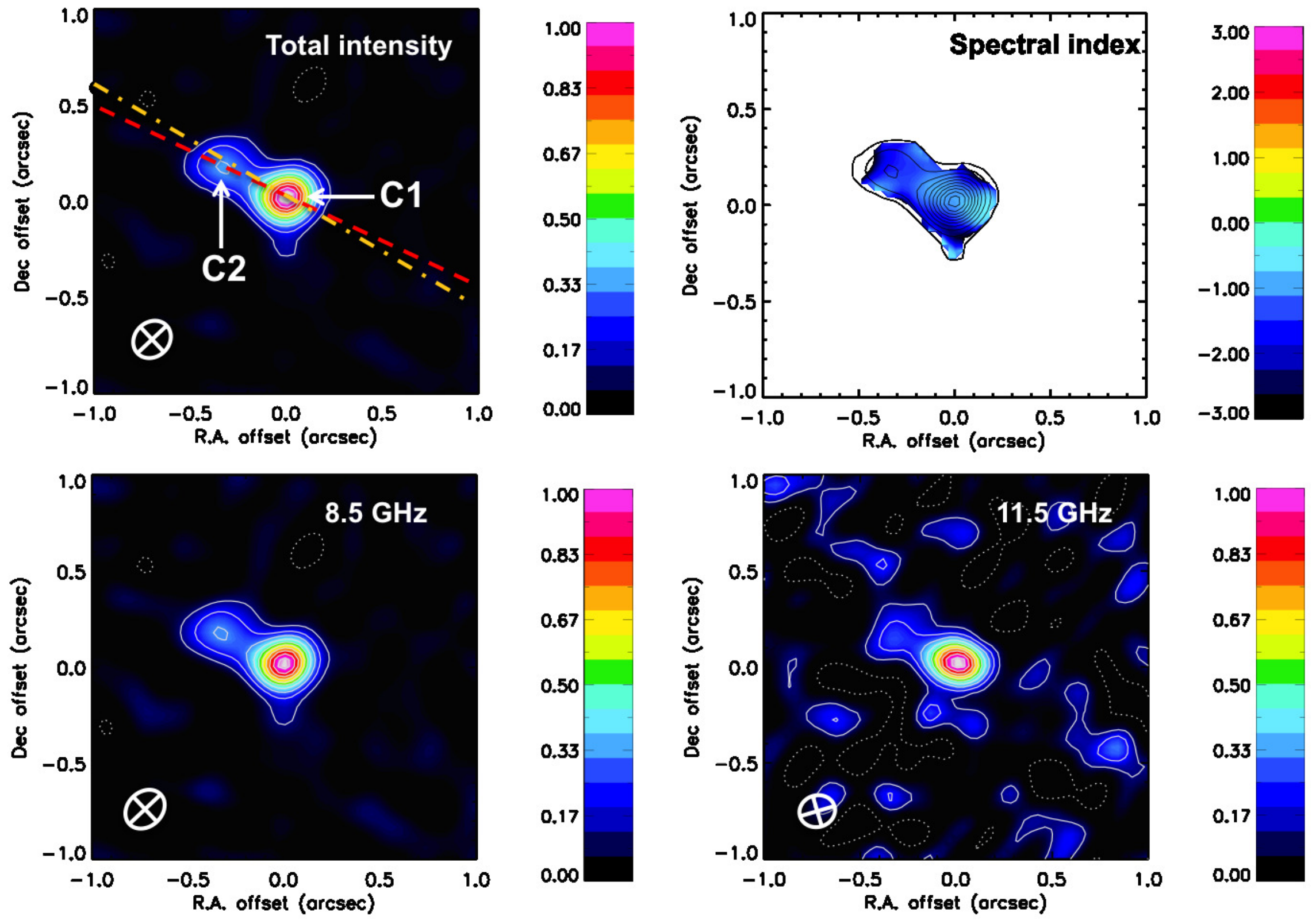}
\caption{VLA images with contours of J$0002+0045$ and spectral index image. The contours are set at $-10$, $+10$, $+20$, $+30$, $+40$, $+50$, $+60$, $+70$, $+80$, and $+90\%$ of the peak flux density. This corresponds to 45 times the rms noise in the total intensity image (which has a value of $6.7\,\mu$Jy beam$^{-1}$), 44 times the rms at 8.5 GHz, and 17 times the rms at 11.5 GHz (native spatial resolution, Table 5). The dotted white lines represent negative contours and the solid white lines positive contours. The restoring VLA beam is the crossed ellipse shown on the lower left in each image. The images are centered at the location of the strongest peak of emission in the total intensity image. The red dashed line indicates the position angle of the photometric major axis of the galaxy, and the yellow dotted-dashed line the position angle of the two peaks of [O III] emission in the long-slit spectra. For the spectral index image, the images were masked at $10\%$ of the peak of emission in the total intensity image, and only positive contours are shown. The origin of the double-peaked narrow AGN emission lines in this galaxy is ambiguous. All images are oriented so that North is up and East is to the left.
\label{fig1}}
\end{figure*}


\section{Sample and Observations}\label{sect2}

\subsection{The Sample}\label{sample}

Comerford et al. (2012) present long-slit spectroscopy obtained at the Lick, Palomar, and MMT observatories of a subsample of 81 SDSS galaxies with double-peaked narrow AGN emission lines at $0.03<z<0.36$ and $14.5<r<18.5$. These galaxies were selected from a parent sample of 340 double-peaked AGNs derived from the three catalogs that have identified active galaxies in SDSS with double-peaked [O III] emission lines: Wang et al. (2009), Liu et al. (2010a), and Smith et al. (2010). 

From the 81 galaxies presented in Comerford et al. (2012), we have chosen as targets the full sample of 18 SDSS galaxies with double-peaked narrow emission lines that have (1) VLA FIRST detections, indicating they should be detectable with the VLA; and (2) optical long-slit spectra that exhibit two relatively compact AGN emission components with angular separations of at least $0.2\arcsec$, so that the VLA A-configuration at frequencies $\geq8$ GHz can resolve double radio cores spatially without difficulty. All galaxies are Type 2 AGNs with redshifts between $0.04<z<0.22$ (Table 1). 
The measured angular separations of the two optical emission components on the sky
for the sample are $0.21\arcsec<D_{\mathrm{[O III]}}<0.87\arcsec$, which correspond to expected physical separations of $0.35 - 2.3$ kpc. 
Details on the long-slit observations and data reduction can be found in Comerford et al. (2012). 

\begin{figure*}
\epsscale{.99}
\plotone{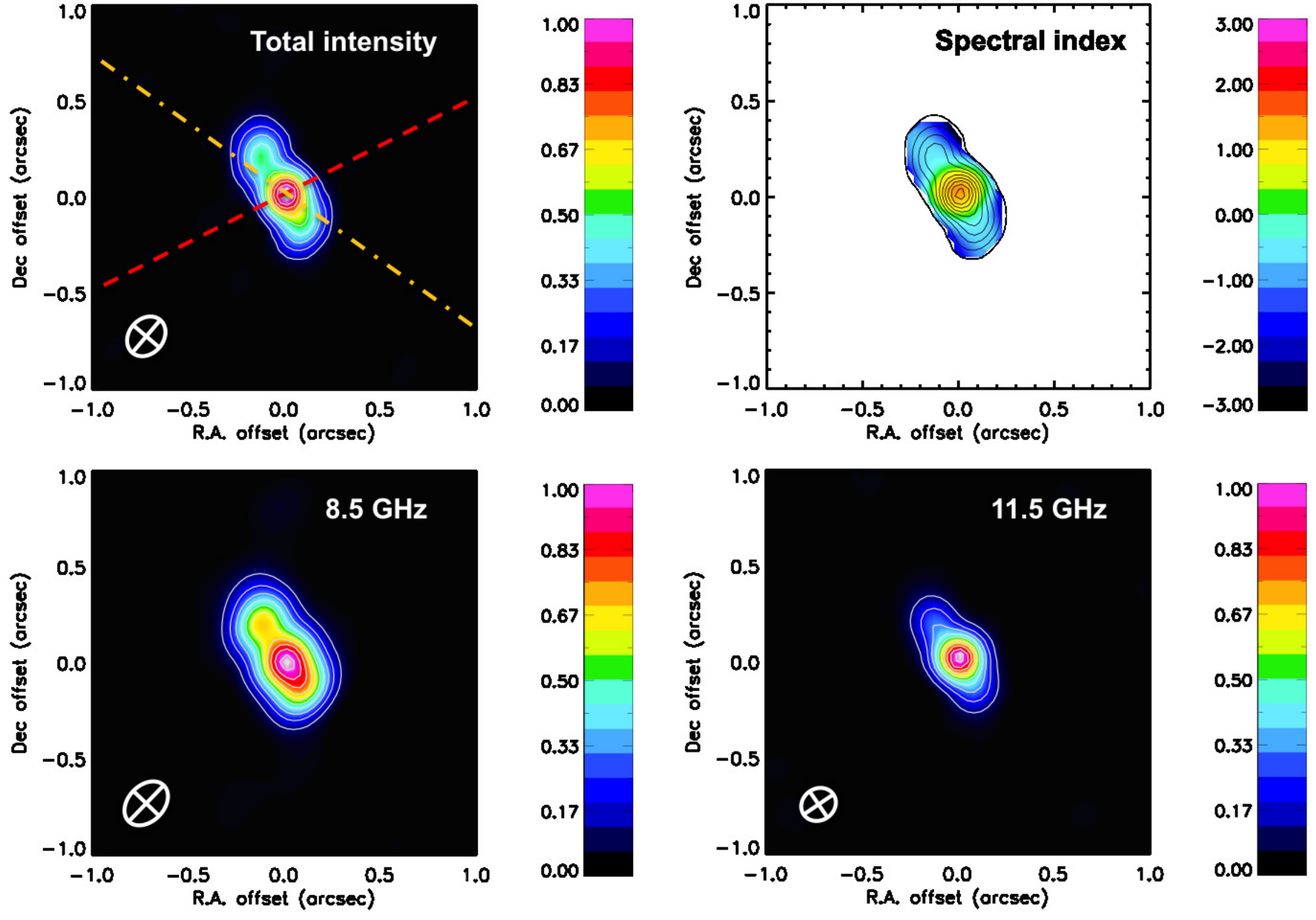}
\caption{Same as Fig. 1, but for J$0009-0036$. In this case, the peak flux density in the total intensity image corresponds to $450\times$rms (rms$=13.7\,\mu$Jy beam$^{-1}$), $442\times$rms at 8.5 GHz, and $529\times$rms at 11.5 GHz (Table 3). This galaxy has a two-sided radio jet that suggests a radio jet-driven outflow produces the double-peaked narrow emission lines. 
\label{fig2}}
\end{figure*}

\begin{figure*}
\epsscale{.99}
\plotone{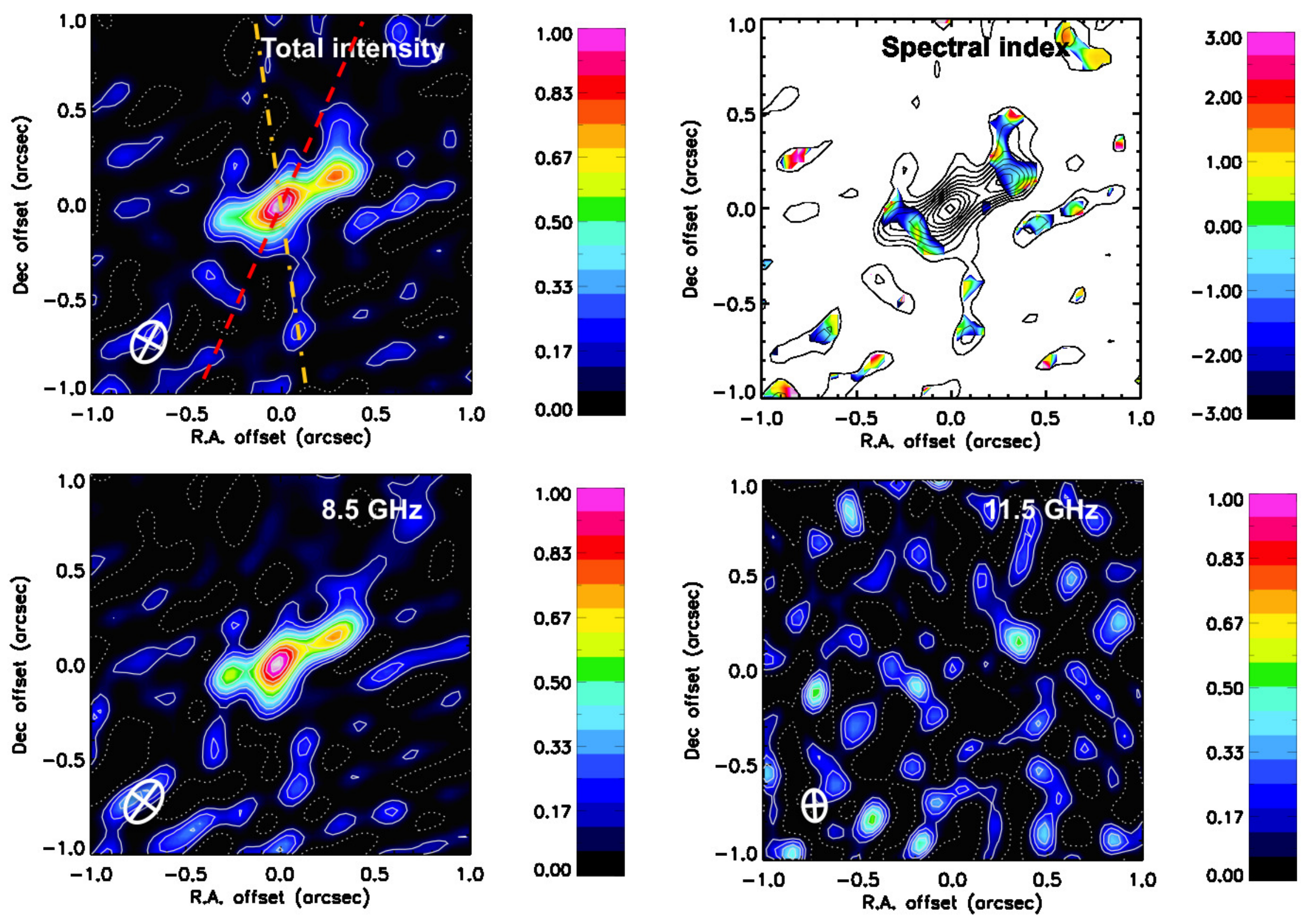}
\caption{Same as Fig. 1, but for J$0731+4528$. In this case, the peak flux density in the total intensity image corresponds to $18\times$rms (rms$=4.6\,\mu$Jy beam$^{-1}$), $16\times$rms at 8.5 GHz, and $6\times$rms at 11.5 GHz (Table 3). This galaxy has a two-sided radio jet that is not aligned with the ionized gas emission. The optical data suggest that an AGN wind-driven outflow produces the double-peaked narrow emission lines. 
\label{fig3}}
\end{figure*}

\begin{figure*}
\epsscale{.99}
\plotone{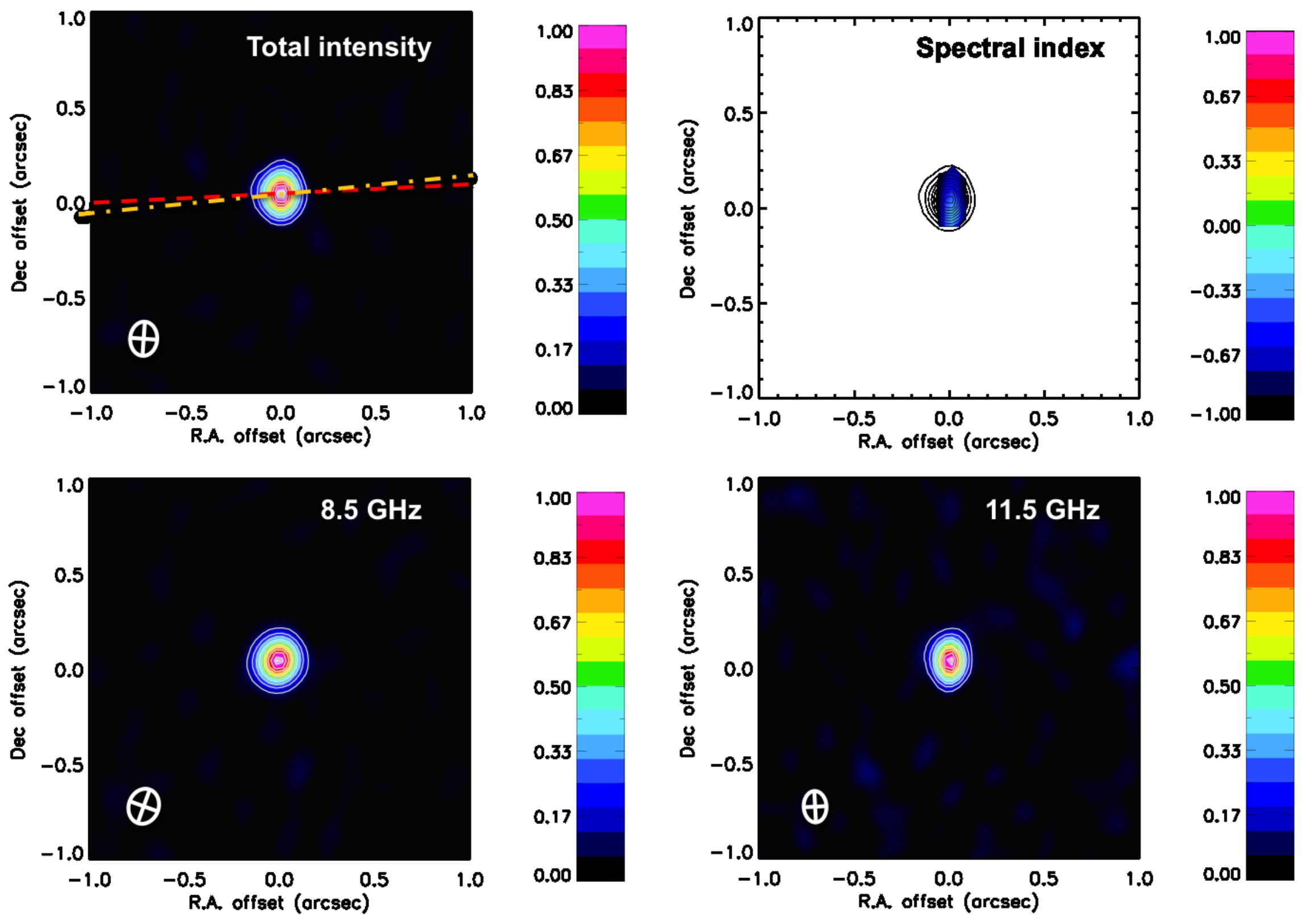}
\caption{Same as Fig. 1, but for J$0736+4759$. In this case, the peak flux density in the total intensity image corresponds to $98\times$rms (rms$=9.2\,\mu$Jy beam$^{-1}$), $97\times$rms at 8.5 GHz, and $56\times$rms at 11.5 GHz (Table 2). This galaxy is classified as a single AGN with a rotating NLR. 
\label{fig4}}
\end{figure*}

\begin{figure*}
\epsscale{.99}
\plotone{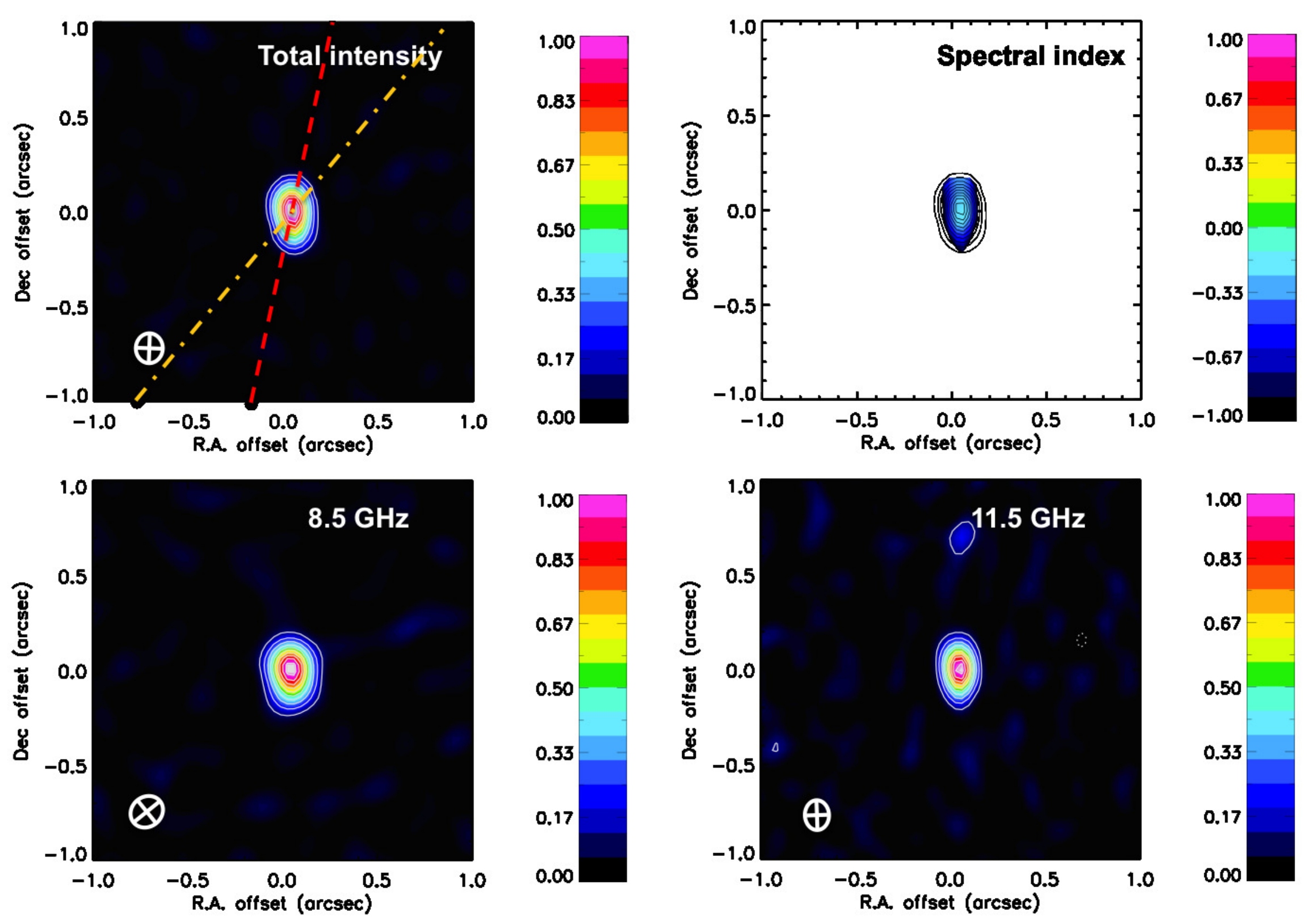}
\caption{Same as Fig. 1, but for J$0802+3046$. In this case, the peak flux density in the total intensity image corresponds to $74\times$rms (rms$=6.6\,\mu$Jy beam$^{-1}$), $68\times$rms at 8.5 GHz, and $44\times$rms at 11.5 GHz (Table 2). This galaxy is classified as a single AGN with an AGN wind-driven outflow. 
\label{fig5}}
\end{figure*}

\begin{figure*}
\epsscale{.99}
\plotone{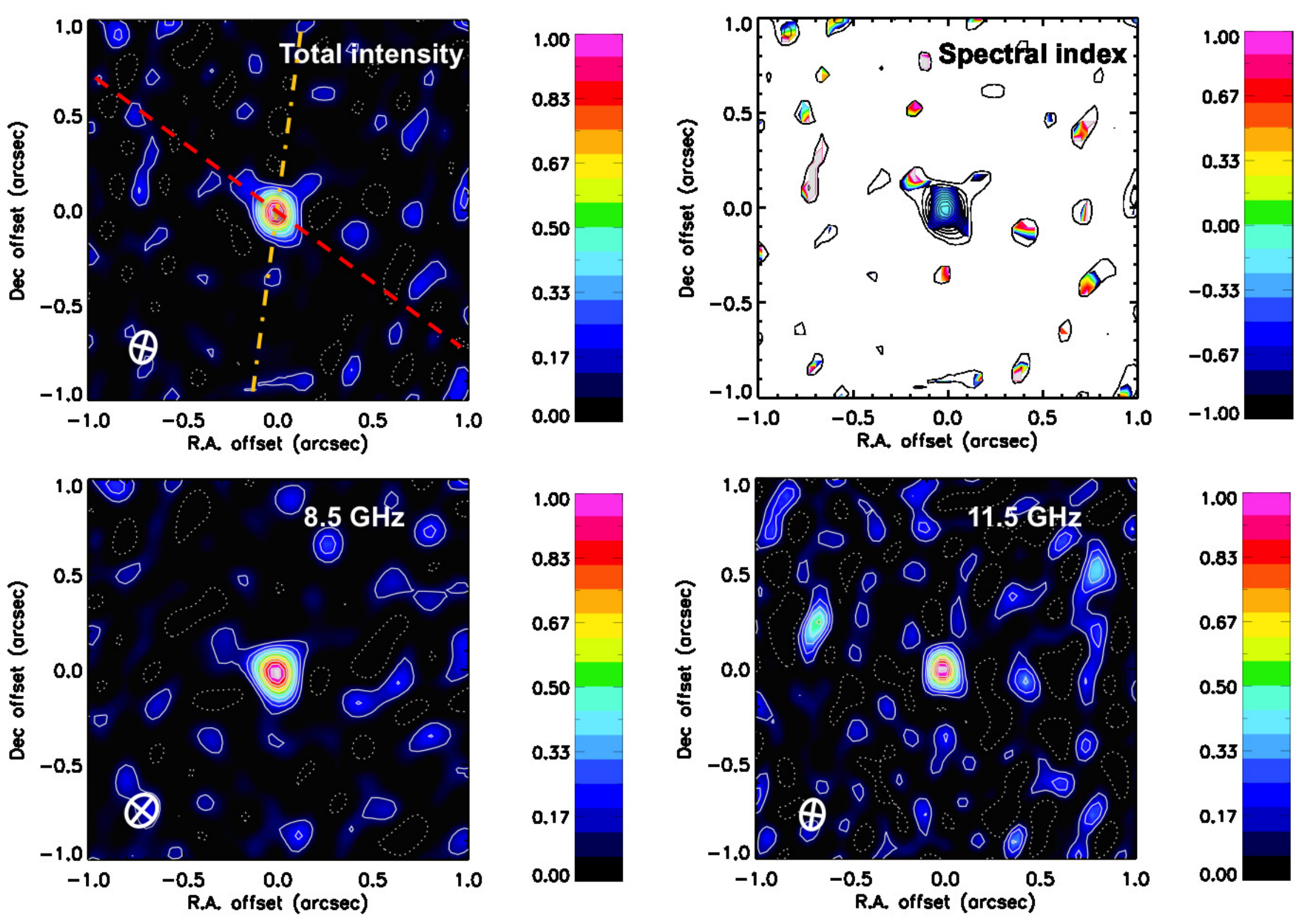}
\caption{Same as Fig. 1, but for J$0846+4258$. In this case, the peak flux density in the total intensity image corresponds to $21\times$rms (rms$=6.9\,\mu$Jy beam$^{-1}$), $20\times$rms at 8.5 GHz, and $13\times$rms at 11.5 GHz (Table 2). This galaxy is classified as a single AGN with an AGN wind-driven outflow. 
\label{fig6}}
\end{figure*}

\begin{figure*}
\epsscale{.99}
\plotone{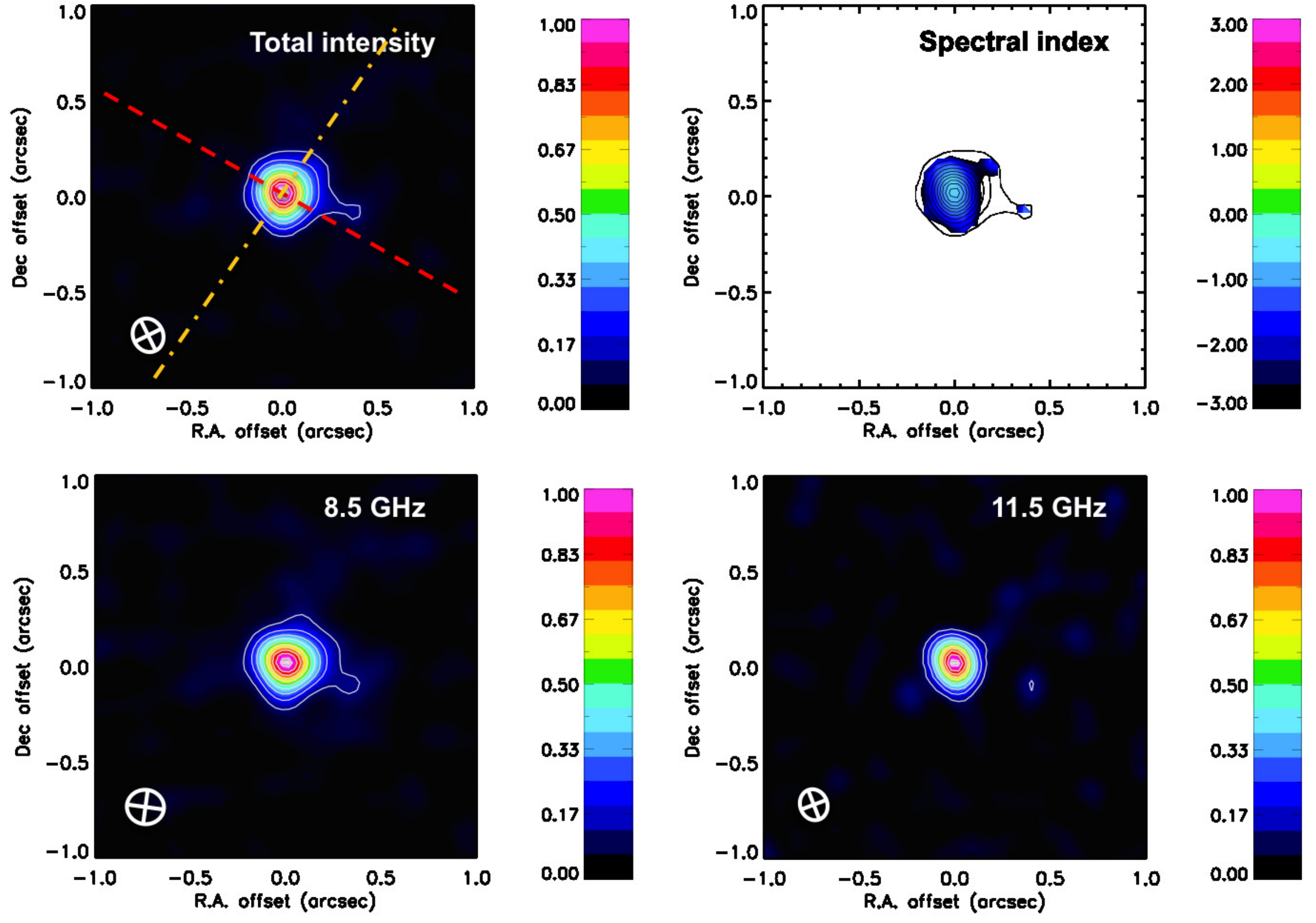}
\caption{Same as Fig. 1, but for J$0858+1041$. In this case, the peak flux density in the total intensity image corresponds to $95\times$rms (rms$=6.7\,\mu$Jy beam$^{-1}$), $98\times$rms at 8.5 GHz, and $55\times$rms at 11.5 GHz (Table 3). This galaxy has a one-sided radio jet that suggests a radio jet-driven outflow produces the double-peaked narrow emission lines. 
\label{fig7}}
\end{figure*}

\begin{figure*}
\epsscale{.99}
\plotone{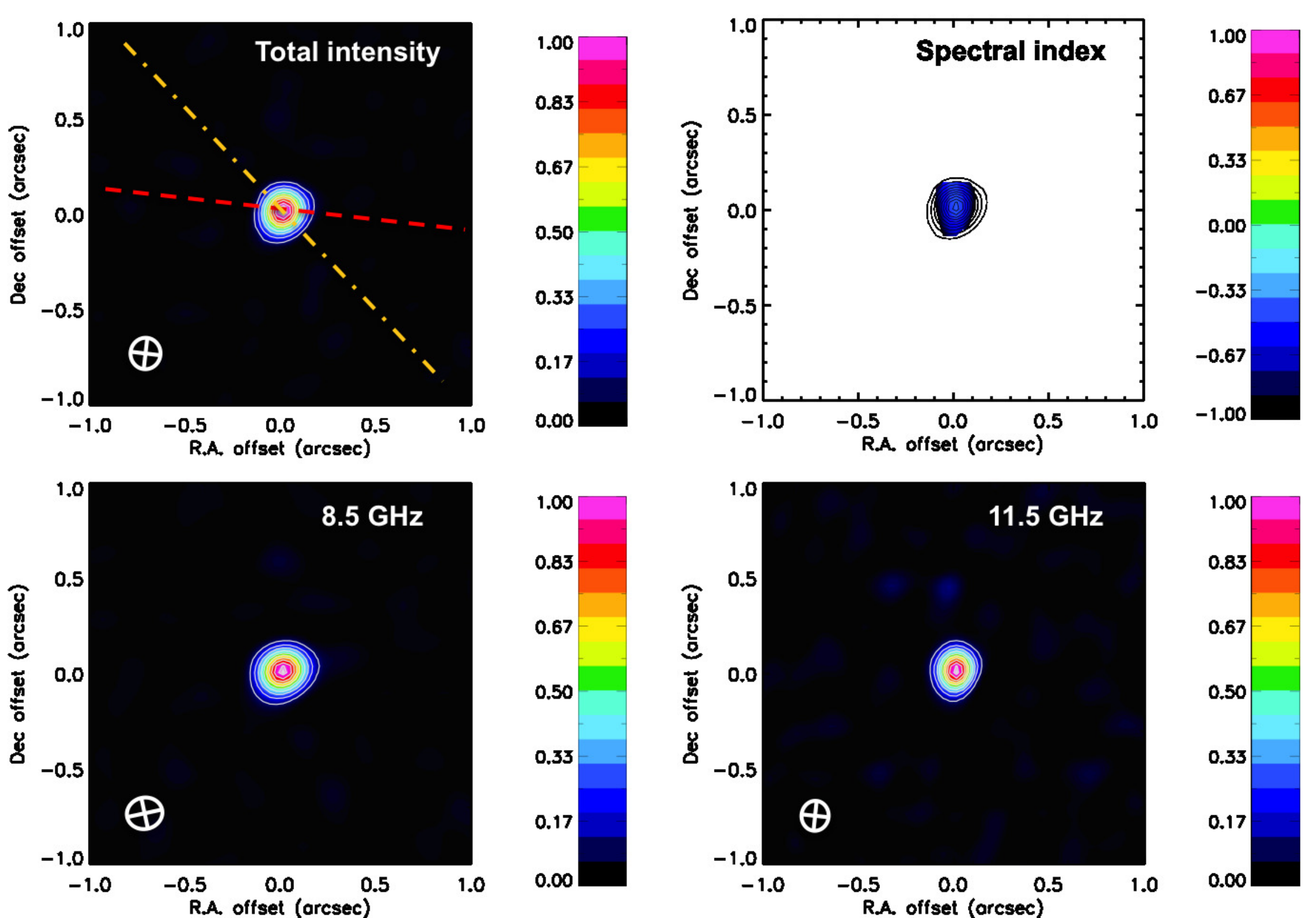}
\caption{Same as Fig. 1, but for J$0916+2835$. In this case, the peak flux density in the total intensity image corresponds to $130\times$rms (rms$=7.7\,\mu$Jy beam$^{-1}$), $138\times$rms at 8.5 GHz, and $74\times$rms at 11.5 GHz (Table 2). This galaxy is classified as a single AGN with an AGN wind-driven outflow. 
\label{fig8}}
\end{figure*}

\begin{figure*}
\epsscale{.99}
\plotone{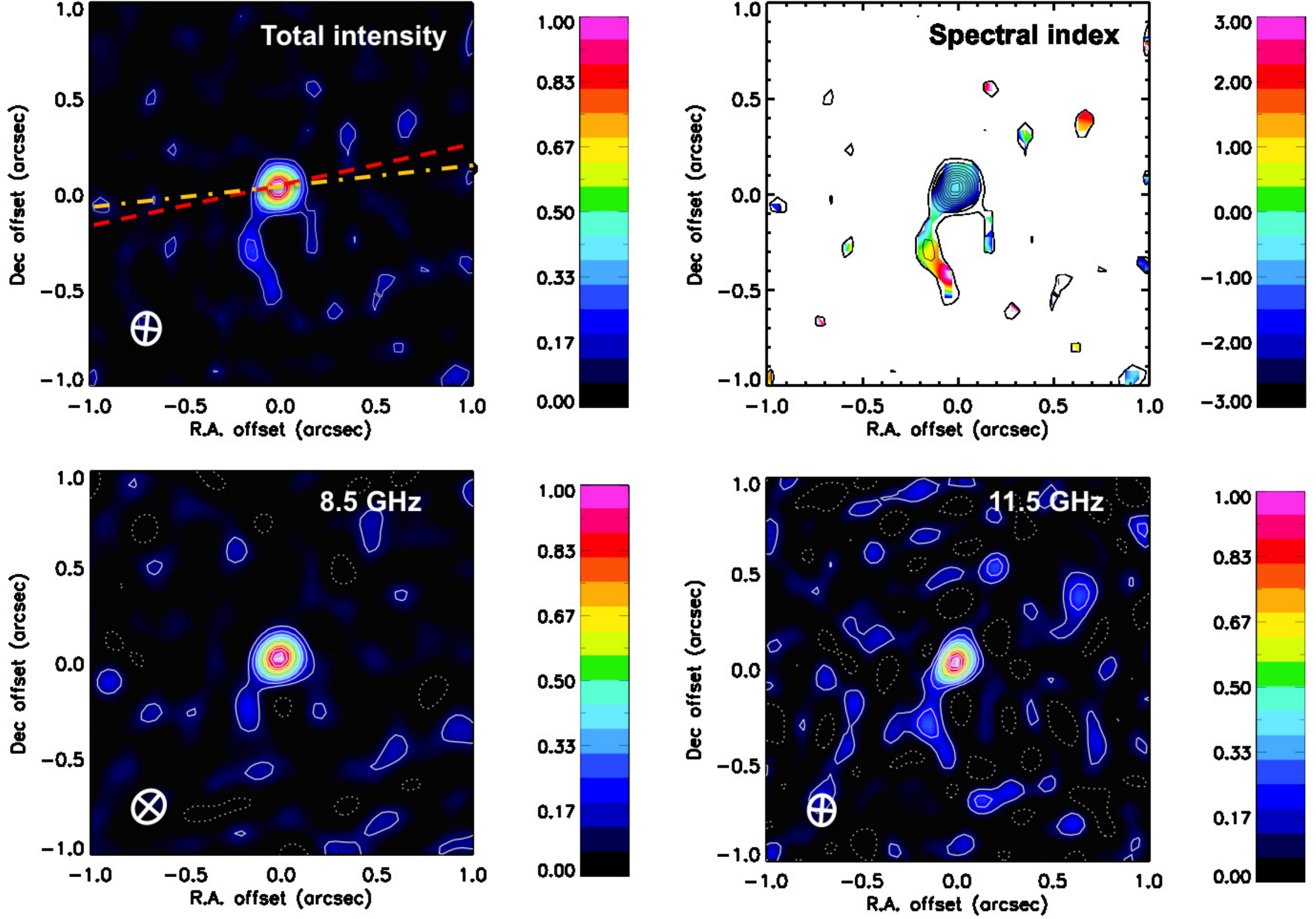}
\caption{Same as Fig. 1, but for J$0930+3430$. In this case, the peak flux density in the total intensity image corresponds to $25\times$rms (rms$=5.9\,\mu$Jy beam$^{-1}$), $27\times$rms at 8.5 GHz, and $17\times$rms at 11.5 GHz (Table 3). This galaxy has a one-sided radio jet that is not aligned with the ionized gas emission. The optical data suggest that an AGN wind-driven outflow produces the double-peaked narrow emission lines. 
\label{fig9}}
\end{figure*}

\begin{figure*}
\epsscale{.99}
\plotone{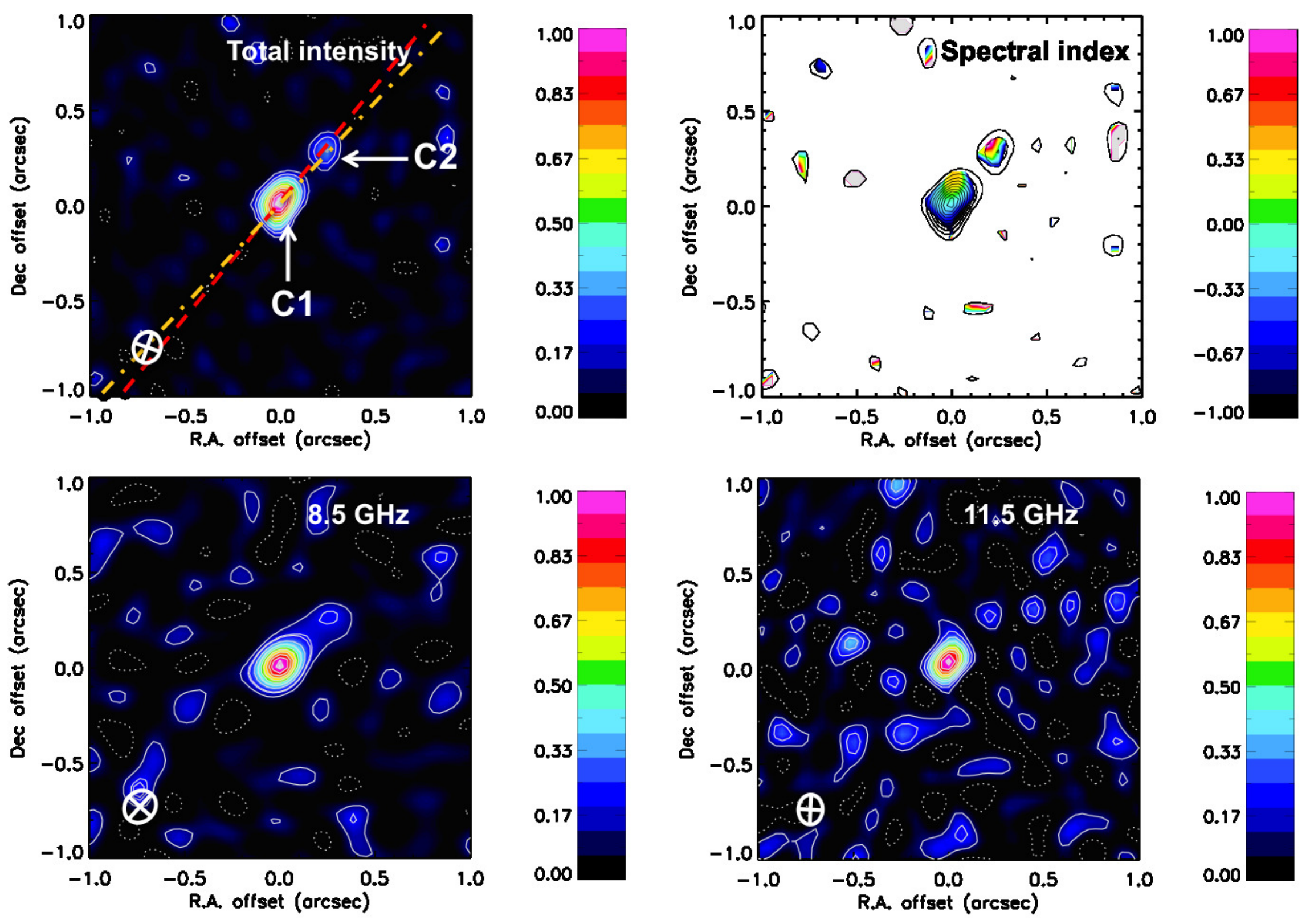}
\caption{Same as Fig. 1, but for J$1023+3243$. In this case, the peak flux density in the total intensity image corresponds to $23\times$rms (rms$=6.3\,\mu$Jy beam$^{-1}$), $22\times$rms at 8.5 GHz, and $17\times$rms at 11.5 GHz (Table 4). This galaxy is classified as a dual AGN system. 
\label{fig10}}
\end{figure*}

\begin{figure*}
\epsscale{.99}
\plotone{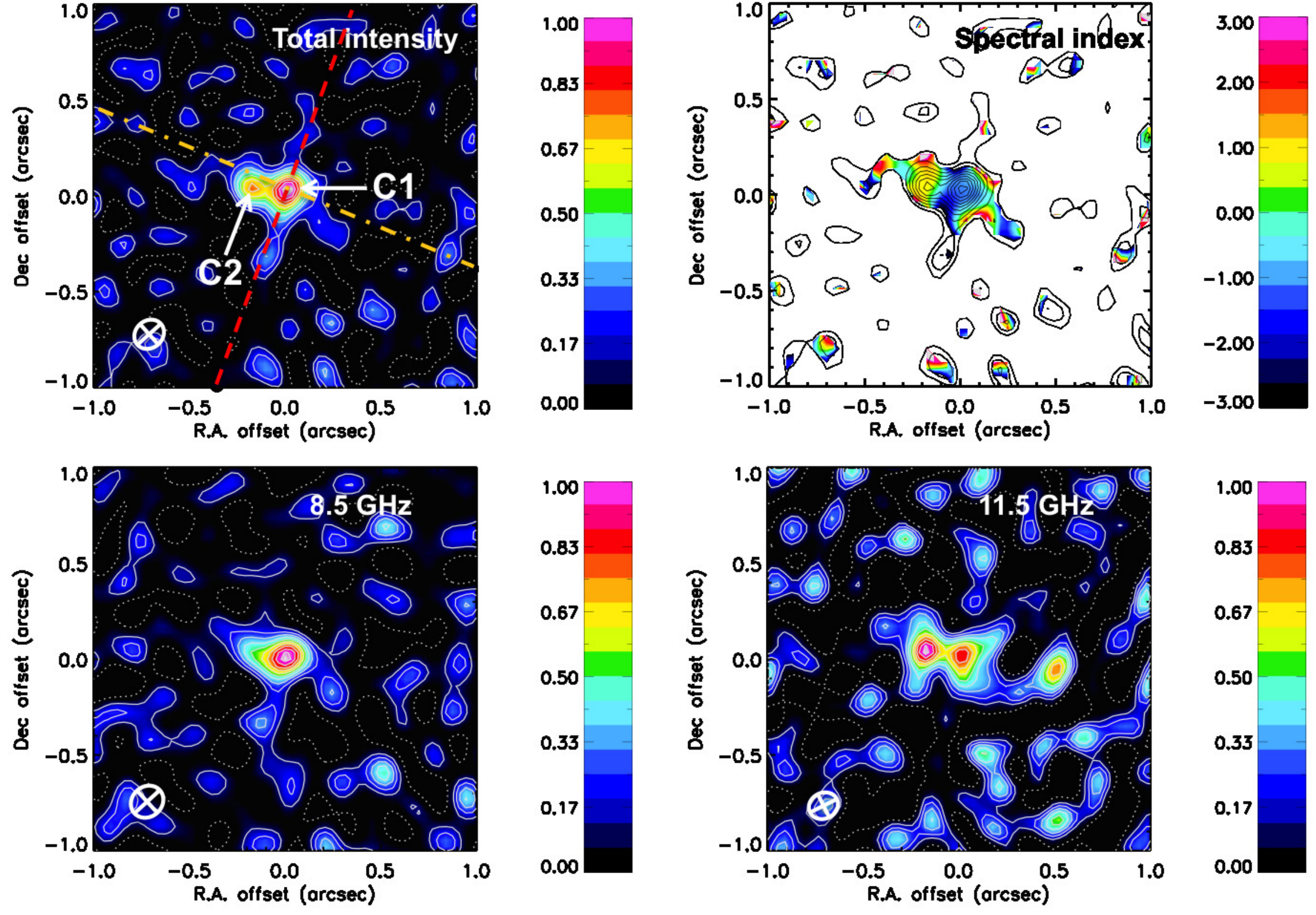}
\caption{Same as Fig. 1, but for J$1027+3059$. In this case, the peak flux density in the total intensity image corresponds to $11\times$rms (rms$=7.0\,\mu$Jy beam$^{-1}$), $12\times$rms at 8.5 GHz, and $7\times$rms at 11.5 GHz (Table 5). The origin of the double-peaked narrow AGN emission lines in this galaxy is ambiguous. 
\label{fig11}}
\end{figure*}

\begin{figure*}
\epsscale{.99}
\plotone{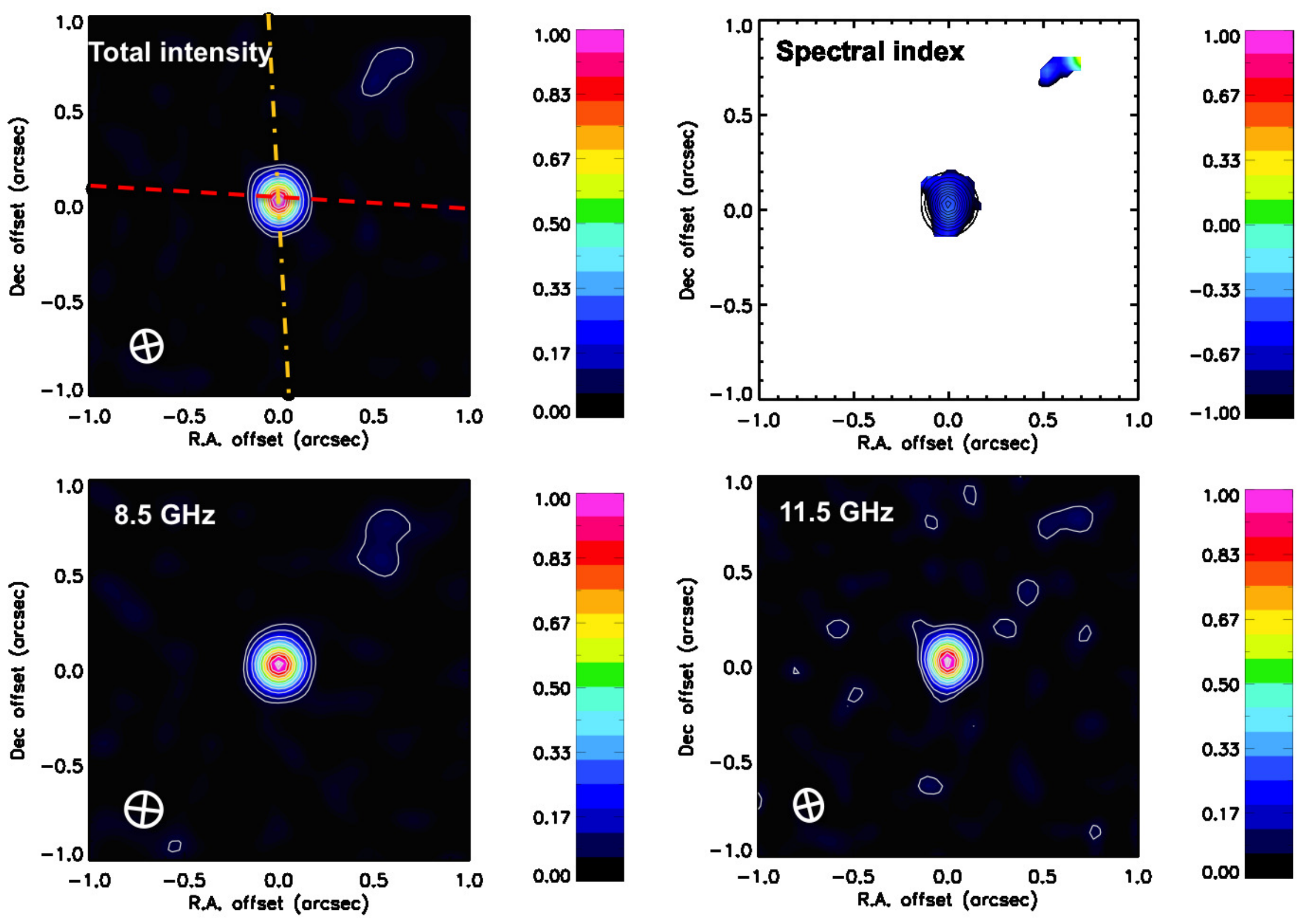}
\caption{Same as Fig. 1, but for J$1112+2750$. In this case, the peak flux density in the total intensity image corresponds to $91\times$rms (rms$=7.0\,\mu$Jy beam$^{-1}$), $93\times$rms at 8.5 GHz, and $61\times$rms at 11.5 GHz (Table 2). This galaxy is classified as a single AGN with an AGN wind-driven outflow. 
\label{fig12}}
\end{figure*}

\begin{figure*}
\epsscale{.99}
\plotone{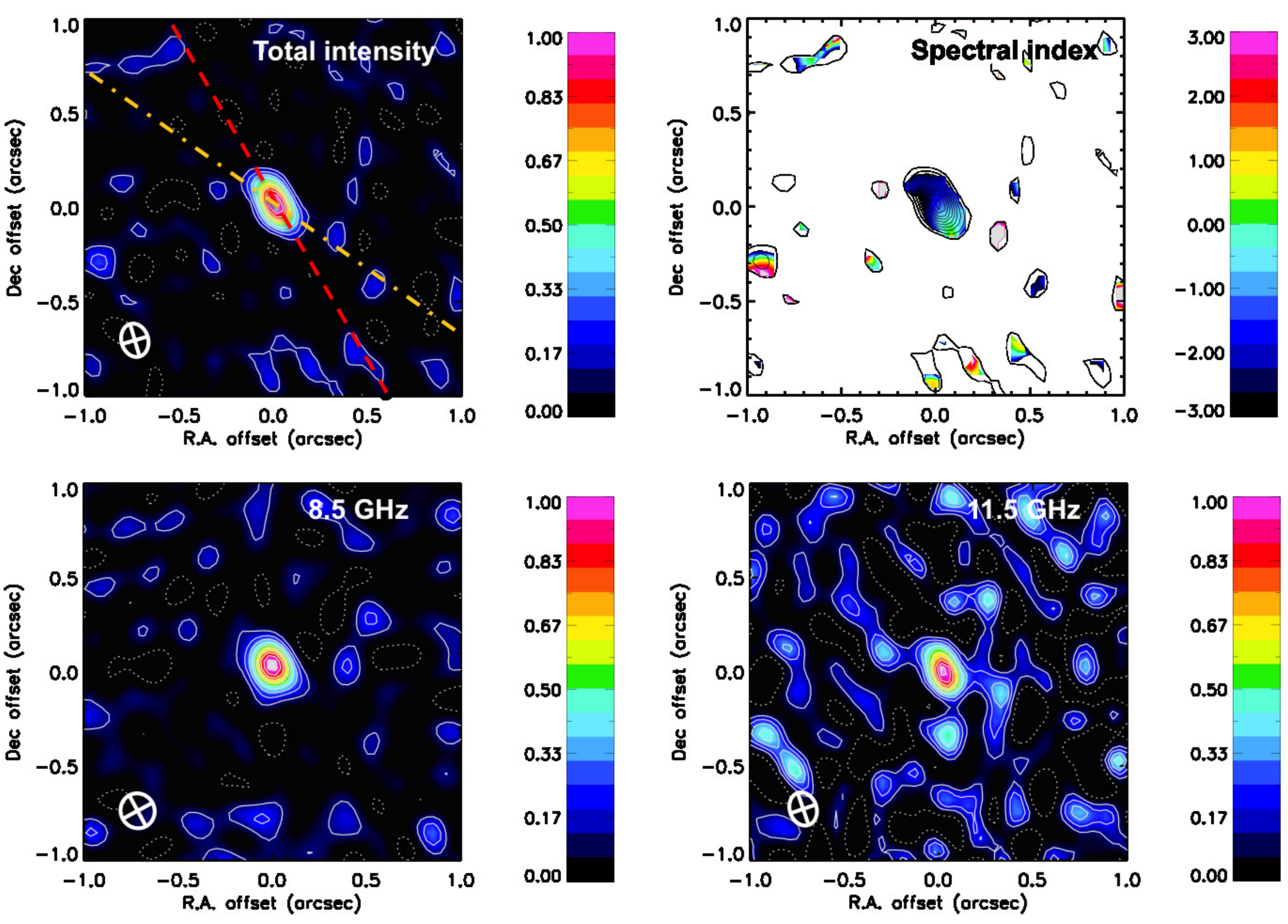}
\caption{Same as Fig. 1, but for J$1152+1903$. In this case, the peak flux density in the total intensity image corresponds to $19\times$rms (rms$=6.3\,\mu$Jy beam$^{-1}$), $20\times$rms at 8.5 GHz, and $10\times$rms at 11.5 GHz (Table 3). This galaxy has a two-sided radio jet that suggests a radio jet-driven outflow produces the double-peaked narrow emission lines. 
\label{fig13}}
\end{figure*}

\begin{figure*}
\epsscale{.99}
\plotone{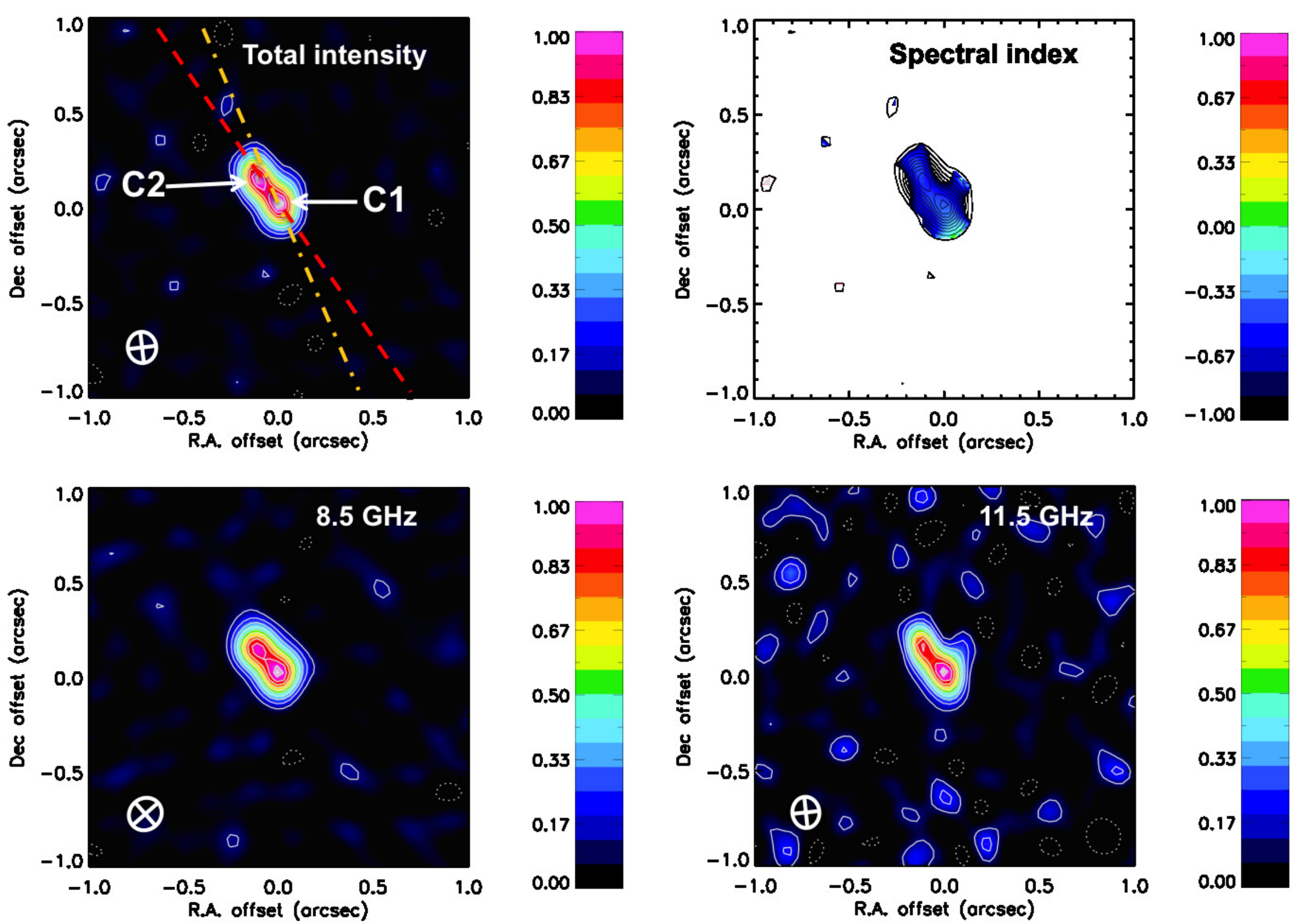}
\caption{Same as Fig. 1, but for J$1158+3231$. In this case, the peak flux density in the total intensity image corresponds to $34\times$rms (rms$=6.6\,\mu$Jy beam$^{-1}$), $36\times$rms at 8.5 GHz, and $23\times$rms at 11.5 GHz (Table 4). This galaxy is classified as a dual AGN system. 
\label{fig14}}
\end{figure*}

\begin{figure*}
\epsscale{.99}
\plotone{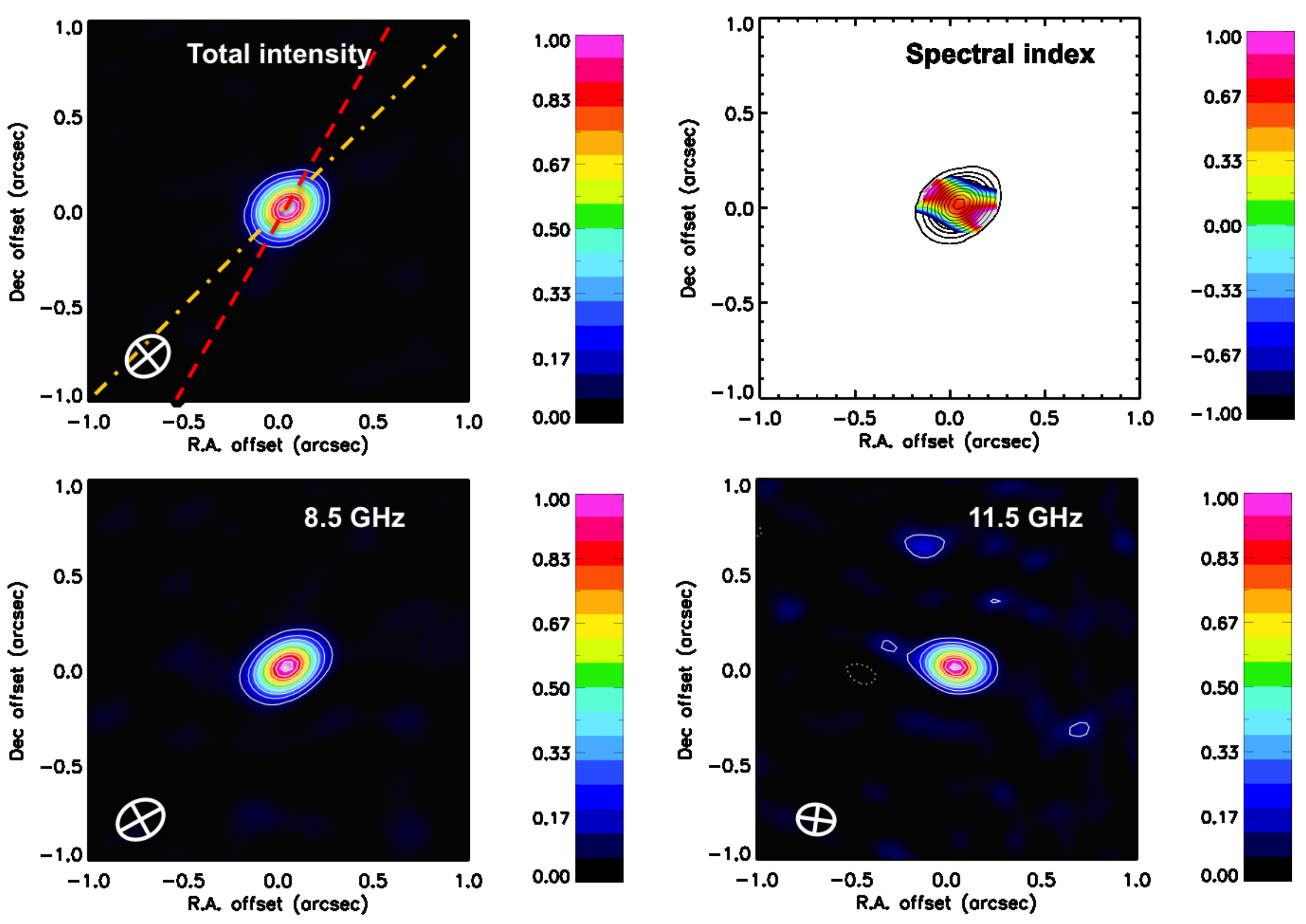}
\caption{Same as Fig. 1, but for J$1556+0948$. In this case, the peak flux density in the total intensity image corresponds to $140\times$rms (rms$=5.9\,\mu$Jy beam$^{-1}$), $77\times$rms at 8.5 GHz, and $52\times$rms at 11.5 GHz (Table 2). This galaxy is classified as a single AGN with an AGN wind-driven outflow. 
\label{fig15}}
\end{figure*}

\begin{figure*}
\epsscale{.99}
\plotone{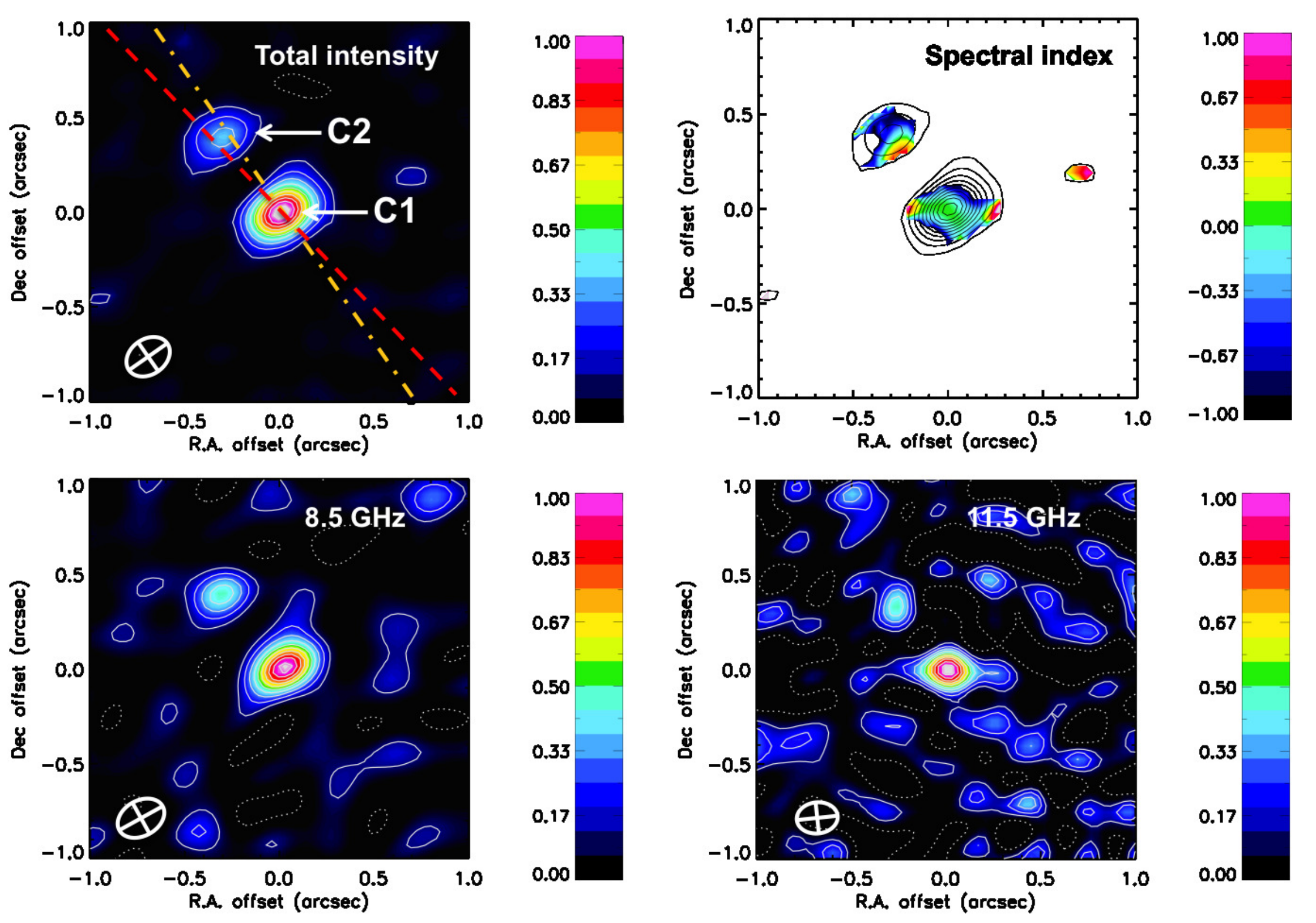}
\caption{Same as Fig. 1, but for J$1623+0808$. In this case, the peak flux density in the total intensity image corresponds to $40\times$rms (rms$=5.2\,\mu$Jy beam$^{-1}$), $22\times$rms at 8.5 GHz, and $17\times$rms at 11.5 GHz (Table 4). This galaxy is classified as a dual AGN system. 
\label{fig16}}
\end{figure*}

\begin{figure*}
\epsscale{.99}
\plotone{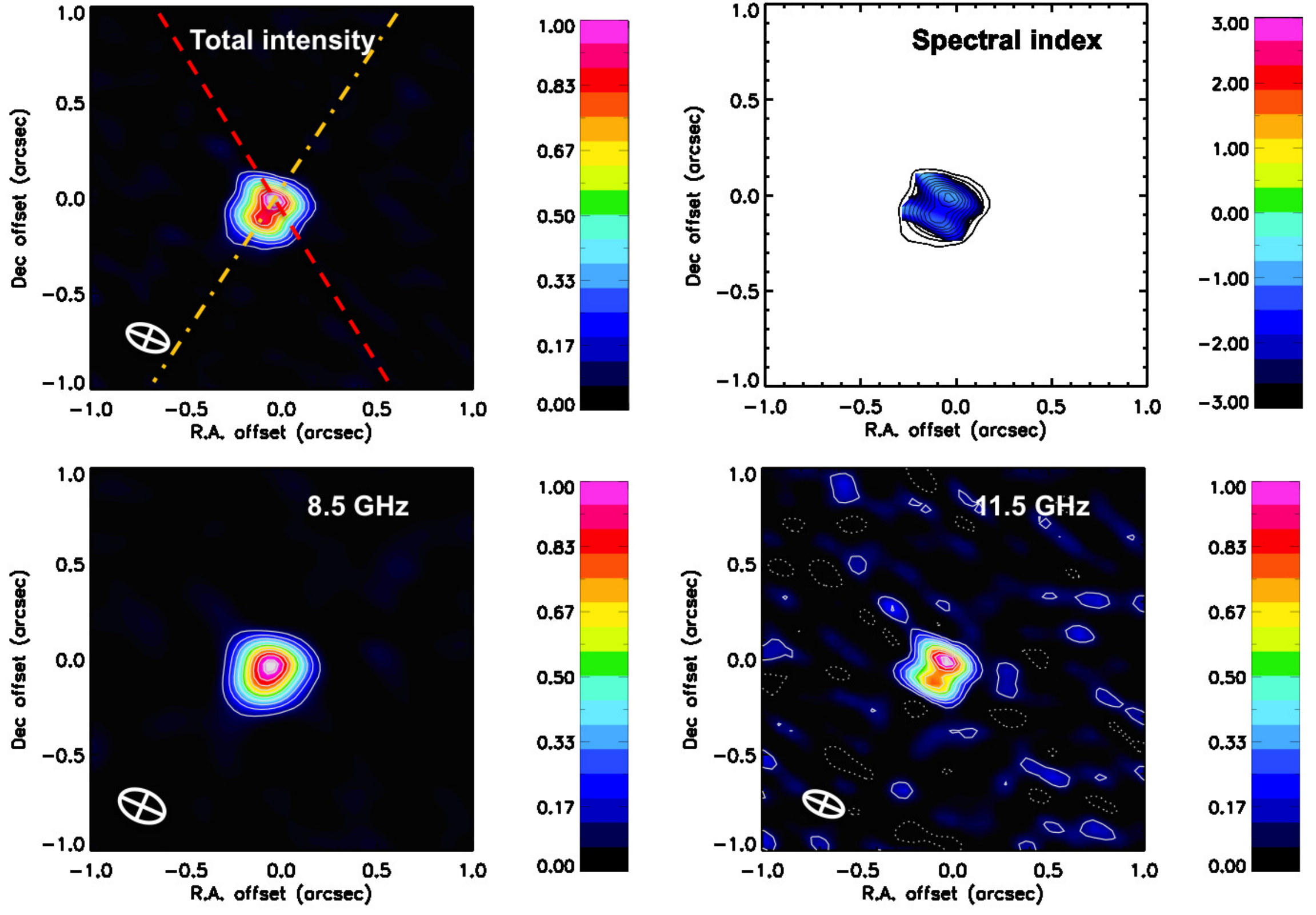}
\caption{Same as Fig. 1, but for J$1715+6008$. In this case, the peak flux density in the total intensity image corresponds to $75\times$rms (rms$=11.7\,\mu$Jy beam$^{-1}$), $110\times$rms at 8.5 GHz, and $35\times$rms at 11.5 GHz (Table 3). This galaxy has a one-sided radio jet that suggests a radio jet-driven outflow produces the double-peaked narrow emission lines. 
\label{fig17}}
\end{figure*}

\begin{figure*}
\epsscale{.99}
\plotone{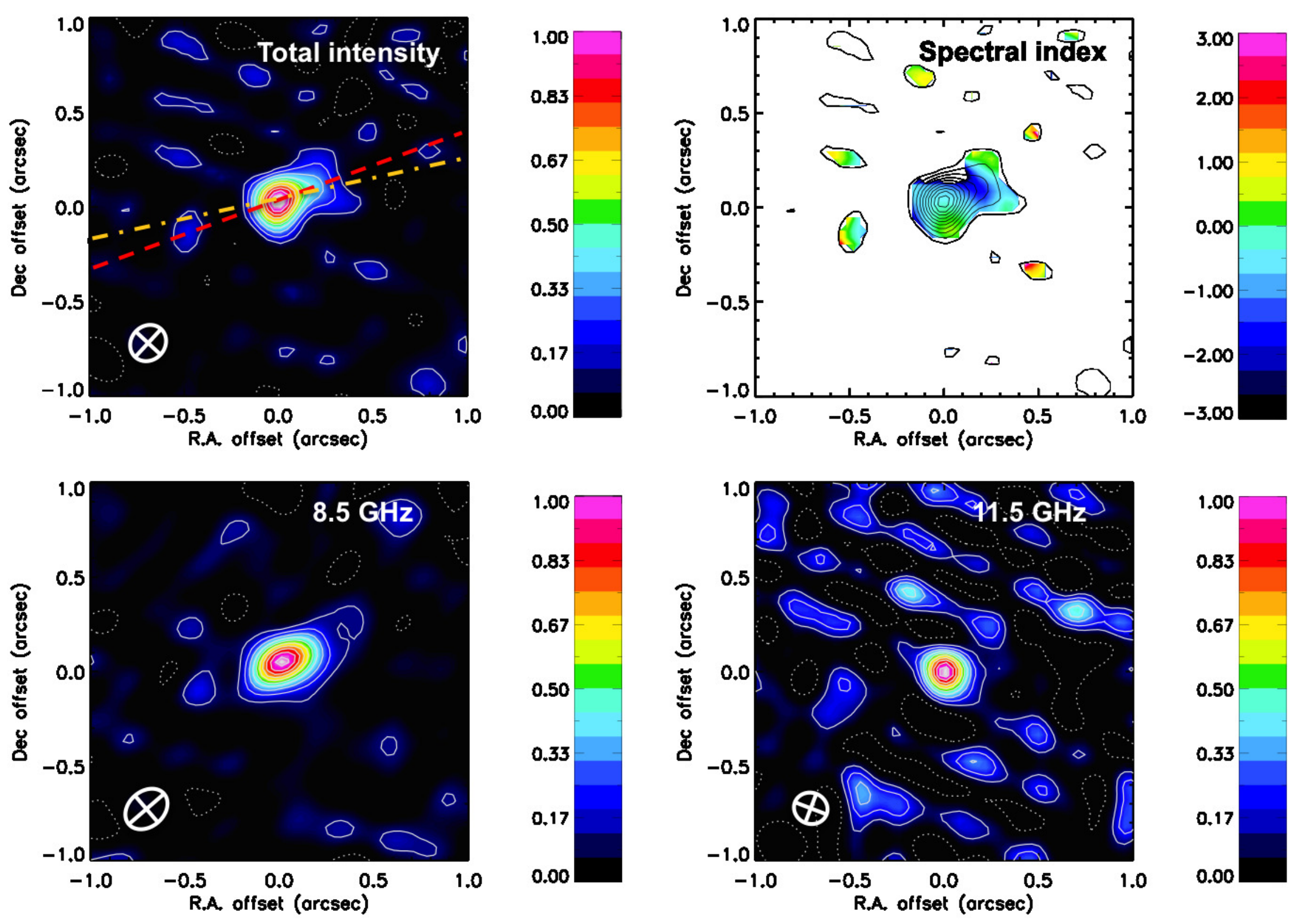}
\caption{Same as Fig. 1, but for J$2254-0051$. In this case, the peak flux density in the total intensity image corresponds to $25\times$rms (rms$=6.5\,\mu$Jy beam$^{-1}$), $25\times$rms at 8.5 GHz, and $12\times$rms at 11.5 GHz (Table 3). This galaxy has a one-sided radio jet that suggests a radio jet-driven outflow produces the double-peaked narrow emission lines. 
\label{fig18}}
\end{figure*}

\subsection{VLA Observations and Processing}\label{observations}


We obtained NRAO VLA observations of 18 galaxies with double-peaked, spatially-offset narrow AGN emission lines at $X-$band ($8-12$ GHz). The observations were carried out in the A configuration for 12 hours with two $1024$ MHz wide basebands centered at 8.5 and 11.5 GHz. The correlator was configured with eight spectral windows, each of which contained 64 channels with a frequency resolution of 2 MHz. The observations consisted of $7-8$ minute scans of the target sources, interspersed with $1-1.5$ minute scans of nearby phase reference sources. Switching angles between target and reference pairs ranged between $0.5-4\degr$. Therefore, over the 12 hr period, between $28-32$ min of integration time was obtained for each of the 18 targets. The 18 galaxies, together with the flux density and phase calibrators used for each galaxy, are listed in Table 1. The uncertainty of the flux density calibration is typically 3\% at these frequencies (Perley \& Butler 2013). 



We flagged, calibrated, and imaged the $X-$band data with the Common Astronomy Software
Applications (CASA) package. Phase-referenced (Stokes I) images of each galaxy were produced by
applying the CLEAN task with a Briggs robust parameter of $-0.5$ to achieve good spatial resolution with only a modest loss of sensitivity. For the analysis, we smoothed the upper 11.5 GHz baseband image to match the resolution of the lower 8.5 GHz baseband using the CASA task IMSMOOTH. These two images were then combined to generate an image of the spectral
index distribution across the source. This observational approach avoids spectral index uncertainty that results from intrinsic source variability. 
We also generated a total intensity image by integrating all the spectral windows in the two basebands. 
The radio properties of each galaxy at 8.5 GHz and 11.5 GHz (both in native resolution and smoothed to match the resolution of the 8.5 GHz image) are given in Tables $2-5$, and radio images with contours are shown in Figures $1-18$. 
The images show the central $2\arcsec\times2\arcsec$ of each galaxy to display its detailed morphology. 
None of the galaxies show structures above the $3\sigma$ rms level outside this region. In some galaxies, two compact sources are detected in the total intensity image. We label these sources `C1' and `C2', with C1 being the brightest amongst the two.

\subsection{Reanalysis of Published Optical Long-Slit Spectroscopy}\label{reanalysis}
 
In Comerford et al. (2012) three morphological parameters from the long-slit spectra were used in an attempt  to constrain what mechanisms 
produce double-peaked narrow AGN emission lines: the projected spatial separation between the two [O III] emission features ($D_{\mathrm{[O III]}}$), the true position angle of the two emission components on the sky (PA$_{\mathrm{[O III]}}$), and the spatial extent of the [O III] emitting region. While a quantitatively accurate analysis of $D_{\mathrm{[O III]}}$ and PA$_{\mathrm{[O III]}}$ was performed, the spatial extent of the emission was only measured by eye. 
The visual method used in Comerford et al. (2012) gives two qualitative classifications: compact and extended. While this approach is useful to identify very extended emission line regions (likely produced by AGN outflows), it is less helpful in moderately extended/compact regions (like the 18 galaxies discussed in this paper, as indicated by Comerford et al. 2012). First, it does not provide a quantitative measurement of the size of the emitting line region, so sources classified as compact might be in reality extended (especially if the emission is faint), and second, the detection of `compact' [O III] emitting regions should not be taken as evidence for or against AGN outflows or dual AGNs (especially when the real size in parsecs is not known), because extended emission line regions ($>10$ kpc) have been observed in confirmed dual AGNs (such as NGC 6240 and Mrk 266) and compact emission line regions ($<1$ kpc) have been observed in single AGNs with outflows (M\"uller-S\'anchez et al. 2011).  

In order to provide the base for a quantitative analysis, we have measured the size of the [O III] emitting region using the Akaike information criterion (AIC), defined by Akaike (1974). Briefly, the method works as follows (see Fig. 19): we first select a 2D window with a 100 pixel width in the spectral direction centered on [O III] so that it contains the entire emission line plus continuum on both sides of the line, and a $20\arcsec-30\arcsec$ width on the spatial direction. Then, for each row in the spatial direction, we fit both a 2 parameter and 5 parameter model. These models correspond to a line with a given slope (2 parameters) and a single Gaussian combined with a line with a given slope (5 parameters). Then, we used the AIC to determine which model is a better fit. The AIC is defined as: AIC$=\chi^2+2k$, where $k$ is the number of parameters. 
It penalizes for the use of more parameters when a fit with more variables is unnecessary. We used the corrected AIC$_c$ for finite sample size: AIC$_c=$AIC$+2k(k+1)/(n-k-1)$ where $n$ is the sample size. A smaller AIC$_c$ indicates a better relative fit between the two models.  We applied this criterion for each row in the spatial direction and determined the last row where the more complex Gaussian + linear fit has a lower AIC$_c$ statistic (the Akaike width). This corresponds to the row of last significant emission, and therefore, the size of the [O III] emitting region. The uncertainties were estimated using Monte Carlo techniques. The method involves adding noise to the galaxy spectra, and refitting the result to yield a new set of Akaike widths.   
After repeating this 500 times, the standard deviation of the distribution of the Akaike width was used as the uncertainty for the size of the [O III] emitting region. 

With the Gaussian fits in hand, we created surface brightness radial profiles of the [O III] flux distribution for each position angle of the slit, by measuring the area under the Gaussian of each spectrum along the spatial direction. We compared these profiles with the radial profiles of the continuum emission and the PSF (see Fig. 19). The continuum measurements were taken at both sides of the [O III] emission line in windows of twenty pixels along the wavelength axis. We fit a Gaussian in the spatial direction for each continuum window, and then took the average of the two Gaussians. The PSF profiles (Gaussian) were constructed in a similar way, but using standard stars observed closest in time to the science frames and a single spectral window of 40 pixels. For each galaxy, we measured the spatial extent of the continuum and [O III] emission at 5\% and 50\% of the peak flux.  

The results are shown in Table 6. An example of the radial profiles of the three components ([O III] flux, galaxy continuum, PSF) is shown in Fig. 19 for the case of J$1023+3243$. For completeness, we present the measurements obtained at each position angle of the slit in Table 6, but the discussion presented here is based only on the PA$_{\mathrm{slit}}$ exhibiting the most extended emission (column 4 in Table 6). It is also important to point out that these measurements are position angle dependent, and therefore the results presented here represent a lower limit on the size of the emitting region, except in those cases where one of the PA$_{\mathrm{slit}}$ is consistent (within the errors of $\pm15\degr$) with PA$_{\mathrm{[O III]}}$ (the position angle of the most extended emission, Comerford et al. 2012). 

The continuum emission is resolved (FWHM$_{\mathrm{cont}}>$FWHM$_{\mathrm{PSF}}$) in all cases, but the [O III] emission is resolved in only 10 galaxies (FWHM$_{\mathrm{[O III]}}>$FWHM$_{\mathrm{PSF}}$). In fact, most of the [O III] emission consists of a marginally resolved/unresolved core and extended emission out to several kiloparsecs. Also, in all cases the FWHM of the continuum is larger than the FWHM of the [O III] emission. However, in three galaxies, the spatial extent of the [O  III] emission at 5\% of the peak flux is larger than the size of the continuum at the same percentage of the peak value. One of them (J$1715+6008$) has in fact the most extended [O III] emitting region of the whole sample ($\sim30$ kpc). In order to compare the extent of the [O III] emission in all galaxies, we have plotted in Fig. 20 the surface brightness radial profiles of the [O III] flux distribution. The results will be discussed in Section~\ref{size}.

\begin{table*}
\caption[Summary of Radio Properties]{Summary of Radio Properties of the Single AGNs.}
\begin{center}
{\scriptsize
\begin{tabular}{l c c c | c c | c c c | c}
\hline
\hline \noalign{\smallskip}
 & & & & \multicolumn{2}{c}{Beam parameters} \vline & \multicolumn{3}{c}{Source parameters} \vline & \\
 \hline
SDSS name & Source & Frequency & rms & $\theta_M \times \theta_m$ & PA & $\theta_M \times \theta_m$ & 
PA & Flux Density & $\alpha$ \\
  &  & (GHz) & ($\mu$Jy beam$^{-1}$) & ($\arcsec$) & ($\degr$E of N) & ($\arcsec$) & ($\degr$E of N) & (mJy) & \\
\hline \noalign{\smallskip}
\multirow{3}{*}{J0736+4759} & 1 & 8.5 & $10.2$ & $0.19\times0.16$ & $-19.9$ & $0.18\times0.17$ & 
$3.8$ & $1.0$ & \multirow{3}{*}{$-0.61\pm0.14$} \\
& 1 & 11.5 & $12.5$ & $0.17\times0.12$ & $-178.0$ & $0.18\times0.12$ & $1.8$ & $0.79$ & \\
& 1 & 11.5 & $11.3$ & $0.19\times0.16$ & $-19.9$ & $0.18\times0.17$ & $1.7$ & $0.83$ & \\
\hline
\multirow{3}{*}{J0802+3046} & 1 & 8.5 & $7.5$ & $0.18\times0.16$ & $-51.2$ & $0.23\times0.18$ & 
$-8.6$ & $0.53$ & \multirow{3}{*}{$-0.19\pm0.16$} \\
& 1 & 11.5 & $9.3$ & $0.16\times0.13$ & $1.7$ & $0.21\times0.13$ & $-8.1$ & $0.45$ & \\
& 1 & 11.5 & $9.5$ & $0.18\times0.16$ & $-51.2$ & $0.23\times0.14$ & $-7.7$ & $0.50$ & \\
\hline
\multirow{3}{*}{J0846+4258} & 1 & 8.5 & $7.8$ & $0.18\times0.15$ & $-39.7$ & $0.21\times0.18$ & 
$-5.0$ & $0.17$ & \multirow{3}{*}{$-0.41\pm0.18$} \\
& 1 & 11.5 & $10.1$ & $0.16\times0.12$ & $-9.1$ & $0.16\times0.12$ & $-6.5$ & $0.12$ & \\
& 1 & 11.5 & $10.5$ & $0.18\times0.15$ & $-39.7$ & $0.18\times0.15$ & $-4.5$ & $0.15$ & \\
\hline
\multirow{3}{*}{J0916+2835} & 1 & 8.5 & $8.9$ & $0.19\times0.17$ & $-77.4$ & $0.2\times0.18$ & 
$107.1$ & $1.20$ & \multirow{3}{*}{$-0.66\pm0.13$} \\
& 1 & 11.5 & $9.9$ & $0.16\times0.14$ & $-5.5$ & $0.17\times0.14$ & $5.1$ & $0.90$ & \\
& 1 & 11.5 & $9.5$ & $0.19\times0.17$ & $-77.4$ & $0.18\times0.15$ & $5.7$ & $0.98$ & \\
\hline
\multirow{3}{*}{J1112+2750} & 1 & 8.5 & $7.7$ & $0.19\times0.18$ & $84.0$ & $0.19\times0.19$ & 
$0.8$ & $0.72$ & \multirow{3}{*}{$-0.34\pm0.14$} \\
 & 1 & 11.5 & $8.4$ & $0.17\times0.15$ & $14.0$ & $0.17\times0.15$ & $178.4$ & $0.52$ & \\
& 1 & 11.5 & $8.8$ & $0.19\times0.18$ & $84.0$ & $0.19\times0.16$ & $-1.5$ & $0.65$ & \\
\hline
\multirow{3}{*}{J1556+0948} & 1 & 8.5 & $8.8$ & $0.25\times0.19$ & $-62.5$ & $0.26\times0.19$ & 
$-61.0$ & $0.80$ & \multirow{3}{*}{$0.73\pm0.14$} \\
& 1 & 11.5 & $17.9$ & $0.20\times0.16$ & $82.0$ & $0.22\times0.16$ & $79.5$ & $0.96$ & \\
& 1 & 11.5 & $17.5$ & $0.25\times0.19$ & $-62.5$ & $0.25\times0.19$ & $77.7$ & $1.0$ & \\
\hline
\hline
\end{tabular}
}
\end{center}
\tablecomments{The uncertainties in the size and PA of the sources are $0.03-0.05\arcsec$ and $4-7\degr$, respectively, from IMFIT in CASA. Errors on the measured flux densities are approximately $3.5\%$, dominated by the uncertainty in the flux density scale. It corresponds to the sum of the error reported by IMFIT in CASA and the $3\%$ calibration error, added in quadrature. 
}
\label{table2}
\end{table*}

\begin{table*}
\caption[Summary of Radio Properties]{Summary of Radio Properties of the Extended Radio Sources.}
\begin{center}
{\scriptsize
\begin{tabular}{l c c c | c c | c c c | c}
\hline
\hline \noalign{\smallskip}
 & & & & \multicolumn{2}{c}{Beam parameters} \vline & \multicolumn{3}{c}{Source parameters} \vline & \\
 \hline
SDSS name & Source & Frequency & rms & $\theta_M \times \theta_m$ & PA & $\theta_M \times \theta_m$ & 
PA & Flux Density & $\alpha$ \\
  &  & (GHz) & ($\mu$Jy beam$^{-1}$) & ($\arcsec$) & ($\degr$E of N) & ($\arcsec$) & ($\degr$E of N) & (mJy) & \\
\hline \noalign{\smallskip}
\multirow{3}{*}{J0009-0036} & 1 & 8.5 & $11.0$ & $0.27\times0.19$ & $-44.0$ & $0.30\times0.20$ & 
$30.0$ & $7.8$ & \multirow{3}{*}{$0.36\pm0.13$} \\
 & 1 & 11.5 & $16.0$ & $0.19\times0.17$ & $-56.1$ & $0.20\times0.18$ & $57.6$ & $8.2$ & \\
& 1 & 11.5 & $16.2$ & $0.27\times0.19$ & $-44.0$ & $0.27\times0.19$ & $56.7$ & $8.7$ & \\
\hline
\multirow{3}{*}{J0731+4528} & 1 & 8.5 & $5.2$ & $0.25\times0.17$ & $-37.1$ & $0.38\times0.21$ & 
$138.7$ & $0.17$ & \multirow{3}{*}{$--$} \\
& 1 & 11.5 & $9.6$ & $0.17\times0.12$ & $-177.3$ & $--$ & $--$ & $--$ & \\
& 1 & 11.5 & $8.5$ & $0.25\times0.17$ & $-37.1$ & $--$ & $--$ & $--$ & \\
\hline
\multirow{3}{*}{J0858+1041} & 1 & 8.5 & $7.4$ & $0.2\times0.18$ & $81.9$ & $0.22\times0.2$ & 
$85.6$ & $0.90$ & \multirow{3}{*}{$-1.23\pm0.14$} \\
& 1 & 11.5 & $8.4$ & $0.17\times0.14$ & $25.3$ & $0.19\times0.17$ & $172.3$ & $0.59$ & \\
& 1 & 11.5 & $7.9$ & $0.2\times0.18$ & $81.9$ & $0.2\times0.18$ & $172.9$ & $0.62$ & \\
\hline
\multirow{3}{*}{J0930+3430} & 1 & 8.5 & $6.6$ & $0.18\times0.16$ & $-41.0$ & $0.18\times0.17$ & 
$106.6$ & $0.20$ & \multirow{3}{*}{$-0.35\pm0.16$} \\
& 1 & 11.5 & $7.6$ & $0.16\times0.13$ & $-3.9$ & $0.19\times0.13$ & $144.7$ & $0.17$ & \\
& 1 & 11.5 & $8.0$ & $0.18\times0.16$ & $-41.0$ & $0.19\times0.15$ & $147.9$ & $0.18$ & \\
\hline
\multirow{3}{*}{J1152+1903} & 1 & 8.5 & $7.6$ & $0.19\times0.18$ & $29.7$ & $0.24\times0.18$ & 
$30.9$ & $0.20$ & \multirow{3}{*}{$-1.18\pm0.17$} \\
& 1 & 11.5 & $7.8$ & $0.18\times0.14$ & $19.1$ & $0.24\times0.15$ & $27.6$ & $0.11$ & \\
& 1 & 11.5 & $8.0$ & $0.19\times0.18$ & $29.7$ & $0.24\times0.16$ & $26.5$ & $0.14$ & \\
\hline
\multirow{3}{*}{J1715+6008} & 1 & 8.5 & $9.2$ & $0.25\times0.15$ & $66.4$ & $0.31\times0.25$ & 
$139.3$ & $1.12$ & \multirow{3}{*}{$-1.1\pm0.13$} \\
& 1 & 11.5 & $18.8$ & $0.22\times0.11$ & $67.5$ & $0.26\times0.21$ & $140.5$ & $0.66$ & \\
& 1 & 11.5 & $18.5$ & $0.25\times0.15$ & $66.4$ & $0.27\times0.22$ & $138.6$ & $0.80$ & \\
\hline
\multirow{3}{*}{J2254-0051} & 1 & 8.5 & $7.3$ & $0.25\times0.19$ & $-49.1$ & $0.29\times0.19$ & 
$114.8$ & $0.20$ & \multirow{3}{*}{$-0.95\pm0.17$} \\
& 1 & 11.5 & $8.8$ & $0.18\times0.17$ & $71.1$ & $0.21\times0.19$ & $70.5$ & $0.15$ & \\
& 1 & 11.5 & $8.6$ & $0.25\times0.19$ & $-49.1$ & $0.25\times0.19$ & $67.2$ & $0.15$ & \\
\hline
\hline
\end{tabular}
}
\end{center}
\tablecomments{As Table 2, but for the extended radio sources.}
\label{table3}
\end{table*}

\section{Results}\label{results}


Our primary interest is to identify bona fide kpc-scale dual AGNs, which motivated our original long-slit spectroscopy of double-peaked AGNs in the SDSS database (Comerford et al. 2012). 
The method we propose to discriminate between dual AGNs, radio jet-driven outflows, AGN wind-driven outflows\footnote{The various mechanisms to push the gas out from the nucleus include a radiation pressure-driven wind, a thermally-driven wind, or a magnetohydrodynamic (MHD) wind (Crenshaw et al. 2010, M\"uller-S\'anchez et al. 2011).}, and rotating disks
combines high resolution multi-band radio observations and spatially resolved optical spectroscopy. 
First, the nature of the radio emission in double-peaked AGNs is investigated. Then, in Section 3.2 we combine these results with our previous long-slit optical spectroscopy to identify the source of the double-peaked narrow emission lines in each system. 
It is important to point out that in most cases, the use of either one of these observational techniques alone does not provide sufficient information to interpret the origin of the double-peaked, spatially-offset narrow AGN emission lines (see Section~\ref{limitations}).   




\subsection{Nature of the Radio Sources}\label{step1}

Four measurements are used to characterize the radio emission of the sample galaxies: radio morphology, size of the central bright sources, spectral index value of the compact sources, and spatial distribution of the spectral index (spectral index image). The radio morphologies of the double-peaked AGNs in our sample comprise three categories: (1) galaxies presenting a single radio source, (2) galaxies with extended radio features above the $3\sigma$ level, and (3) galaxies showing two independent radio sources (two bright peaks of emission). We used the task IMFIT in CASA to fit a single elliptical Gaussian to each of the compact sources detected in each galaxy (except for J$1158+3231$ in which we fitted simultaneously two elliptical Gaussian components, see Fig. 14). The parameters of the Gaussians (flux density, size, and position angle) are summarized in Tables $2-5$. A source is considered compact if it is unresolved or its size is comparable to the size of the beam (within the errors). 

From the flux densities at 8.5 and 11.5 GHz, we obtained an $X-$band spectral index $\alpha$ (where $S \propto\nu^\alpha$) for each of the Gaussian components. In general, compact flat- and slightly steep-spectrum radio sources ($\alpha\gtrsim-0.8$) are interpreted as AGN\footnote{Typically, a source is considered flat when $\alpha>-0.5$. However, several studies indicate that compact steep-spectrum radio sources are mostly powerful AGNs (O'Dea 1998). Two notable examples are the studies by Rodriguez et al. (2006) and Fu et al. (2011a) who found AGN with spectral indices $-0.6$ and $-0.9$, respectively. In this work compact sources with $\alpha\gtrsim-0.8$ are interpreted as AGNs. This value corresponds to the upper limit of the typical values found for radio jets ($\alpha$ between $-0.7$ and $-0.8$, Pushkarev et al. 2012, Hovatta et al. 2014).}.  
Spectral index images of AGNs typically show a flat spectrum in the region of the core, though the spectrum can be inverted if significant synchrotron self-absorption is occurring (Edwards et al. 2000). These sources are described as young AGNs in the early stages of development, before expansion to a larger size (the spectral peak of an expanding radio source moves towards lower frequencies, Vollmer et al. 2008), and are common among Type 2 quasars (Lal \& Ho 2010). 
Jet components usually have steeper spectra ($\alpha\lesssim-0.8$) due to aging of the electron population, as higher energy electrons lose energy via synchrotron radiation faster than those at lower energies (e.g., Readhead et al. 1979, Fomalont 1981, Marscher 1988, Pushkarev et al. 2012).  
Thus, neglecting edge effects (the spectral index images become more uncertain with decreasing continuum brightness), in AGNs where the core dominates (no evidence for radio jets down to the limits of sensitivity and spatial resolution), only a constant flat spectral index is observed in the region of the radio core, with typical values in the range $-0.8\lesssim\alpha\lesssim1$. 
On the other hand, spatial variations of the spectral index (an overall steepening of the spectra with distance, sometimes even within the size of the VLA beam) is a common characteristic of radio jets (Hovatta et al. 2014). 

We also measured the position angle of the extended radio emission from the lowest $3\sigma$ contours in the total intensity images (PA$_{\mathrm{radio}}$, Table 7). 
For the galaxies showing two compact radio cores, this corresponds to the position angle of the vector between the two sources on the sky. This angle was not measured in galaxies presenting only a single radio core, since this corresponds to the position angle of the elliptical Gaussian (Table 2).

Finally, we estimated the possible contribution from star formation to the radio flux density of our sources. 
Following Kauffmann et al. (2003), typical H$\alpha$ star formation rates for our sample are about 1 M$_\sun$ yr$^{-1}$, leading to an expected contribution of only about 0.02 mJy at 1.4 GHz (Murphy et al. 2011). Assuming a flat spectrum between $-0.3$ to $-0.1$, this corresponds to $\lesssim0.01$ mJy at 11.5 GHz. This is less than $10\%$ of the flux density of each of the sources. The only exception is Source 2 in J1023+3243 in which the 11.5 GHz flux density due to star formation is over a factor of 3 lower than our measured 11.5 GHz flux of 0.03 mJy. Nevertheless, this suggests that an additional emission source, such as a weak AGN, would be needed to provide the missing flux density. Therefore, in our sample galaxies, a contribution from star formation is expected to be negligible. Furthermore, all the galaxies in our sample have optical line ratios that are consistent with AGN-dominated emission rather than star formation (Wang et al. 2009, Liu et al. 2010a). 

Based on the VLA data alone, we classify the 18 galaxies in four categories\footnote{When we add in the optical data, our classifications become more specific. This mostly affects the galaxies classified as single AGNs, which are then divided into single AGNs with wind-driven outflows or single AGNs with rotating NLRs (Section 3.2).}:  

\begin{itemize} 
	\item {\bf Single AGNs:} The VLA images show a single compact radio source. 
The spectral index is flat or inverted ($\alpha\gtrsim-0.8$) and constant in the spectral index images (neglecting edge effects). These are typical characteristics of an AGN. Six galaxies are included in this category (Table 2). 

	\item  {\bf Radio Jets:} Extended jet-like radio structures are detected above the $3\sigma$ level in 7 galaxies (Table 3). The morphologies mostly consist of a small-diameter radio component (in most cases unresolved or marginally resolved) and extended linear features that emanate from one side, or from opposite sides of the radio core. The small central radio components have a steep spectrum $\alpha<-0.9$ (Table 3), which probably corresponds to combined emission from the radio core (the AGN) and the radio jet. The only exceptions are J$0009-0036$ (which shows an inverted spectrum $\alpha=0.36$), and J$0930+3430$ (which has a flat spectrum $\alpha=-0.35$). Therefore, in these two cases, the radio emission of the core is dominated by the AGN. In all cases, the spectral index images show variations of the spectral index with distance along the linear features. It is obvious from the difference in the core and jet spectral indices that some steepening in the index is occurring along the jet, with values of the core and jet components in the range $-3\lesssim\alpha\lesssim2$, typical of jets in AGNs (Hovatta et al. 2014). 
Note that the spectral index image for J$0731+4528$ could not be obtained, since the source was not detected at 11.5 GHz. However, this indicates that the source has a very steep spectrum and is consistent with the jet interpretation. 
	
	\item {\bf Dual AGNs:} Three galaxies show two spatially-separated radio sources in our VLA images (Table 4, Figs. 10, 14 and 16). The properties of each of the sources are consistent with those of single AGNs as described above (they are compact and the spectral index is flat and constant within the FWHM of the sources), which confirm the presence of two AGNs in these galaxies. There is no evidence of extended radio emission, or diffuse radio emission between the sources, which suggests that the morphological structures most likely are not associated with radio jets. This fact and the lack of axial symmetry between the two sources suggest that they are not symmetric hotspots of an obscured AGN\footnote{The two compact sources in J$1158+3231$ exhibit axial symmetry. However, the spectral index of the two sources is flat ($\alpha\sim-0.47$) and constant in the spectral index image, which would be very atypical for symmetric hotspots of an obscured AGN.}
(An \& Baan. 2012, Wrobel et al. 2014b). This is supported by the spectral index analysis. For high-power compact symmetric objects (CSOs), spectral indices steepen with source separation (An \& Baan 2012). Such trends have not yet been investigated for low-power CSOs. But if a similar trend exists, the spectral indices shown in Table 4 would then be very atypical for the source separations of $0.6-1.6$ kpc. Therefore, based on the radio data alone, we conclude that three galaxies in our sample host dual AGNs.\footnote{Note that this does not necessarily imply that the optical double-peaked narrow emission lines are produced by the dual AGNs, as will be discussed in Section 3.2.1.}

	\item {\bf Ambiguous:} The total intensity images show two compact radio sources, but the properties of one of them can not be measured either at 8.5 GHz or 11.5 GHz. The two sources are embedded in general diffuse emission, which suggests the presence of radio jets. Therefore, the two sources in the total intensity images could be interpreted either as knots or hotspots in radio jets, or dual AGNs. Two galaxies exhibit these characteristics (Table 5, Figs. 1 and 11), and will be discussed in detail in Section~\ref{ambiguous}. 
\end{itemize}

These four categories represent our best effort to interpret these galaxies based on the current VLA data. If additional structures exist in the central $3\arcsec\,\times3\arcsec$ of these galaxies (the diameter of the SDSS fiber), 
then they would have to have radio emission below the detection limit in our images (the rms image noises were typically $10\,\mu$Jy beam$^{-1}$, giving a $3\sigma$ detection threshold of $\sim30\,\mu$Jy beam$^{-1}$), or projected separations smaller than our angular resolution ($\sim0.18\arcsec$). Keeping this in mind, we adopt this classification scheme in our following discussion.


\begin{figure*}
\epsscale{.99}
\plotone{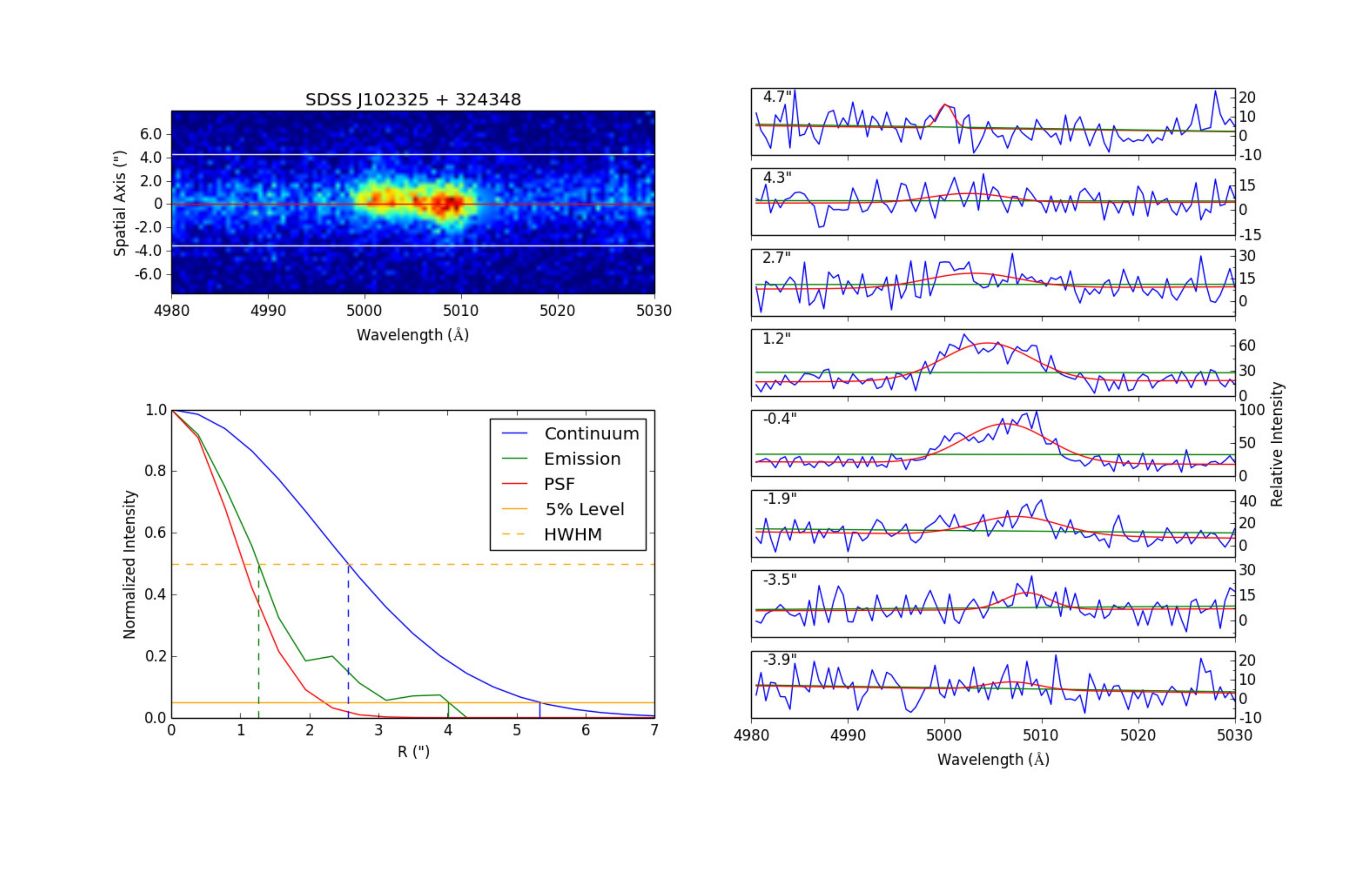}
\caption{Demonstration of the method used to measure the size of the [O III] emitting region in J$1023+3243$.
$Top$ $left$: 2D spectrum of the galaxy centered at the rest wavelength of [O III] $\lambda5007$. The red solid line indicates the center of the continuum. The white lines correspond to the spatial extent of the [O III] emission, defined by the last rows of significant emission (in this case from $-3.9\arcsec$ to $4.3\arcsec$, see Section~\ref{reanalysis} for details). 
$Right$: 1D spectra extracted along the spatial axis. The data are shown in blue and the fit model (Gaussian+linear function) is shown in red. 
The model does not provide a significantly better fit than a single linear function (green line) at distances larger than $4.3\arcsec$ and $-3.9\arcsec$ (this is shown as an example in the 1D spectra at $4.7\arcsec$ and $-3.9\arcsec$). Inside this region the model provides a good fit to the data. 
$Bottom$ $left$: Radial profiles of the [O III] emission, the continuum and the PSF model (Gaussian). The horizontal dashed line indicates the HWHM and the horizontal solid line the flux at 5\% of the peak of emission. 
\label{fig19}}
\end{figure*}

\subsection{Identifying the Sources of Double-Peaked Optical Emission}\label{step2}

Combining the observed radio properties and the measurements obtained from our optical long-slit spectroscopy (Comerford et al. 2012), we now discuss the scenarios that may give rise to the double-peaked narrow emission lines in our AGN sample. 

Our method makes use of morphological parameters derived from optical data\footnote{In some specific cases, it is necessary to incorporate kinematic parameters to distinguish between the different scenarios (see Sections \ref{winds2} and \ref{disks}).}.  
Two parameters are crucial for the identification process: the position angle of the [O III] emission (PA$_{\mathrm{[O III]}}$) and the photometric position angle of the major axis of the galaxy (PA$_{\mathrm{gal}}$). 
In galaxies exhibiting two compact radio sources in the total intensity images (dual AGNs and ambiguous cases, Section~\ref{step1}), the projected spatial separation between the two [O III] emission features ($D_{\mathrm{[O III]}}$) is also relevant.\footnote{As mentioned in Section~\ref{reanalysis}, in general the size of the NLR cannot be used to classify the nature of the double-peaked
[O III] emission lines in SDSS spectra. We discuss the implications of these measurements for each of the scenarios that give rise to the double-peaked emission lines in Section~\ref{size}.}

PA$_{\mathrm{[O III]}}$ and $D_{\mathrm{[O III]}}$ are taken directly from Comerford et al. (2012). 
Since we observed each galaxy at two position angles, the full spatial separation of the two emission components on the sky as well as their position angle could be determined (see details in Comerford et al. 2012). We used SDSS photometry to derive PA$_{\mathrm{gal}}$, except in J$0002+0045$, J$0736+4759$ and J$1715+6008$, where we used $HST$ F106W images (Fig. 21, P.I. Liu). 
The main results of the identification process 
are summarized in Table 7, and each case is discussed individually in the next subsections.

\begin{figure*}
\epsscale{.99}
\plotone{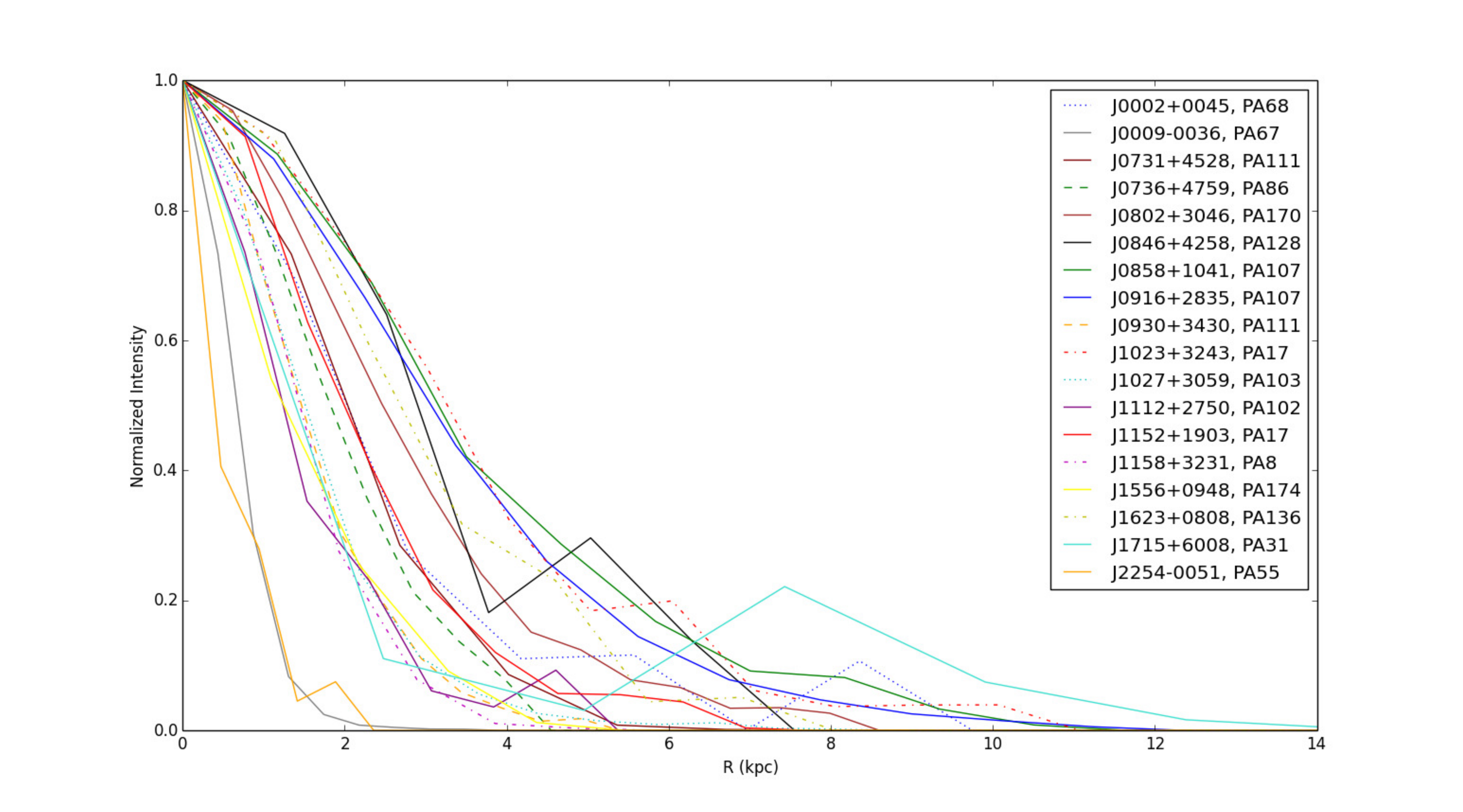}
\caption{Comparison of the radial profiles of [O III] emission for the galaxies observed with the VLA as indicated in the legend. Only shown for PA$_{\mathrm{slit}}$ with most extended emission (Table 6). 
\label{fig20}}
\end{figure*}

\subsubsection{Optical Spectral Signature of Dual AGNs}\label{dual2} 

The radio properties of J$1023+3243$, J$1158+3231$ and J$1623+0808$ confirm the presence of dual AGNs in these galaxies (Table 4). However, this does not necessarily imply that the double-peaked narrow emission lines observed in these objects are produced by the relative motion of the two AGNs. For example, a composite NLR, whose kinematics is dominated by outflows, has been observed in the dual AGN galaxies NGC 6240 and Mrk 266 (Engel et al. 2010, Mazzarella et al. 2012). 

The necessary condition for observing double-peaked narrow emission lines produced by dual AGNs is that the double radio cores have to be spatially coincident with the double [O III] emission components.
In J$1158+3231$ and J$1623+0808$, the two compact radio sources have the same orientation as the double optical emission components in the long-slit spectra, within a typical error of $15\degr$ (PA$_{\mathrm{[O III]}}\sim$PA$_{\mathrm{radio}}$, Table 7), 
and the spatial separation between the two radio cores  is consistent, within the errors, with the spatial separation between the two [O III] emission features ($D_{\mathrm{radio}}\sim\,D_{\mathrm{[O III]}}$, Table 4). These results strongly support the hypothesis that the double-peaked narrow emission lines are produced by the relative motion of the two AGNs in these two SDSS galaxies. In J$1023+3243$, PA$_{\mathrm{[O III]}}\sim$PA$_{\mathrm{radio}}$ but $D_{\mathrm{[O III]}}$ is slightly larger than $D_{\mathrm{radio}}$ (Table 4). 
This indicates the presence of an additional kinematic component, probably associated with outflows.  
Outflowing clouds drive shocks into the surrounding material, with an environmental impact that may be enough to produce peaks of ionized gas emission at different locations along the NLR (see e.g., Mazzalay et al. 2013, Meyer et al. 2015), 
such that the two radio cores do not spatially coincide with the double optical emission components. This effect appears to be occurring close to the dual AGNs in J$1023+3243$. 
$D_{\mathrm{[O III]}}$ is greater than $D_{\mathrm{radio}}$ by only a few tenths of parsecs ($<0.05\arcsec$) and the radio and optical components have the same orientation. 

Interestingly, in these three galaxies
the dual AGNs are aligned with the major axis of the galaxy (Figs. 10, 14 and 16), i.e. they are located in the plane of the galaxy. As noted by Comerford et al. (2012), dual AGNs should preferentially reside in the plane of the host galaxy, since the dynamics of the AGNs are dominated by the host galaxy potential. The implications of this result will be discussed further in Section~\ref{mergers}. 

\begin{figure*}
\epsscale{.99}
\plotone{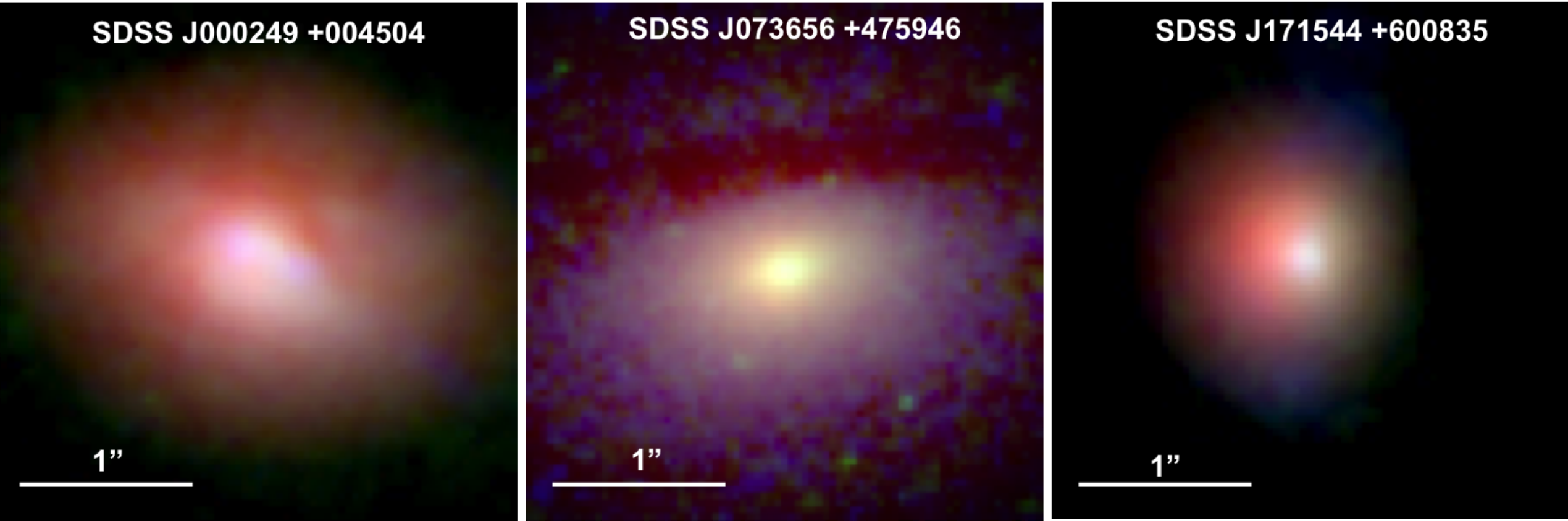}
\caption{Three-color composite images of $HST$ WFC3 medium (F547M and F621M) and wide filters (F106W) of the central $3\arcsec\times3\arcsec$ of J$0002+0045$, J$0736+4759$, and J$1715+6008$. The continuum (F106W) is shown in red, H$\alpha$ (F621M) in green, and the [O III] emission (F547M) in blue. In the three panels North is up and East is to the left. The three images confirm our interpretations based on VLA images and optical long-slit spectroscopy. Note the outflow of ionized gas in J$1715+6008$ and the rotating disk in J$0736+4759$. J$0002+0045$ is an ambiguous case (see Section~\ref{ambiguous} for details). 
\label{fig21}}
\end{figure*}

\subsubsection{Radio Jet-Driven Outflows}\label{jets} 

Radio jet-driven outflows are a convenient way of explaining the kinematics observed in double-peaked AGNs (they push away the ISM in the host galaxy creating complex emission-line profiles), and are easy to identify with our data. If the jet is aligned with the emission-line gas visible in the long-slit spectra (PA$_{\mathrm{[O III]}}\sim$PA$_{\mathrm{radio}}$), then that is confirmation of a radio jet-driven AGN outflow.\footnote{This classification is independent of the orientation of the galaxy disk (PA$_{\mathrm{gal}}$).   
Investigations of the distribution of the angle between the major axis of the galaxy and the radio jet suggest a homogeneously wide distribution in the range $0-90\degr$ (Ulvestad \& Wilson 1984, Kinney et al. 2000, Schmitt et al. 2002). Therefore, the directions of the radio jets are consistent with being completely uncorrelated with the planes of the host galaxies (see also Nagar \& Wilson 1999).} 

In five galaxies with extended radio emission (J$0009-0036$, J$0858+1041$, J$1152+1903$, J$1715+6008$, and J$2254-0051$; Table 3), the PA of the radio jet 
is consistent, within the errors, with the PA of the ionized gas (PA$_{\mathrm{[O III]}}\sim$PA$_{\mathrm{radio}}$; Table 7). 
Moreover, we find that the AGN emission components in the long-slit spectra are spatially extended ($>12$ kpc, except in J$0009-0036$ and J$2254-0051$, see Section~\ref{reanalysis} and Fig. 20), 
which is characteristic of powerful outflows (e.g., Greene et al. 2012, Liu et al. 2013, Hainline et al. 2013). 
These results indicate that 
the double-peaked narrow emission lines in these five SDSS galaxies are produced by the motion of radio jets across the ISM.   

It is important to point out some additional characteristics of one of the galaxies in this category: J$1715+6008$. This galaxy exhibits at 8.5 GHz a central bright source and extended emission. We made cleaned images of the field at 8.5 GHz using several robust parameters and the morphology was always similar to the one shown in Fig. 17. Therefore, the Gaussian fits at 8.5 GHz and the smoothed image at 11.5 GHz indicate an extended source with a steep spectrum, characteristic of radio jets. However, the native resolution image at 11.5 GHz shows a central bright core, and a secondary peak of emission at $\sim0.1\arcsec$ Southeast from the nucleus. Furthermore, the spatial distribution of the spectral index shows an interesting morphology. Two regions of relatively flat spectra are observed, coincident with the two peaks of emission at 11.5 GHz (which are also noticeable in the total intensity image). They appear to have the same shape of the beam, which suggests that they might be two compact radio sources, but the $\sim0.1\arcsec$ separation is unresolved by these observations. However, observations at optical and X-ray wavelengths do not show evidence for a close AGN pair. The separation of the two peaks of radio emission at 11.5 GHz ($D_{\mathrm{radio}}=0.11\pm0.03\arcsec$) is not consistent with the separation of the two [O III] emission components and the location of the two X-ray sources in this galaxy  $D_{\mathrm{[O III]}}=0.68\pm0.16\arcsec$ (Comerford et al. 2011, 2012), which indicates that the double peaks are likely due to an outflow of ionized gas. 
This interpretation is supported by recent $HST$ observations of this galaxy. It can be seen in Fig. 21 that the [O III] emission (in blue) seems to be emanating from the plane of this galaxy as traced by the F106W filter (in red), and is spatially extended. This morphology suggests an outflowing structure of ionized gas. Since the radio emission has the same PA as the [O III] emission (Fig. 17 and Table 7), the outflow appears to be driven by the radio jet. We cannot completely rule out the possibility of a close SMBH binary as indicated by the radio analysis (further observations with higher spatial resolution would be useful to confirm the nature of this object). This, however, does not appear to be responsible for the double peaks in the SDSS spectrum.  


\subsubsection{AGN Wind-Driven Outflows}\label{winds} 

The radio images of the galaxies classified as single AGNs do not provide sufficient information to identify the origin of the double-peaked, spatially-offset narrow AGN emission lines (Table 2). In the absence of extended radio emission and/or a secondary radio core, the most likely scenarios would be either an AGN outflow that is not radio jet-driven or disk rotation. Using morphological information from our optical long-slit data we can 
distinguish between these scenarios. Disk rotation is ruled out when the emission-line gas visible in the long-slit spectra is not located in the plane of the galaxy (PA$_{\mathrm{[O III]}}$$\nsim$PA$_{\mathrm{gal}}$). However, in the cases where PA$_{\mathrm{[O III]}}$$\sim$PA$_{\mathrm{gal}}$, the situation is more complicated since the double-peaked narrow emission lines could be produced either by disk rotation or by an AGN outflow that is intersecting the plane of the galaxy (M\"uller-S\'anchez et al. 2011). In these cases a more detailed analysis of the gas kinematics is necessary to distinguish between these two scenarios.

In five galaxies classified as single AGNs (J$0802+3046$, J$0846+4258$, J$0916+2835$, J$1112+2750$ and J$1556+0948$), the two components of [O III] emission are located at relatively large angles from the galactic disk\footnote{In J$0802+3046$ the misalignment is only $\sim20\degr$ (Table 7), slightly larger than the typical error of $\sim15\degr$.}, indicating that the ionized gas is not rotating in the plane of the galaxy (PA$_{\mathrm{[O III]}}$$\nsim$PA$_{\mathrm{gal}}$). Therefore, we conclude that the double-peaked narrow emission lines in these objects are produced by an AGN wind-driven outflow (Table 7). 

J$0731+4528$ (Table 3) shows an extended jet-like feature along a position angle of $\sim108\degr$, almost perpendicular to the position angle of the two [O III] emission components ($\sim6\degr$, Fig. 3). The distinct orientations of the two structures (the radio jet and the [O III] gas) suggest that the radio jet is not influencing the kinematics of the ionized gas on scales of hundreds of parsecs (PA$_{\mathrm{[O III]}}$$\nsim$PA$_{\mathrm{radio}}$). Therefore, we conclude that the double-peaked narrow emission lines in this galaxy are not produced by radio-jet driven outflows. Since PA$_{\mathrm{[O III]}}$$\nsim$PA$_{\mathrm{gal}}$, the gas is not rotating in the plane of the galaxy, and therefore the most plausible explanation is that the double-peaked [O III] lines are produced by an AGN outflow that is not radio jet-driven. 

As was the case for J$1715+6008$ in the previous section, we now discuss in more detail two galaxies in this category which show additional weak radio components (J$1112+2750$), or additional components at other wavelengths (J$0916+2835$).

In J$0916+2835$ a secondary stellar component has been detected at near-IR wavelengths $1.23\arcsec$ North of the nucleus of the galaxy (Fu et al. 2012). This component, however, does not have a counterpart in [O III] emission (Fu et al. 2012), is very weak at 8.5 GHz ($\sim2\sigma$ level in our 8.5 GHz image), and is not present in our 11.5 GHz image. Furthermore, the size of the [O III] emitting region is $\sim20$ kpc, which is one of the most extended regions in our sample (Fig. 20 and Table 7). Deeper radio observations may enable a firm detection of this secondary source and confirm the system as dual AGNs. However, based on the data we have, we consider J$0916+2835$ as a galaxy with a single radio core, and the optical long-slit spectroscopy suggests an AGN-driven outflow. 

In J$1112+2750$, a second source of radio emission is detected at $\sim0.9\arcsec$ NW from the central radio core (Fig. 12). This source, however, is very weak ($3.4\sigma$ at 8.5 GHz and $2.2\sigma$ at 11.5 GHz). At this sensitivity, the source is not compact, which argues against the presence of a secondary radio core and instead suggests a hotspot-like radio feature. Since its radio properties could not be well measured, we consider this object as a galaxy with a single radio core. The central source is compact and shows a flat spectrum consistent with an AGN (Table 2).

\subsubsection{J$0930+3430$: An AGN wind-driven outflow in the plane of the galaxy}\label{winds2} 

The 8.5 GHz image of J$0930+3430$ shows a one-sided radio jet that is almost perpendicular to the plane of the galaxy and the [O III] emission (Fig. 9). In the 11.5 GHz image, the flux density of this structure is comparable to the residuals in the cleaning process, making its association with the radio jet uncertain. If this structure is not real, then J$0930+3430$ would exhibit the same radio properties of single AGNs (Section~\ref{step1}). But even if the jet is real, it is not influencing the kinematics of the ionized gas, as indicated by the difference in position angles (PA$_{\mathrm{[O III]}}\nsim$PA$_{\mathrm{radio}}$). 

Since PA$_{\mathrm{[O III]}}$$\sim$PA$_{\mathrm{gal}}$ (Table 7, Fig.9), the origin of the double-peaked narrow emission lines in this galaxy is in the plane of the galaxy. As mentioned in Section~\ref{winds}, in order to differentiate between disk rotation and AGN outflows intersecting the plane of the galaxy, we need to study in more detail the kinematics of the gas in our long-slit spectra. For the purpose of this paper, we look for evidence of disturbed kinematics in the spectral profiles of the emission lines\footnote{A detailed analysis of the kinematics (including multi-Gaussian fits, velocity curves, dispersion curves and kinematic modeling of the emission lines) will be presented elsewhere (Nevin et al. 2015, in prep.).}. 
Multiple kinematic components in narrow emission lines are usually associated with outflows (see e.g., Crenshaw et al. 2010 and references therein). Since all our galaxies show double-peaked spectral profiles, we consider more than two kinematic components to be evidence for an outflow\footnote{We deblend the two velocity components in the emission lines by fitting the spectrum with double Gaussian functions and a linear continuum model. The profiles show overall discernible velocity shifts of the two peaks, so we fitted simultaneously the two Gaussians to each part of the spectrum (blue and red).}. High radial velocities also are ideal tracers of outflows as they discount the likelihood that the observed kinematics could be due to rotation, where the line-of-sight velocity is rarely greater than 400 km s$^{-1}$ (M\"uller-S\'anchez et al. 2011). Asymmetric wings (either blueshifted or redshifted) are also indicative of outflow kinematics (Crenshaw et al. 2010, Harrison et al. 2014). Finally, the presence of broad components (broader than forbidden lines from the NLR, typically $\sim500$ km s$^{-1}$ FWHM; Osterbrock \& Pogge 1987), also suggest the presence of outflows, particularly when they are located at some distance from the nucleus (M\"uller-S\'anchez et al. 2011). 

The spectral profiles of J$0930+3430$ show three of the characteristics mentioned above\footnote{The rest of the galaxies studied in this paper also exhibit at least three of these characteristics in their long-slit spectra (Nevin et al. 2015, in prep.). The only exception is J$0736+4759$, which do not show any of these characteristics (Section~\ref{disks} and Fig. 23)}: multiple components, asymmetric wings and  broad components. As can be seen in Fig. 22, the two-Gaussian fits in the central region (from $-0.4\arcsec$ to $0.4\arcsec$) do not provide a good match to the observed profiles, indicating the presence of a third component. Furthermore, the profiles at distances larger than $\sim-0.7\arcsec$ show a broad asymmetric blue wing ( $\sim600$ km s$^{-1}$), which on the other side of the nucleus (at distances larger than $\sim0.7\arcsec$) becomes the dominant component. At these locations, an asymmetric redshifted wing appears. All these characteristics provide enough evidence to conclude that J$0930+3430$ hosts an AGN outflow in the plane of the galaxy. 

\subsubsection{J$0736+4759$: A Rotating Disk}\label{disks}

As is the case for all of the galaxies in Table 2, in J$0736+4759$ we detect only a single flat-spectrum radio source. The ionized gas emission is spatially coincident with the galaxy's major axis (Fig. 4), suggesting that the source of the double-peaked narrow emission lines is in the plane of the galaxy (PA$_{\mathrm{[O III]}}$$\sim$PA$_{\mathrm{gal}}$). 
Recent $HST$ observations of this galaxy support this interpretation. As can be seen in Fig. 21 the [O III] emission and the near-IR continuum are spatially coincident. 
As was the case for J$0930+3430$, a more detailed analysis of the gas kinematics is needed in order to distinguish disk rotation from AGN outflows in the plane of the galaxy. 

The spectral profiles of J$0736+4759$ do not show evidence for disturbed kinematics as described in Section~\ref{winds2}. 
Two Gaussian components provide a good fit to the data (Fig. 23). Furthermore, the two components are narrow (FWHM$<500$ km s$^{-1}$), have low radial velocities ($<150$ km s$^{-1}$) and are symmetric. 
Although there might be a configuration in which an outflow produces narrow symmetric double-peaked profiles (Crenshaw et al. 2010), disk rotation provides a simpler explanation for the observed properties of J$0736+4759$\footnote{A simple rotating disk illuminated by a single AGN can not produce the observed double-peaked narrow emission lines at different locations along the long-slit. Partial obscuration of the galaxy disk or dynamical processes in the disk could explain the kinematics of this galaxy. This will be investigated in a forthcoming publication (Nevin et al. 2015, in prep.).}. 
This is supported by the work of Smith et al (2012), who studied a sample of double-peaked galaxies with symmetrical profiles and nearly similar components (like J$0736+4759$). Based on their statistical sample they argued that a sample of double-peaked galaxies with nearly symmetrical line components have diagnostics in agreement with rotational disk models. We therefore conclude that the most likely origin of the double-peaked narrow emission lines in J$0736+4759$ is disk rotation.

\subsubsection{Ambiguous Cases}\label{ambiguous}


\begin{center}
{\bf {\small J$0002+0045$: Dual AGNs or a Radio Jet-driven Outflow?}}
\end{center}

The radio continuum at 11.5 GHz and the spectral index image of this galaxy suggest the presence of a single AGN with a one-sided radio jet. However, the total intensity image and the radio continuum at 8.5 GHz suggest the presence of two spatially separated radio sources (C1 and C2, Fig. 1 and Table 5). Radio continuum at 11.5 GHz was detected at the location of C2, but the emission is not consistent with a compact source (the elliptical Gaussian that we fit did not converge at that position).  Therefore, the 11.5 GHz emission in this region is extended and the spectral index image shows a steep spectrum ($\alpha\sim-1.3$). Deeper observations at 11.5 GHz would allow us to better measure the properties of this source and might reveal a secondary compact radio source. 

An $HST$ WFC3 [O III] image of this galaxy shows the presence of two compact sources of ionized gas that are spatially coincident with the two radio sources (Fig. 21), as in the galaxies where dual AGNs have been found (see Section~\ref{dual2} and Fu et al. 2011a). 
Furthermore, as can be seen in Figs. 1 and 21, the two sources are located in the plane of the galaxy, which also supports the hypothesis of dual AGNs. 

In summary, the two possible scenarios for the origin of the double-peaked narrow emission lines in this galaxy are the following: (i) dual AGNs with rather unusual radio properties for source 2 (a steep spectrum and a relatively large linear size), or (ii) a one-sided radio jet-driven outflow, which is aligned with the plane of the galaxy. 

 
 \begin{center}
{\bf {\small J$1027+3059$: Dual AGNs with Two Radio Jets or a One-sided Radio Jet?}}
\end{center}
 
This galaxy potentially exhibits the most complex radio morphology of all, but we think this apparent complexity is spurious and mainly stems from residual calibration errors (Fig. 11). As a consequence, we confine our attention only to the two well-detected sources, C1 and C2 (which are detected above $5\sigma$ level).

The properties of C1 are consistent with a compact radio source with flat spectrum, which indicates emission driven by an AGN. C2 has an inverted spectrum. As discussed in Section~\ref{step1}, the current working theory describes these sources as young active galactic nuclei. 
Therefore, we might be seeing in this galaxy dual AGNs, in which one of the nuclei (C1) has been active for a longer period of time than the other (C2), which was activated recently. The morphology of the total intensity image suggests the presence of four linear structures which might be connected to these two AGNs. If these structures are real (and not simply a product of calibration or imaging errors), the `X-shaped' morphology arises from two double-sided radio jets coming from each of the central sources. 

In addition to the dual AGN scenario mentioned above, 
it is possible that C1 and/or C2 could be a jet component lit up in a collision with a dense ISM. If C1 is the core, the inverted spectrum of C2 would be difficult to explain in terms of a radio jet, since radio jets usually have steep spectra (Hovatta et al. 2014). On the other hand, if C2 is the core and C1 is a jet component, the spectral index value of C1 would be in the upper limit of the typical values found for radio jets ($\alpha\sim-0.7$, Pushkarev et al. 2012). Furthermore, the brightness and compactness of this source argue against being a knot of emission in the radio jet.
Even so, this scenario might be plausible and therefore, we consider it as an alternative interpretation to the dual AGNs scenario. In this case, J$1027+3059$ would be a single AGN (C2) with a radio jet extended along a position angle of $\sim85$ degrees.
 
The position angle of the ionized gas emission (PA$_{\mathrm{[O III]}}$) is almost perpendicular to PA$_{\mathrm{gal}}$, and it seems to be aligned with sources C1 and C2 (Fig. 11). 
This fact favors the presence of radio-jet driven outflows. Another interesting result is that the spatial separation between the two radio cores $D_{\mathrm{radio}} =0.22\pm0.04\arcsec$ (Table 5) is not consistent with the spatial separation between the two [O III] emission features $D_{\mathrm{[O III]}}=0.31\pm0.02\arcsec$ (Comerford et al. 2012). This also favors radio-jet driven outflows, since we would expect that the double radio sources spatially coincide with the double optical emission components in the
long-slit spectra (Section~\ref{dual2}, Fu et al. 2011a). 
Therefore, we conclude that the double peaks in this galaxy are likely produced by a radio-jet driven outflow.

However, we cannot completely rule out the presence of dual AGNs. It is also possible that C1 and C2 are both accreting SMBHs, and that the double-peaked narrow emission lines are produced not by the dual AGNs but by outflows associated with one of the AGNs. In this scenario, we would have the three sources (C1, C2 and a double-sided radio jet) aligned at a position angle of $\sim85\degr$ (Fig. 11). The jet is perpendicular to the plane of the galaxy and could be associated either with C1 or C2. C2 is not located in the plane of the galaxy, but due to its inverted spectrum (which suggests a young radio source, Section~\ref{step1}), we would not necessarily expect that it is orbiting in the plane of the host galaxy. This might indicate that we are witnessing a merger in which a dual AGN has recently formed. 

\begin{figure*}
\epsscale{.99}
\plotone{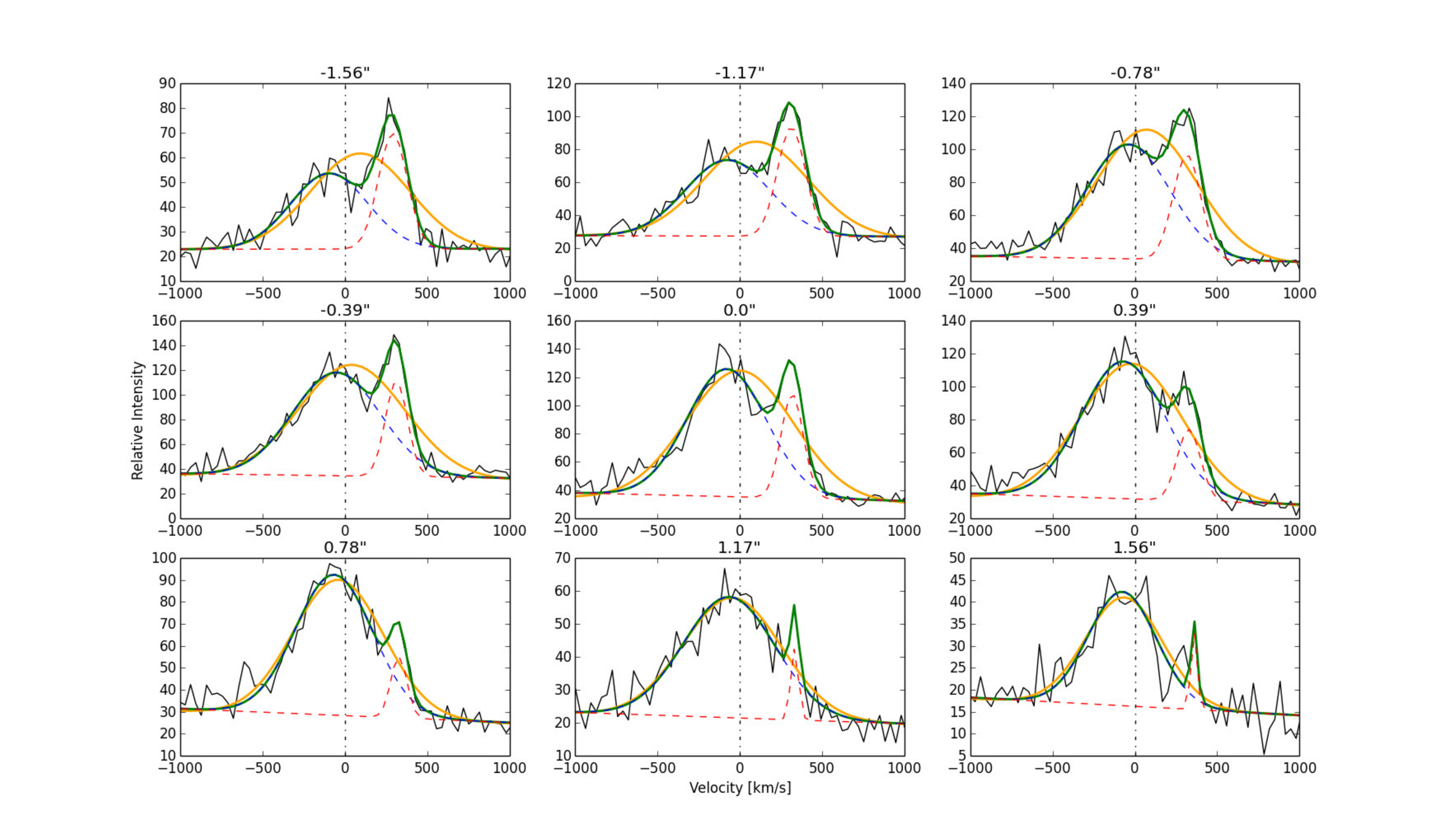}
\caption{Spectral profiles of [O III] emission in J$0930+3430$. The panels show the 1d spectra for slit slices at 9 different spatial locations (flux scale is arbitrary). The number in the upper part of each panel indicates the spatial location along the slit in arcseconds. The dashed-dotted vertical lines correspond to the systemic velocity and divide the emission line into the blueshifted and redshifted components. For the 1d spectrum in each spatial bin, we deblend the two velocity components in the emission lines by fitting the spectrum with double Gaussian functions and a linear continuum model. The profiles show overall discernible velocity shifts of the two peaks, so we fitted simultaneously the two Gaussians to each part of the spectrum (blue and red). The two Gaussian components are represented by blue and red dashed lines, and the sum of the two components is represented by a continuous green line. We also included the single Gaussian fits as described in Section~\ref{reanalysis} (continuous orange line). The two Gaussians provide a good fit to the data (in black), except in the middle row panels (see Section~\ref{winds2} for details).   
\label{fig22}}
\end{figure*}

\begin{figure*}
\epsscale{.99}
\plotone{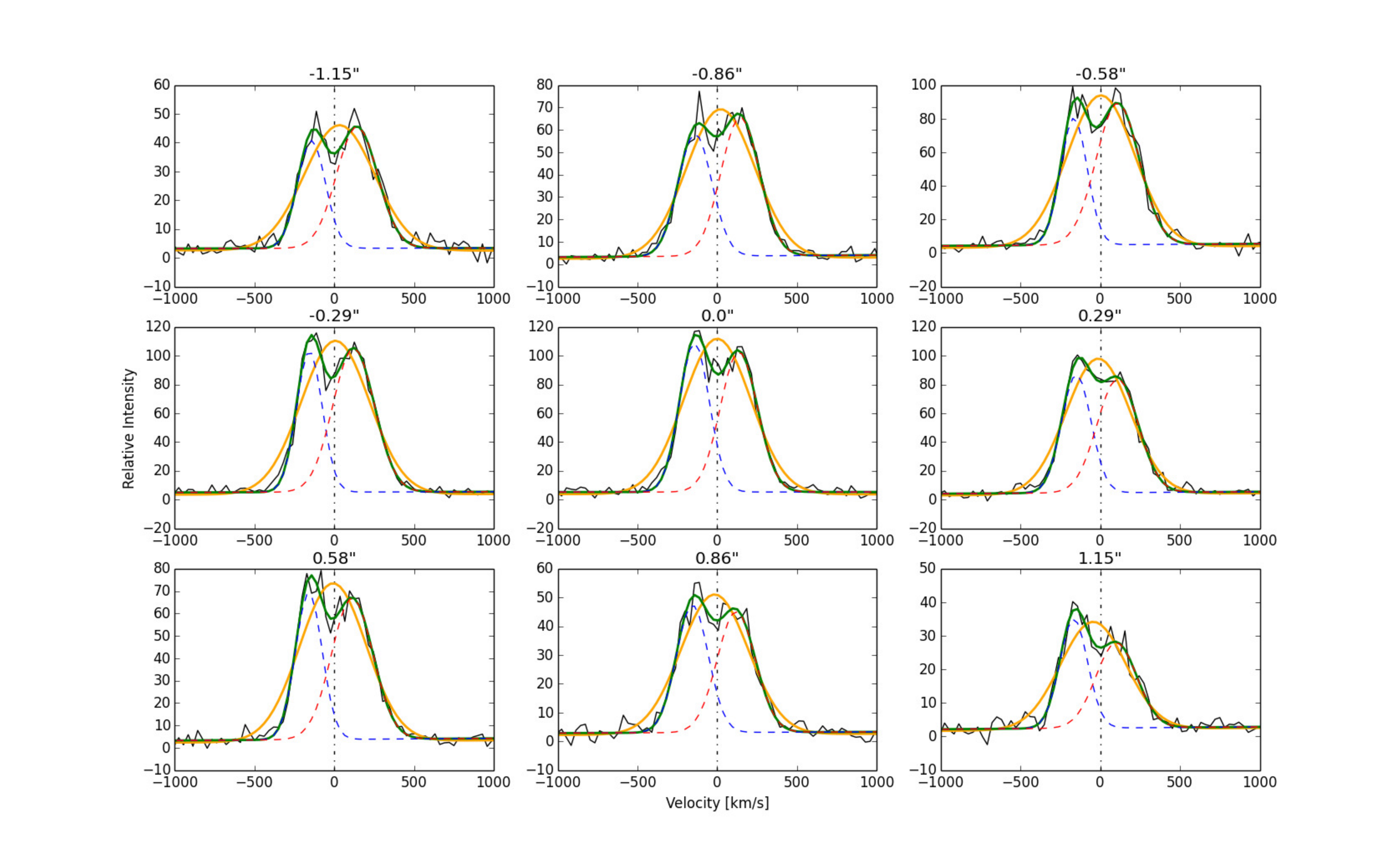}
\caption{Same as Fig. 22 but for J$0736+4759$. Two symmetric Gaussians provide a good fit to the data (in black), suggesting a rotating disk (see Section~\ref{disks} for details). 
\label{fig23}}
\end{figure*}

\begin{figure}
\epsscale{.99}
\plotone{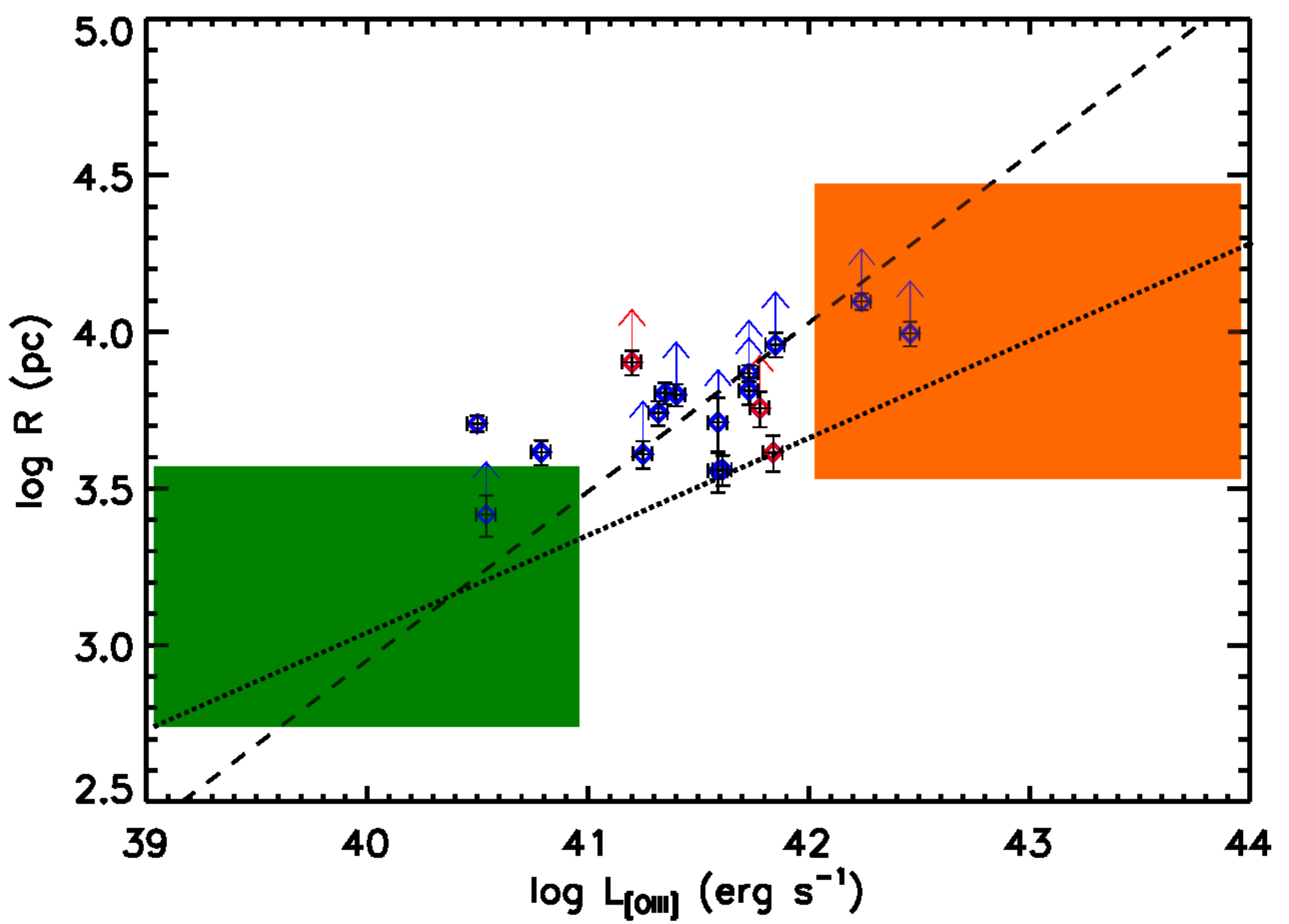}
\caption{Logarithmic plot of NLR size (radius of the [O III]-emitting region) versus [O III] luminosity for the 18 double-peaked AGN studied in this paper. The red points indicate dual AGNs. Points with arrows show lower limits of NLR sizes. The dashed line represents the best linear fit to the data (slope$=0.52\pm0.14$), and the fit from Liu et al. (2013, slope$=0.25\pm0.02$) is shown with a dotted line. The orange rectangle represents the area covered by quasars in this plot, and the green rectangle the area covered by Seyferts (Hainline et al. 2013).
\label{fig24}}
\end{figure}


\begin{turnpage}
\begin{table*}
\caption[Summary of Radio Properties]{Summary of Radio Properties of the Confirmed Dual AGNs.}
\begin{center}
{\scriptsize
\begin{tabular}{l c c c | c c | c c c | c c c}
\hline
\hline \noalign{\smallskip}
 & & & & \multicolumn{2}{c}{Beam parameters} \vline & \multicolumn{3}{c}{Source parameters} \vline & & \\
 \hline
SDSS name & Source & Frequency & rms & $\theta_M \times \theta_m$ & PA & $\theta_M \times \theta_m$ & 
PA & Flux Density & $\alpha$ & $D_{\mathrm{radio}}$ & $D_{\mathrm{[O III]}}$ \\
  &  & (GHz) & ($\mu$Jy beam$^{-1}$) & ($\arcsec$) & ($\degr$E of N) & ($\arcsec$) & ($\degr$E of N) & (mJy) & & ($\arcsec$) & ($\arcsec$) \\
\hline \noalign{\smallskip}
\multirow{6}{*}{J1023+3243} & 1 & 8.5 & $7.2$ & $0.18\times0.16$ & $-43.0$ & $0.22\times0.16$ & 
$131.0$ & $0.25$  & \multirow{3}{*}{$0.25\pm0.16$} & $0.45\pm0.05$ & \multirow{3}{*}{$0.58\pm0.03$}\\
 & 1 & 11.5 & $8.3$ & $0.16\times0.13$ & $-2.2$ & $0.23\times0.16$ & $148.3$ & $0.27$ & & $0.42\pm0.06$ \\
& 1 & 11.5 & $8.7$ & $0.18\times0.16$ & $-43.0$ & $0.22\times0.16$ & $146.6$ & $0.27$ & & $0.42\pm0.06$ \\
& 2 & 8.5 & $7.2$ & $0.18\times0.16$ & $-43.0$ & $0.26\times0.25$ & $178.0$ & $0.06$ & \multirow{3}{*}{$-0.29\pm0.19$} & $0.45\pm0.05$ & \multirow{3}{*}{$0.58\pm0.03$}\\
& 2 & 11.5 & $8.3$ & $0.16\times0.13$ & $-2.2$ & $0.15\times0.13$ & $59.0$ & $0.03$  & & $0.42\pm0.06$ \\
& 2 & 11.5 & $8.7$ & $0.18\times0.16$ & $-43.0$ & $0.22\times0.16$ & $121.8$ & $0.055$ & & $0.42\pm0.06$ \\
\hline
\multirow{6}{*}{J1158+3231} & 1 & 8.5 & $8.0$ & $0.18\times0.17$ & $-49.0$ & $0.2\times0.18$ & 
$35.4$ & $0.3$ & \multirow{3}{*}{$-0.47\pm0.17$} & $0.22\pm0.04$ & \multirow{3}{*}{$0.27\pm0.01$}\\
 & 1 & 11.5 & $8.5$ & $0.17\times0.14$ & $-174$ & $0.19\times0.14$ & $37.8$ & $0.25$ & & $0.22\pm0.04$ \\
& 1 & 11.5 & $8.8$ & $0.18\times0.17$ & $-49.0$ & $0.2\times0.18$ & $30.6$ & $0.26$ & & $0.22\pm0.04$ \\
& 2 & 8.5 & $8.0$ & $0.18\times0.17$ & $-49.0$ & $0.19\times0.17$ & 
$35.4$ & $0.29$ & \multirow{3}{*}{$-0.49\pm0.17$} & $0.22\pm0.04$ & \multirow{3}{*}{$0.27\pm0.01$}\\
& 2 & 11.5 & $8.5$ & $0.17\times0.14$ & $-174.0$ & $0.18\times0.15$ & $37.8$ & $0.23$ & & $0.22\pm0.04$ \\
& 2 & 11.5 & $8.8$ & $0.18\times0.17$ & $-49.0$ & $0.19\times0.17$ & $30.6$ & $0.25$ & & $0.22\pm0.04$ \\
\hline
\multirow{6}{*}{J1623+0808} & 1 & 8.5 & $8.7$ & $0.27\times0.19$ & $-62.1$ & $0.28\times0.20$ & 
$-60.3$ & $0.24$ & \multirow{3}{*}{$0.26\pm0.16$} & $0.47\pm0.05$ & \multirow{3}{*}{$0.37\pm0.06$}\\
 & 1 & 11.5 & $10.3$ & $0.22\times0.17$ & $-80.7$ & $0.25\times0.18$ & $87.8$ & $0.25$ & & $0.44\pm0.05$ \\
& 1 & 11.5 & $10.8$ & $0.27\times0.19$ & $-62.1$ & $0.27\times0.20$ & $88.6$ & $0.26$ & & $0.47\pm0.05$ \\
& 2 & 8.5 & $8.7$ & $0.27\times0.19$ & $-62.1$ & $0.27\times0.20$ & 
$-67.4$ & $0.10$ & \multirow{3}{*}{$-0.17\pm0.18$} & $0.47\pm0.05$ & \multirow{3}{*}{$0.37\pm0.06$}\\
& 2 & 11.5 & $10.3$ & $0.22\times0.17$ & $-80.7$ & $0.23\times0.17$ & $14.0$ & $0.09$ & & $0.44\pm0.05$ \\
& 2 & 11.5 & $10.8$ & $0.27\times0.19$ & $-62.1$ & $0.27\times0.19$ & $5.9$ & $0.095$ & & $0.47\pm0.05$ \\
\hline
\hline
\end{tabular}
}
\end{center}
\tablecomments{As Table 2, but for the confirmed dual AGNs. The parameter $D_{\mathrm{radio}}$ indicates the distance between the two strongest radio sources detected in each image. The uncertainty in $D_{\mathrm{radio}}$ was  
calculated via standard error propagation of the errors in the positions of the two sources. 
$D_{\mathrm{[O III]}}$ indicates the projected spatial separation between the two [O III] emission features. 
(from Comerford et al. 2012).}
\label{table4}
\end{table*}
\end{turnpage}


\begin{turnpage}
\begin{table*}
\caption[Summary of Radio Properties]{Summary of Radio Properties of the Ambiguous Cases.}
\begin{center}
{\scriptsize
\begin{tabular}{l c c c | c c | c c c | c c c}
\hline
\hline \noalign{\smallskip}
 & & & & \multicolumn{2}{c}{Beam parameters} \vline & \multicolumn{3}{c}{Source parameters} \vline & & \\
 \hline
SDSS name & Source & Frequency & rms & $\theta_M \times \theta_m$ & PA & $\theta_M \times \theta_m$ & 
PA & Flux Density & $\alpha$ & $D_{\mathrm{radio}}$ & $D_{\mathrm{[O III]}}$ \\
  &  & (GHz) & ($\mu$Jy beam$^{-1}$) & ($\arcsec$) & ($\degr$E of N) & ($\arcsec$) & ($\degr$E of N) & (mJy) & & ($\arcsec$) & ($\arcsec$) \\
\hline \noalign{\smallskip}
\multirow{6}{*}{J0002+0045} & 1 & 8.5 & $6.6$ & $0.22\times0.18$ & $-46.0$ & 
$0.24\times0.21$ & $96.1$ & $ 0.35$ & \multirow{3}{*}{$-0.62\pm0.15$} & $0.39\pm0.05$ & \multirow{3}{*}{$0.58\pm0.05$}\\
 & 1 & 11.5 & $10.3$ & $0.19\times0.17$ & $-73.0$ & $0.25\times0.18$ & $77.3$ & $0.25$ & & $--$ \\
& 1 & 11.5 & $10.5$ & $0.22\times0.18$ & $-46.0$ & $0.25\times0.18$ & $77.6$ & $0.29$ & & $--$ \\
& 2 & 8.5 & $6.6$ & $0.22\times0.18$ & $-46.0$ & $0.38\times0.27$ & $72.1$ & $0.22$ & 
\multirow{3}{*}{$--$} & $0.39\pm0.05$ & \multirow{3}{*}{$0.58\pm0.05$}\\
& 2 & 11.5 & $10.3$ & $0.19\times0.17$ & $-73.0$ & $--$ & $--$ & $--$ & & $--$ \\
& 2 & 11.5 & $10.5$ & $0.22\times0.18$ & $-46.0$ & $--$ & $--$ & $--$ & & $--$ \\
\hline
\multirow{6}{*}{J1027+3059} & 1 & 8.5 & $7.4$ & $0.18\times0.16$ & $-47.0$ & $0.26\times0.20$ & 
$93.8$ & $0.18$ & \multirow{3}{*}{$-0.83\pm0.19$} & $--$ & \multirow{3}{*}{$0.31\pm0.02$}\\
 & 1 & 11.5 & $8.0$ & $0.16\times0.13$ & $-64.2$ & $0.28\times0.19$ & $107.2$ & $0.14$ & & $0.22\pm0.04$\\
& 1 & 11.5 & $8.7$ & $0.18\times0.16$ & $-47.0$ & $0.26\times0.19$ & $113.4$ & $0.14$ & & $0.22\pm0.04$\\
& 2 & 8.5 & $7.4$ & $0.18\times0.16$ & $-47.0$ & $--$ & $--$ & $--$ & \multirow{3}{*}{$--$} & $--$ & \multirow{3}{*}{$0.31\pm0.02$}\\
& 2 & 11.5 & $8.0$ & $0.16\times0.13$ & $-64.2$ & $0.17\times0.16$ & $4.6$ & $0.08$ & & $0.22\pm0.04$\\
& 2 & 11.5 & $8.7$ & $0.18\times0.16$ & $-47.0$ & $0.18\times0.16$ & $-4.9$ & $0.08$ & & $0.22\pm0.04$ \\
\hline
\hline
\end{tabular}
}
\end{center}
\tablecomments{As Table 4, but for the ambiguous cases. 
}
\label{table5}
\end{table*}
\end{turnpage}

\begin{table*}[h]
\caption{Summary of results from the analysis of the spatial extent of [O III] emission.} 
\begin{center}
{\scriptsize
\centering
\begin{tabular}{  c c c c c c c c c } 
\toprule 
 \multicolumn{9}{c}{} \\ 
SDSS ID&  PA$_{\mathrm{slit}}$ & PA$_{\mathrm{[O III]}}$  &  Total Size\tablenotemark{a}  &FWHM$_{\mathrm{cont}}$ & FWHM$_{\mathrm{[O III]}}$ & Size 5\% cont &Size 5$\%$ [O III] & FWHM$_{\mathrm{PSF}}$ \\
&&& $\arcsec$(kpc) & $\arcsec$ & $\arcsec$ & $\arcsec$ & $\arcsec$ & $\arcsec$ \\

\midrule 

\midrule
 J0002+0045 &68&63.9& 7.80 (12.64)&4.09&2.27&8.68&7.80&1.59\\
&158&&7.80 (12.64) & 3.25&2.06&6.99&4.58&\\

\midrule
J0009--0036&23&54.2&4.03 (5.60)&1.23&0.77&2.65&1.85&0.94\\ 

&67&&5.18 (7.20)&1.35&0.89&2.83&2.05&\\

\midrule

 J0731+4528&21&6&3.12 (4.90)&4.34&3.13&9.12&3.12&3.67\\

&111& &7.02 (11.02)&4.36&2.36&9.15&5.39&\\

\midrule

 J0736+4759&86&98.7&4.03 (7.17)&2.78&1.85&5.79&4.03&0.91\\

&176&&4.03 (7.17)&1.79&1.20&3.77&3.12&\\
\midrule

J0802+3046&80&150.0&9.34 (13.54)&4.09&3.13&8.52&7.49&3.71\\

&170&&10.11 (14.66)&4.89&3.13&10.17&8.17&\\

\midrule
J0846+4258&38&175.7& 2.88 (10.20)&1.81&1.50&3.81&2.78&0.99\\

&128& &2.88 (10.20)&2.51&1.33&5.23&2.88&\\
\midrule

J0858+1041&17&157.3&5.45 (14.06)&3.37&2.27&7.01&5.20&2.09\\

&107&&7.00 (18.06) &3.25&2.11&6.81&5.96&\\
\midrule

J0916+2835&17&40.8&7.00 (17.50)&2.73&2.14&5.73&5.28&2.09\\

&107&&7.80 (19.50)&2.89&2.12&6.06&5.38&\\

\midrule

J0930+3430&21&105.9&4.67 (5.51) &2.04&1.68&4.33&3.84&3.04\\
&111&&7.00 (8.26)&3.82&2.33&7.98&5.73&\\

\midrule

J1023+3243&17&137.4&7.00 (15.90)&5.11&2.52& 10.66&5.81&2.09\\

&107&&5.72 (13.00)&3.72&2.46&7.75&5.72&\\
\midrule

 J1027+3059&13&76.3&4.61 (10.28)&1.60&1.04 &3.37 &2.43&1.13\\

&103&&5.76 (12.84)&1.55&1.19&3.27&3.00&\\
\midrule

 J1112+2750&12&4.9&10.92 (10.16)&2.98&2.86&6.22&6.87&2.41\\
&102&&10.14 (9.43)&4.14&2.51&8.77&10.07&\\

\midrule
 J1152+1903&17&60.6&7.00 (12.53)&3.85&2.00&8.04&5.79&2.09\\
&107&&5.45 (9.76)&3.20&2.18&6.71&5.45&\\

\midrule
 J1158+3231&8&13.0&2.30 (6.53)&1.24&0.74&2.66&1.90&1.02\\
&98&&2.88 (8.18)&1.05&0.86&2.24&1.97&\\
\midrule

J1556+0948&46&134.7&6.24 (8.11)&2.54&1.74&5.53&6.24&1.98\\
&174&&3.11 (4.04)&2.96&1.82&6.19&3.11&\\

\midrule
J1623+0808&46&28.9&2.88 (9.50) &1.67&1.58&3.48&2.88&1.02\\
&136&&3.46 (11.41)&1.66&1.34&3.46&3.46&\\

\midrule

J1715+6008&31&145.6&10.92 (29.60)&2.57&0.88&5.62&6.89&1.63\\
&121&&4.68 (12.68)&2.47&1.02&5.41&4.68&\\

\midrule

J2254--0051 &35&117.4&2.88 (4.32)&1.25&0.76&2.68&2.25&0.94\\
&55&&3.46 (5.20)&1.74&0.49&3.66&2.42&\\

\midrule 
\bottomrule 
\end{tabular}
}
\tablecomments{
$^\mathrm{a}$Size of the [O III] emitting region (based on the Akaike information criterion, see Section~\ref{reanalysis}). The errors in PA$_{\mathrm{[O III]}}$ and total size are shown in Table 7, but omitted in this Table for compactness.}
\end{center}
\label{table6} 
\end{table*}

\section{Discussion}\label{discussion}

\subsection{Size of the NLR}\label{size}

In Section~\ref{reanalysis} we developed a quantitative technique to revise the previous visual analysis of the spatial extent of sources by Comerford et al. (2012). Here we discuss the implications for each of the scenarios that give rise to the double-peaked narrow emission lines in Section~\ref{step2}. 

The FWHM of the [O III] emission ranges from 1 to 3.5 kpc with a mean FWHM$_{\mathrm{[O III]}}=2.3$ kpc, while the total size of the [O III] emitting region ranges from 5 to 30 kpc with a mean size of 12.2 kpc (Fig. 20 and Table 7). These measurements reflect the size of the compact bright core and the extended component, respectively. Interestingly, galaxies with outflows cover the whole range of sizes (the most compact [O III] emitting region is observed in a galaxy with outflows, J$2254-0051$), while the only galaxy classified as rotating disk is compact (FWHM and total size are smaller than the mean values). The three dual AGNs discovered in this work appear in the three regimes (as compared with the mean value of total size of [O III] emission): compact (J$1158+3231$), moderately extended (J$1623+0808$), and extended (J$1023+3243$). This suggests that searches for dual AGNs should not use the assumption that the [O III] emitting region in dual AGNs is compact, since the region ionized by the combined effect of the two AGNs rotating in a galactic potential is probably larger than in the case of single AGNs. Furthermore, the gas kinematics in this region should also be complex with inflows and outflows in each of the two AGNs (as in the prototypical dual AGN system NGC 6240, Engel et al. 2010). Finally, we note that some of the normalized [O III] radial profiles (Fig. 20) exhibit a bump on the descending part of the curve after the FWHM level. This is only observed in galaxies with outflows (and one dual AGN system, J$1023+3243$), and can be attributed to low surface brightness outflowing clumps which appear as diffuse components in the 2D spectra. The fact that J$1023+3243$ exhibits this component suggests that one or both of the AGNs also has visible emission from gas kinematics or jets. This is supported by the large size of the [O III] emitting region observed in this galaxy, and the fact that $D_{\mathrm{[O III]}}$ is slightly larger than $D_{\mathrm{radio}}$ (Section~\ref{dual2} and Table 4).

\subsubsection{Size-Luminosity Relation}

An important question in the study of AGNs is how the size of the NLR scales with the luminosity of the central source. So far, most of the work on this subject has been concentrated on Seyferts (Schmitt et al. 2003, Fraquelli et al. 2003, Bennert et al. 2006) and nearby quasars (Bennert et al. 2002, Greene et al. 2011, Husemann et al. 2013, Liu et al. 2013, Hainline et al. 2013), but this is an unexplored territory for double-peaked AGNs\footnote{Most of the galaxies in our sample have [O III] luminosities ($L_{\mathrm{[O III]}}$) in the range $10^{41-42}$ erg s$^{-1}$. This would classify them as luminous Seyferts (Schmitt et al. 2003). As can be seen in Fig. 2 of Hainline et al. (2013), there exist very few measurements of the size of the NLR in this range of luminosities. Our data provide the first measurements at these luminosities (see  Fig. 24).}. 

The sample and measurements being used in this paper are ideal to determine if there is a correlation between NLR size and luminosity in double-peaked AGNs\footnote{It is important to point out that in 11 galaxies these measurements represent a lower limit on the size of the emitting region, since PA$_{\mathrm{slit}}$ is not oriented along PA$_{\mathrm{[O III]}}$ (the position angle of the most extended emission, Table 6).}. 
If such a correlation exists for these galaxies, 
we can use this to improve our knowledge about the ionization structure of the NLR and the formation of double-peaked narrow emission lines in galaxies, as well as derive an extra input in photoionization/shock models. 

We plot the derived NLR sizes (Table 7) against $L_{\mathrm{[O III]}}$ (the total [O III] luminosity within the $3\arcsec$ SDSS fiber; Wang et al. 2009, Liu et al. 2010a, Comerford et al. 2012) in Fig. 24. 
This plot also presents a linear fit to the data (dashed line), which gives $R_{\mathrm{NLR}}\propto\,L_{\mathrm{[O III]}}^{0.52\pm0.14}$, with a correlation coefficient of 0.67. 
This relation goes as approximately $R_{\mathrm{NLR}}\propto\,L_{\mathrm{[O III]}}^{1/2}$, similar to the result found for type 2 quasars by Bennert et al. (2002), but steeper than the observed relationship between $R_{\mathrm{NLR}}$ and  $L_{\mathrm{[O III]}}$ for Seyfert 2 galaxies ($R_{\mathrm{NLR}}\propto\,L_{\mathrm{[O III]}}^{0.33}$, Schmitt et al. 2003). These results contrast with the shallow slope of $0.22-0.25$ found for radio-quiet obscured quasars (Greene et al. 2012, Liu et al. 2013)\footnote{We only discuss results for Type 2 AGN, since all of our double-peaked AGNs are also Type 2. Schmitt et al. (2003) found that Seyfert 1's have a steeper slope with a larger uncertainty, which suggests that Type 1 AGNs probably follow a different size-luminosity relation.}. 


The NLR sizes from the galaxies in our AGN sample are slightly larger than what has been measured in other studies of active galaxies for the same range of [O III] luminosities ($L_{\mathrm{[O III]}}$ between $10^{40.5}$ to $10^{42.5}$ erg s$^{-1}$; see Hainline et al. 2013 and references therein). As can be seen in Fig. 24, our galaxies lie above the fit from Liu et al. (2013, slope$=0.25\pm0.02$, dotted line). We attribute this and the steeper slope of the observed double-peaked AGNs to an additional contribution of shocks to the ionized gas emission. All of the distinct scenarios proposed for the origin of double-peaked narrow lines (Section 3.2) suggest the presence of shocks (including dynamical processes in rotating disks and dual AGNs in galaxy mergers, see e.g., Medling et al. 2015). Shock-enhanced emission can alter measurements of physical properties in host galaxies, including size and morphology of the NLR and galaxy's emission line ratios (Allen et al. 2008). It is also possible that since several double-peaked AGN are related to interacting systems, presenting tidal tails and bridges (Liu et al. 2013), these interactions can provide debris at large distances from the nucleus, which can be ionized by the central engine, increasing the size of the NLR. This could be also the case for the QSOs studied in Bennert et al. (2002), which also have a slope of $\sim0.5$ and are usually found in merging systems (Martini 2004). 



\subsection{Compact Symmetric Object Candidates}\label{cso} 

Compact symmetric objects (CSOs) are small ($\sim1$ kpc) radio sources that have symmetric double lobes or jets. 
The compact double components show steep radio spectra, similar to the lobes in large radio galaxies. In some CSOs, the core is clearly visible (Peck \& Taylor 2000; Gugliucci et al. 2005), whereas in most CSOs the central cores are too weak to be detected. The dominant theory for the small size of these objects is that they are young radio sources that could grow into larger radio galaxies. 

The `S-symmetry', sometimes observed in CSOs (Tremblay et al. 2009, Deane et al. 2014), has long been predicted to be associated with the presence of binary black holes (Begelman, Blandford \& Rees 1980). Therefore, these systems are of great interest, since they might be tracers of merger-induced SMBH growth and potential gravitational wave emitters. To date the closest separation SMBH binary system ($\sim7$ pc), 0402+379, has been found in a CSO (Rodriguez et al. 2006). More recently, Deane et al. (2014) re-analyzed archival VLA data of the dual AGN system SDSS J$150243.09+111557.3$ and found that the point-source-subtracted VLA 5GHz residual image contained extended `S-shaped' radio emission in the southern nucleus. 
They interpreted this as evidence for a binary SMBH system and later, through VLBI observations with a spatial resolution of $9\times6$ mas, confirmed the presence of two compact flat-spectrum radio sources, converting this object into a triple SMBH system. This interpretation, however, is being challenged by recent VLBA observations with three times better spatial resolution, which reveal two extended radio sources at the position of the southern nucleus (Wrobel et al. 2014b). This morphology is consistent with a double-hotspot scenario, in which both hotspots are energized by a single obscured SMBH. 

J$0009-0036$, J$0731+4528$ and J$1152+1903$ present morphologies reminiscent of CSOs, and might be included in this class: the compact radio core resides between $<1$ kpc jets that have approximately an `S-symmetry' and a steep-spectrum. Further observations with VLBI would be required to address the putative presence of binary SMBHs in these systems.

\begin{table*}
\caption[Summary of results]{Summary of results.}
\begin{center}
{\scriptsize
\begin{tabular}{l c c c c c c}
\hline
\hline \noalign{\smallskip}
 SDSS name & Radio Classification & PA$_{\mathrm{radio}}$\tablenotemark{a} & PA$_{\mathrm{[O III]}}$\tablenotemark{b} & PA$_{\mathrm{gal}}$\tablenotemark{c} & Size [O III]\tablenotemark{d} & Origin double peaks\tablenotemark{e} \\
  & & ($\degr$E of N) & ($\degr$E of N) & ($\degr$E of N) & (kpc) & \\
 \hline
J$0002+0045$ & ambiguous & $64.8\pm6.4$ & $63.9\pm5.0$ & $64.8\pm3.1$ & $12.64\pm1.29$ & ambiguous \\
J$0009-0036$ & two-sided radio jet & $40.5\pm6.8$ & $51.2\pm3.9$ & $135.3\pm3.9$ & $7.20\pm0.43$ & radio jet-driven outflow \\
J$0731+4528$ & two-sided radio jet & $108.3\pm5.9$ & $6.2\pm6.7$ & $154.5\pm6.7$ & $11.02\pm1.08$ & AGN-driven outflow \\
J$0736+4759$ & single AGN & $--$ & $98.7\pm3.8$ & $94.0\pm3.0$ & $7.17\pm0.74$ & rotating disk \\
J$0802+3046$ & single AGN & $--$ & $150.0\pm1.9$ & $170.5\pm1.9$ & $14.66\pm0.87$ & AGN-driven outflow \\ 
J$0846+4258$ & single AGN & $--$ & $175.7\pm8.6$ & $51.9\pm8.6$ & $10.20\pm1.59$ & AGN-driven outflow \\
J$0858+1041$ & one-sided radio jet & $147.0\pm7.8$ & $157.3\pm3.5$ & $85.3\pm3.5$ & $18.06\pm0.83$ & radio jet-driven outflow \\
J$0916+2835$ & single AGN & $--$ & $40.8\pm1.3$ & $81.4\pm1.3$ & $19.50\pm1.22$ & AGN-driven outflow \\ 
J$0930+3430$ & one-sided radio jet & $177.0\pm9.2$ & $105.9\pm2.2$ & $110.3\pm2.2$ & $8.26\pm0.70$ & AGN-driven outflow \\
J$1023+3243$ & dual AGNs & $141.3\pm8.3$ & $137.4\pm8.5$ & $125.3\pm8.5$ & $15.90\pm1.73$ & dual AGNs\tablenotemark{f} \\ 
J$1027+3059$ & ambiguous & $85.2\pm8.7$ & $76.3\pm2.4$ & $167.5\pm2.4$ & $12.84\pm0.64$ & ambiguous \\ 
J$1112+2750$ & single AGN & $--$ & $4.9\pm6.5$ & $87.4\pm6.5$ & $10.16\pm0.84$ & AGN-driven outflow \\
J$1152+1903$ & two-sided radio jet & $48.6\pm7.7$ & $60.2\pm5.1$ & $37.0\pm5.1$ & $12.53\pm0.99$ & radio jet-driven outflow \\ 
J$1158+3231$ & dual AGNs & $21.2\pm7.1$ & $13.0\pm4.5$ & $8.0\pm4.5$ & $8.18\pm0.9$ & dual AGNs \\ 
J$1556+0948$ & single AGN & $--$ & $134.7\pm6.6$ & $159.8\pm6.6$ & $8.11\pm1.16$ & AGN-driven outflow \\ 
J$1623+0808$ & dual AGNs & $29.5\pm7.5$ & $27.9\pm4.8$ & $46.4\pm4.8$ & $11.41\pm0.9$ & dual AGNs \\ 
J$1715+6008$ & one-sided radio jet & $147.5\pm7.0$ & $145.6\pm6.8$ & $33.9\pm3.5$ & $29.60\pm1.48$ & radio jet-driven outflow \\ 
J$2254-0051$ & one-sided radio jet & $110.2\pm7.5$ & $117.4\pm5.3$ & $114.8\pm5.3$ & $5.20\pm0.64$ & radio jet-driven outflow \\
\hline
\hline
\end{tabular}
}
\tablecomments{
$^\mathrm{a}$Position angle of the extended radio emission. This was measured from the lowest $3\sigma$ contours in the total intensity images. For the dual AGNs, this corresponds to the position angle of the vector between the two compact radio cores on the sky. The typical error is $\sim8\degr$. This angle was not measured in galaxies presenting only a single radio core, since this corresponds to the PA of the elliptical Gaussian (Table 2). \newline
$^\mathrm{b}$Position angle of the two components of [O III] emission, from Comerford et al. (2012). The typical error in PA$_{\mathrm{[O III]}}$ is $\sim7\degr$ (Comerford et al. 2012). \newline
$^\mathrm{c}$Position angle of the photometric major axis of the galaxy, from SDSS photometry (with the same error as in Column 4, Comerford et al. 2012), except in J$0002+0045$, J$0736+4759$ and J$1715+6008$, where we used $HST$ F106W images (Fig. 21). \newline
$^\mathrm{d}$Total extent of the [O III] emitting region from Table 6. \newline
$^\mathrm{e}$``AGN-driven outflows'' refers to the cases which are not radio-jet driven. Radiation pressure, magnetic fields, and thermal winds from the AGN could all power the outflow (e.g., M\"uller-S\'anchez et al. 2011). See Section~\ref{ambiguous} for details on the ambiguous cases. \newline
$^\mathrm{f}$An outflowing velocity component is present in this galaxy and might have a significant impact on the velocity offsets of the two peaks (Sections 3.2.1 and 4.1).
}
\end{center}
\label{table7}
\end{table*}

\subsection{Dual AGNs in Systematic Searches of Double-Peaked AGNs}\label{mergers}

As can be seen in Table 8, the three dual AGNs identified in this work share several common characteristics: they are located at similar redshifts ($z\sim0.13-0.2$), they exhibit peak velocity splittings in the range $370-470$ km s$^{-1}$ ($\sim420$ km s$^{-1}$), and the projected spatial separation of the two AGNs is in the range $0.62-1.55$ kpc ($\sim1$ kpc). Furthermore, in all three cases the dual AGNs are located in the plane of the galaxy (PA$_{\mathrm{radio}}\sim$PA$_{\mathrm{gal}}$). Finally, while J$1023+3243$ and J$1623+0808$ appear to be single non-interacting galaxies in SDSS images, J$1158+3231$ has a companion at a projected separation of $\sim10\arcsec$ ($\sim28$ kpc). However, there is no evidence for tidal tails or arcs in any of the three galaxies. 
All of these characteristics suggest that the three systems are in the same evolutionary stage of a merger event. 


The observational evidence for the evolution of merging SMBHs is still rather sparse. Most reported double SMBHs are quasar pairs with separations 10 to 100 kpc, which are in the initial stage of their merging interaction (e.g., Lotz et al. 2008, and references therein). 
Direct imaging of active SMBH pairs in spiral galaxies undergoing a major merger is exemplified by the dual AGN separated by $\sim6$ kpc in Mrk 266 (Mazzarella et al. 2012), while the double radio nuclei with 7.3 pc separation in the elliptical galaxy 0402+37921 suggest the late evolution of a major merger (Rodr\'igues et al. 2006).

With a separation of $\sim1$ kpc between the two active SMBHs, J$1023+3243$, J$1158+3231$ and J$1623+0808$ appear to be in an intermediate phase between Mrk 266 and 0402+37921.  
The host galaxies of J$1023+3243$ and J$1158+3231$ are regular spirals, and J$1623+0808$ appears to be an elliptical galaxy with a disk inside it, 
suggesting in the three cases the later stages of a major merger or a minor merger. 
This is supported by recent hydrodynamic simulations of galaxy mergers (Blecha et al. 2013), which suggest that dual AGNs with separations $0.1-1$ kpc, exhibit peak velocity splittings of $\sim500$ km s$^{-1}$ and are located in the plane of the galaxy.  All of these pieces of evidence indicate that the three dual AGNs identified in this work are in a short-lived stage (a few hundreds of Myr) when the nuclei are about to enter the final phase of coalescence.  


Along with J$1023+3243$, J$1158+3231$ and J$1623+0808$, the following double-peaked AGNs have also been confirmed to contain dual AGNs based on the detection of two compact sources of hard X-rays, or two compact flat-spectrum radio sources\footnote{
These dual AGN confirmations occurred through follow-up observations of double-peaked AGNs in SDSS spectra. Therefore, they are the result of systematic searches, rather than serendipitous discoveries like in NGC 6240.} (see also Section~\ref{limitations}): 
SDSS J110851+065901, SDSS J112659+294442, SDSS J114642+511029, and SDSS J150243+111557 (Fu et al. 2011a, Liu et al. 2013, Comerford et al. 2015). In all cases the SDSS images show disturbed morphologies and tidal features, suggesting on-going mergers (the major axis of the galaxy is not easily recognizable, except in J112659+294442). In J114642+511029 it is still possible to distinguish in $HST$ images the morphologies of the two interacting galaxies. While in J114642+511029 and J150243+111557 the projected spatial separation of the two nuclei is $\sim7$ kpc, in J110851+065901 and J112659+294442 the separation is $\sim2.5$ kpc. The velocity separation of the two peaks is $\sim300$ km s$^{-1}$ in J110851+065901, J112659+294442 and J114642+511029 (Liu et al. 2010a), and $\sim720$ km s$^{-1}$ in J$150243+111557$ (which is rather high for a separation of $\sim7$ kpc, Smith et al. 2010). All of these characteristics suggest that these galaxies belong to a similar evolutionary stage
as Mrk 266, in which the progenitor galaxies are still recognizable (or still interacting before they finally merge), and the separation of the two nuclei is in the range $1-10$ kpc. Note that J112659+294442 is a minor merger in progress (Comerford et al. 2015). 

\begin{table*}
\caption{Properties of the dual AGNs identified in this work.}
\begin{center}
{\small
\begin{tabular}{c c c c c c}
\hline
\hline \noalign{\smallskip} 
Source & $z$ & $\Delta\,v$\tablenotemark{a}  & $L_{\mathrm{[O III], blue}}$\tablenotemark{b} & $L_{\mathrm{[O III], red}}$\tablenotemark{c} & 
$D_{\mathrm{AGN}}$\tablenotemark{d}  \\
  & & (km s$^{-1}$) & ($\times10^{40}$ erg s$^{-1}$) & ($\times10^{40}$ erg s$^{-1}$) & (kpc) \\ 
\hline
SDSS J$102325+324348$ & $0.1271$ & $371\pm9$ & $6.85\pm0.86$ & $8.94\pm0.86$ & $1.02\pm0.16$ \\ 
SDSS J$115822+323102$ & $0.1658$ & $441\pm3$ & $59.31\pm2.22$ & $9.91\pm1.53$ & $0.62\pm0.11$ \\ 
SDSS J$162345+080851$ & $0.1992$ & $469\pm5$ & $17.05\pm1.90$ & $42.56\pm1.79$ & $1.55\pm0.17$ \\ 
\hline
\hline
\end{tabular}
}
\tablecomments{
$^\mathrm{a}$Velocity offset  between the two peaks of [O III] emission in the SDSS spectrum. \newline
$^\mathrm{b}$Estimated luminosity of the blueshifted component of the [O III] line in the SDSS spectrum. \newline
$^\mathrm{c}$Estimated luminosity of the redshifted component of the [O III] line in the SDSS spectrum. \newline
$^\mathrm{d}$Physical projected separation between the two compact radio sources detected in the total intensity images. This is consistent in all cases with $D_{\mathrm{radio}}$ at 8.5 GHz in Table 4. 
}
\label{table8}
\end{center}
\end{table*}

\subsection{Successes and Limitations of our Method for the Confirmation of Dual AGNs}\label{limitations}  

Double-peaked AGNs have gained popularity in the past few years because of the possibility of them being a signature for dual AGNs. Several candidates have been identified using optical images and spectroscopy. However, dual AGNs confirmations are scarce. This is mainly due to the fact that at optical/near-IR wavelengths the indicators of AGN-related activity are often ambiguous, or cannot be differentiated from the NLR of a single AGN. Indicators such as optical line ratios, optical colors, or the presence of highly ionized gas in most cases can be explained as part of the NLR of a single AGN. Photoionized gas by an AGN does not imply that there is an AGN at that specific location. Because of the proximity between the nuclei in all of these systems, it remains ambiguous whether there are one or two ionizing sources (i.e., AGNs) in each system (see e.g., Fu et al 2012 and references therein).  
For example, the two nuclei of the prototypical dual AGNs NGC 6240 (Komossa et al. 2003) are both heavily obscured in the optical, with emission-line ratios characteristic of LINERs (Lutz et al. 1999); another example is Mrk 266 (Mazzarella et al. 2012), whose northern nucleus is optically classified as a composite yet does contain a strong X-ray AGN.

Dual AGNs can only be confirmed by spatially resolving two AGN sources in X-ray or radio observations. Most of the confirmed cases have
been discovered serendipitously with $Chandra$ 
($\sim80\%$ of all of the dual AGNs confirmations). Despite this considerable success, detections of closely separated AGNs on scales of hundreds of parsecs to a few kpc are severely constrained by the limited angular resolution of X-ray telescopes ($Chandra$ has the highest resolution and can only resolve dual AGNs with separations $>0.5\arcsec$). Furthermore, the majority of SDSS galaxies with double-peaked narrow emission lines have estimated X-ray fluxes that are too faint to be detected with reasonable integration times on $Chandra$ (especially considering the evidence for a systematically smaller X-ray-to-[O III] luminosity ratio in dual AGNs, Liu et al. 2013, Comerford et al. 2015). Finally, as noted by Liu et al. (2013), X-ray confirmations of dual AGN candidates can be challenging and ambiguous, due to the complex nature of X-ray obscuration in mergers.


The VLA in its A-configuration can resolve double radio cores with separations $>0.2\arcsec$ and does not need a pre-selection based on optical AGN diagnostics to boost the success rate (e.g., high [O III] luminosities). As an example, 
each of the three dual AGNs discovered in this work would need more than 200 ks with $Chandra$ to detect $>10$ counts from each of the two AGNs in the system (Heckman et al. 2005, Table 8), and they would not be resolved, since the projected spatial separations of the two AGNs are less than $0.5\arcsec$ (Tables 4 and 8). However, in some cases, the VLA resolution is not enough to reveal the true radio nature of the sources, particularly in galaxies showing single radio cores or very small projected separations between compact sources (such as the case of J$1715+6008$). In these cases, the VLA images alone do not provide sufficient information to identify the origin of the double-peaked narrow emission lines. We used the long-slit spectra to distinguish between the different scenarios (Section 3.2). However, the crucial question that remains here is the following: Is there a radio structure within the VLA beam that could be responsible for the double-peaked, spatially-offset narrow AGN emission lines? Follow-up observations with VLBI techniques would be ideal to answer this question.

The second limitation to our approach is linked to weak radio sources, particularly at higher frequencies. For example, in J$0002+0045$ the ambiguities arise because there appear to be two compact radio cores in the total intensity image, but the properties of one them could not be measured at 11.5 GHz. A similar situation occurs in J$1112+2750$ and J$0916+2835$ (both galaxies classified as single AGNs, Table 2 and Section~\ref{winds}).

\section{Conclusions}\label{conclusions}

Systematic searches for double-peaked AGNs (e.g., Wang et al. 2009, Liu et al. 2010a, Barrows et al. 2013), and many follow-up observations (Liu et al. 2010b, Fu et al. 2011a, McGurk et al. 2011, Rosario et al. 2011, Shen et al. 2011,Tingay \& Wayth 2011, Comerford et al. 2011, Greene et al. 2012, Fu et al. 2012, Liu et al. 2012, Barrows et al. 2012, Liu et al. 2013, Comerford et al. 2015), have underscored the difficulty in pinpointing the origin of the double-peaked emission features. With the new VLA observations presented here, we have critically examined the nature of the ionizing sources of 18 optically identified double-peaked AGNs in SDSS. It is now clear that spatially resolved spectroscopy and high-resolution radio observations, when combined, can identify and distinguish between disk rotation, dual AGNs, and AGN outflows in these systems. Our main results and conclusions can be summarized as follows:

\begin{itemize}
	\item The nature of the radio emission in the 18 galaxies is varied. We detect two compact flat-spectrum radio cores with projected spatial separations on the sky between 0.6-1.6 kpc in three galaxies (J$1023+3243$, J$1158+3231$ and J$1623+0808$), which confirms the presence of dual AGNs in these objects. 	
	In the other 15 galaxies we identified six single AGN, four AGN with one-sided radio jets, three AGN with double-sided jets (CSO candidates), and two ambiguous cases. 
	\item The three confirmed dual AGNs in our sample account for only $\sim15\%$ of the double-peaked [O III] AGNs. Gas kinematics produce $\sim75\%$ of the double-peaked narrow emission lines, distributed in the following way: $\sim40\%$ AGN wind-driven outflows, $\sim30\%$ radio jet-driven outflows, and $5\%$ rotating NLRs. The remaining $10\%$ are ambiguous cases. 
	\item  In the two ambiguous cases (J$0002+0045$ and J$1027+3059$), the ambiguities arise because there appear to be two compact radio cores in the total intensity images, but the properties of one of the cores could not be measured either at 8.5 GHz or 11.5 GHz. Nevertheless, they still represent strong dual AGN candidates. Deeper VLA (or VLBI) observations would be helpful to reveal the true nature of these sources.
	\item The `S-symmetry' observed in the three CSO candidates in our sample might be associated with the presence of binary SMBHs with separations of a few hundred parsecs or less. Further observations with VLBI techniques would be required to test this hypothesis. 
	\item We have also investigated the size of the [O III] emitting region for each galaxy. We find that outflows cover the whole range of sizes (from 5-30 kpc), rotating disks are always compact ($<8$ kpc), and dual AGNs can be classified as moderately extended/extended (between 8-16 kpc). 
	\item We find that double-peaked AGNs also follow a size-luminosity relation in the form $R_{\mathrm{NLR}}\propto\,L_{\mathrm{[O III]}}^{1/2}$, steeper than the observed relationship between $R_{\mathrm{NLR}}$ and  $L_{\mathrm{[O III]}}$ for Seyfert 2 galaxies and radio-quiet obscured quasars. This suggests the presence of shocks and/or galaxy interactions. 
\end{itemize}



The results of this program demonstrate the efficacy of this new approach based on VLA imaging and optical long-slit spectroscopy for identifying the sources of double-peaked narrow emission lines in AGNs. Radio observations of more double-peaked AGNs (already in progress) will put our results on a firmer statistical ground. 
Building a much larger sample would also enable the exploration of radio properties of kpc-scale dual AGNs and radio-jet driven outflows as a function of separation and host galaxy properties. 

\acknowledgments


We thank J. Wrobel for help with the analysis of the VLA data; J. Darling, M. Eracleous, J. Newman and K. Nyland for interesting and valuable discussions; and the anonymous referee for helpful suggestions that have improved the clarity and strength of this work. 
Images from $HST$ Program 12521 (P.I. Liu) were used in this paper. 
The National Radio Astronomy Observatory is a facility of the National Science Foundation operated under cooperative
agreement by Associated Universities, Inc.


Facilities: \facility{VLA}, \facility{Shane (Kast Double spectrograph)}, \facility{Hale (Double spectrograph)}, \facility{MMT (Blue Channel spectrograph)}

\end{document}